\documentclass[useAMS,usenatbib]{mn2e}

\usepackage[dvips]{graphicx}
\usepackage{amsmath}
\usepackage{amssymb}

\newcommand\omicron{o}

\title{Carbon and hydrogen radio recombination lines from the cold clouds towards Cassiopeia A}
\author[J.~B.~R. Oonk et al.]{J.~B.~R. Oonk$^{1,2}$\thanks{E-mail:
oonk@astron.nl},  R.~J. van Weeren$^{3}$, P. Salas$^{1}$, F. Salgado$^{1}$, L.~K. Morabito$^{1}$, \newauthor M.~C. Toribio$^{1}$, A.~G.~G.~M. Tielens$^{1}$, H.~J.~A R\"ottgering$^{1}$\\ \\
$^{1}$Leiden Observatory, Leiden University, P.O. Box 9513, NL-2300 RA Leiden, The Netherlands\\
$^{2}$Netherlands Institute for Radio Astronomy (ASTRON), Postbus 2, 7990 AA Dwingeloo, The Netherlands\\
$^{3}$Harvard-Smithsonian Center for Astrophysics, 60 Garden Street, Cambridge, MA 02138, USA\\}
\begin{document}

\date{Accepted 0000. Received 0000; in original form 0000}

\pagerange{\pageref{firstpage}--\pageref{lastpage}} \pubyear{0000}

\maketitle

\label{firstpage}

\begin{abstract}
We use the Low Frequency Array to perform a systematic high spectral resolution investigation of the low-frequency 33-78~MHz spectrum along the line of sight to Cassiopeia A. We complement this with a 304-386~MHz Westerbork Synthesis Radio telescope observation. In this first paper we focus on the carbon radio recombination lines. 

We detect Cn$\alpha$ lines at -47 and -38~km~s$^{-1}$ in absorption for quantum numbers n=438-584 and in emission for n=257-278 with high signal to noise. These lines are associated with cold clouds in the Perseus spiral arm component. Hn$\alpha$ lines are detected in emission for n=257-278. In addition, we also detect Cn$\alpha$ lines at 0~km~s$^{-1}$ associated with the Orion arm.

We analyze the optical depth of these transitions and their line width. Our models show that the carbon line components in the Perseus arm are best fit with an electron temperature 85~K and an electron density 0.04~cm$^{-3}$ and can be constrained to within 15\%. The electron pressure is constrained to within 20\%. We argue that much of these carbon radio recombination lines arise in the CO-dark surface layers of molecular clouds where most of the carbon is ionized but hydrogen has made the transition from atomic to molecular. The hydrogen lines are clearly associated with the carbon line emitting clouds, but the low-frequency upper limits indicate that they likely do not trace the same gas. Combining the hydrogen and carbon results we arrive at a firm lower limit to the cosmic ray ionization rate of 2.5$\times$10$^{-18}$~s$^{-1}$, but the actual value is likely much larger.
\end{abstract}

\begin{keywords}
ISM: clouds -- radio lines : ISM -- ISM: individual objects: Cassiopeia A
\end{keywords}

\section{Introduction}
Spectral lines resulting from atoms recombining with electrons in diffuse, ionised plasma are potentially important diagnostics to probe the conditions of the emitting and absorbing gas. At low quantum numbers recombination gives rise to the well-known optical and near-infrared recombination lines. At higher quantum numbers the energy spacing between subsequent quantum levels decreases and a recombination line transition will emit a photon at radio wavelengths. The associated lines for high quantum numbers are therefore called Radio Recombination lines (RRL).

RRLs can be used to obtain a wealth of information on the properties of the emitting gas \citep*[e.g.][]{Go09}. Emitting in the radio domain, these lines are unbiased by dust obscuration. At low radio frequencies ($<$1~GHz) RRLs provide us with a method to obtain temperature, density and ionisation of the cold neutral medium \citep*[e.g.][]{Sh75,Sh76,Sh76b,So87,Pa89,Oo15}. This information can not easily be obtained by other means, such as 21~cm neutral hydrogen measurements. 

Our own Galaxy is a copious emitter of RRLs. These come in two flavours; (i) {warm, dense gas RRLs (sometimes in the literature referred to as classical or discrete RRLs), and (ii) diffuse RRLs. Warm, dense gas RRLs are associated with common HII regions and dense photodissociation regions (PDRs). Here recombination lines from hydrogen, helium and carbon are seen \citep*[e.g.][]{Pa67,Ro89,Na94,Wy97,Ka98b,Ko05}. These are predominantly observed at frequencies above 1~GHz as the hydrogen and helium recombination lines trace the warm (T$_{e}\sim$10$^{4}$~K), high-density (n$_{e}>$10~cm$^{-3}$) fully ionized gas, while the carbon recombination lines trace the warm ($\sim$500~K), dense (n$_{H}\sim$10$^{3-6}$~cm$^{-3}$) gas in PDRs bordering compact HII regions or associated with reflection nebulae.

Diffuse RRLs, are associated with the lower density, colder interstellar medium \citep[e.g.][Salgado et al. 2016a,b]{Ko80,Bl80,Pa89,Er95,Ka01,Oo14,Oo15}. Here typically only recombination lines from carbon (CRRL) are observed as the ionisation levels are too low to produce observable hydrogen and helium lines. Diffuse CRRLs are best observed at radio frequencies below 1~GHz due to stimulated emission and absorption. Whereas warm, dense gas RRLs have been studied in great detail, the properties of the cold gas associated with diffuse RRLs in our Galaxy is not well determined. Furthermore these diffuse RRLs provide us with a complementary tracer of the physical conditions in the cold neutral medium (CNM) of the Milky Way.

So far, the only line of sight studied in some detail for CRRLs is the one towards the bright supernova remnant Cassiopeia~A (Cas~A). This is because Cas~A is one of the brightest low frequency radio sources in the sky \citep*[e.g.][]{Ba65,Br68,Pa68}, thus serving as a dominating background source, and it shows relatively bright CRRLs in both emission and absorption \citep[e.g.][]{Pa89}. The sightline towards Cas~A cuts through the Milky Way at a galactic longitude l~=~112 degrees and latitude b~=~-2 degrees. Cas~A itself is located in the 2nd Galactic quadrant in the Perseus spiral arm at a distance 3.4~kpc from the Sun and at a Galactocentric radius of about 10.5~kpc. HI 21~cm line observations show both emission and absorption in the range from +30 to -120~km~s$^{-1}$ \citep*[e.g.][]{Me75,Bi91,Sc97}. The strongest HI absorption features are found around -47, -38 and 0~km~s$^{-1}$. These are associated with foreground clouds in the Perseus (-47 and -38~km~s$^{-1}$) arm and the Orion (0~km~s$^{-1}$) spur. These clouds are also observed in other cold gas tracers such as CI (492~GHz), CO (J=2-1), OH, H$_{2}$CO and NH$_{3}$ \citep*[e.g][]{Ki14,Mo06,Li99,An94,Ba84,Ba83,dJ78}.

The CRRLs along this line of sight have been studied by e.g. \citet{Pa89}, \citet{An94}, \citet{Ka98}, \citet{Go09} and \citet{As13}. It was found that the CRRLs show a good correspondence with HI 21~cm absorption, both in velocity and in distribution, and somewhat less good with CO (J=2-1) emission. Attempts were made at modeling the CRRL line properties as a function of quantum number n to derive the physical parameters of the CRRL emitting gas \citep[e.g.][]{Pa89,Ka98}. These investigations showed that the velocity averaged CRRLs from the Perseus arm favor warmer, lower density models (electron temperature T$_{e}\sim$75~K and electron density n$_{e}\sim$0.02~cm$^{-3}$) over colder and denser models (T$_{e}\sim$30~K and n$_{e}\sim$~0.05). However, a clear discrimination was not possible, as the models presented by \citet{Pa89} and \citet{Ka98} were not able to simultaneously fit the $>$150~MHz CRRL emission in combination with the $<$150~MHz CRRL absorption. Furthermore the results they obtain from the line widths differed from those obtained from the optical depths. This is likely due to a number of factors; (i) the limited validity of the CRRL models used at the time, (ii) the difficulty in determining the total line profile at low frequencies and (iii) averaging over multiple velocity components with potentially different physical gas conditions. 

Here we revisit the Cas~A CRRL line of sight making use of new high quality, high spectral resolution interferometric data from the Low Frequency Array (LOFAR) and the Westerbork Synthesis Radio telescope (WSRT) to perform a velocity resolved study of the CRRLs. In addition we make use of our new CRRL models (Salgado et al. 2016a,b) to derive the physical conditions of the associated gas. 

The data presented in this paper are part of the LOFAR Cas~A Spectral Survey (LCASS). This ongoing survey is a dedicated (Directors discretionary time) programme aimed at performing the first detailed low frequency, high spectral resolution, interferometric LOFAR study of the cold interstellar medium along the well studied Cas~A line of sight. The survey, when complete, will cover the entire frequency range accessible to LOFAR, i.e. 10-80~MHz, 110-190~MHz and 200-250~MHz, with a velocity resolution ranging from 11~km~s$^{-1}$ at the lowest frequency to 1~km~s$^{-1}$ at the highest frequency. The primary goal of LCASS is to provide a high signal-to-noise spectral line atlas and spatial maps of low-frequency CRRLs. In addition we will also search the low frequency spectrum for line emission and absorption from other atoms and molecules (e.g. OH and NO). The search for non-RRL lines will be presented in a future paper. In this paper we present the LCASS RRL results for the 33-78~MHz range.

The paper is structured as follows. In Sect.~\ref{s_obs_red} we discuss the first LOFAR observations taken for the LCASS survey and the WSRT observations. The results are presented in Sect.~\ref{s_results}. In Sect.~\ref{s_rrl_model} we fit our new CRRL models to the observations. We discuss the results in Sect.~\ref{s_discuss} and present our conclusions in Sect.~\ref{s_conclus}.

\section[]{Observations and Reduction}\label{s_obs_red}

\subsection{LOFAR (33-78 MHz)}
We obtained LOFAR LBA observations on December 27, 2011, from 10:00 to 20:30 UTC and October 31, 2013, from 11:55 to 21:55 UTC (Table~\ref{t_obs}). For each of these observations LOFAR's multi-beaming capabilities were used to place half of the available instantaneous bandwidth on Cas~A, 122 subbands each 0.1953 MHz wide, totaling about 24~MHz. The other 122 subbands were pointed towards Cyg A which served as a calibrator \citep{Oo14}. For both pointing centers we obtained complete frequency coverage between 33--57~MHz (2011) and 55--78~MHz (2013), although about two dozen subbands were corrupted due to issues with the LOFAR offline storage system. The LBA\_OUTER configuration was used for the LBA stations. In this case 48 (of 96) LBA antennas are used, located mostly in the outer part of the 87~m diameter stations. All four linear correlation products were recorded (XX, XY, YX, YX) and each sub-band was subdivided into 512 frequency channels. The integration time was 2~s. 

For the 2011 observation we used nine remote and 22 core stations providing baselines between 90~m and 80~km. A first step in the data processing is the automatic flagging of radio frequency interference (RFI) with the AOFlagger \citep{Of10}. We slightly decreased the default flagging thresholds to avoid flagging good data as Cas~A and Cyg~A have flux densities $> 10^4$~Jy in the observed frequency range. Typically, a few percent of the data was flagged due to RFI. After flagging we averaged the data to 4~s time steps to reduce its size. The data were calibrated with the BlackBoard Selfcal (BBS) software system \citep{Pa09}. We used high-resolution 10 arcsec clean components models of Cas~A (Fig.\ref{f_cas_model}) and Cyg~A (McKean et al. in prep.) for calibration. These models were obtained from previous LOFAR observations around 70~MHz.

As the observations are pointed towards the two brightest sources on the sky, either Cas A or Cyg A  dominates the total signal on all baselines. We made a copy of the 4~s data and averaged further down from 512 to 1 channel per subband. We then obtained gain solutions for all four correlations with BBS on a 4~s timescale. We assume the sources are unpolarized over the observed frequency range. The gain solutions found were then applied to the 512~frequency channel data and a final round of flagging was carried out with the AOFlagger. Channel cubes were made with \emph{casapy}, imaging and cleaning each channel individually. The first 25 and last 25 channels of the data were ignored as they are too noisy. We chose Briggs weighting \citep{Br95} with a robust value of 0.5 to create images with a resolution ranging between 30x40~arcsec$^{2}$ and 40x60~arcsec$^{2}$. We then convolved all images from all subbands to a common resolution of 45$\times$65~arcsec$^{2}$ and created an image cube for each subband.

For the 2013 observation we used 24 core stations and 14 remote stations. The data reduction was performed in the same way as for the 2011 data set. Due to a clock problem only 18 core stations were used in the final analysis of the observations. We chose Briggs weighting with a robust value of 0.5 to create images with a resolution ranging between 215x255~arcsec$^{2}$ and 310x360~arcsec$^{2}$, the lower resolution being a consequence of using only 18 core stations. We then convolved all images from all subbands to a common resolution of 350$\times$400~arcsec$^{2}$ and created an image cube for each subband.

\subsection{WSRT (304-386 MHz)}
We obtained WSRT P-band observations on January 28, 2012, from 08:29 to 20:28 UTC (Table~\ref{t_obs}). The observations were carried out in the Maxi-short configuration and Doppler tracking was turned off. We observed in frequency switching mode with 10~s sampling and 6 simultaneous 1.25~MHz subbands (IVC bands) each having 2048 channels (using re-circulation) and 2 polarizations (XX,YY). Each subband is centered near an expected CRRL line frequency and we observe 3 CRRL lines per setup where each line is covered twice with a different central frequency setting (typically offset by 0.3-0.4~MHz). In total we specified 8 spectral setups of 6 subbands and cover a total of 24 lines (all 22 $\alpha$ lines within the observed frequency range and 2 additional $\beta$ lines). Each spectral setup observed 10 minutes on-source and then changed to the next setup and after the last setup is done we return to the first setup. This way we cycled the spectral setups through the full 12~h observation and created similar UV and time coverage for each subband. The total observing time per subband was about 1.5~h.

The first step in the data reduction was the automatic flagging of RFI with the AOFlagger. A dedicated WSRT P-band flagging strategy was developed for this purpose. The data were then averaged down and calibrated with CASA \citep{Mc07} using the high resolution 10\arcsec clean components model of Cas~A (Fig.\ref{f_cas_model}). Gain solutions were obtained for both polarizations on  10~s timescale. The gain solutions found were then applied to the 2048~frequency channel data and a final round of flagging was carried out with the AOFlagger. Channel cubes were made with casapy, imaging and cleaning each channel individually. We chose Briggs weighting (Briggs 1995) with a robust value of 0.5 to create images with a resolution ranging between 60x65~arcsec$^{2}$ and 75x95~arcsec$^{2}$. We then convolved all images from all subbands to a common resolution of 80$\times$100~arcsec$^{2}$ and created an image cube for each subband.

\subsection{Spectral analysis and line stacking}\label{s_spc_stack}
From the WSRT 300-390~MHz and LBA 33-57~MHz image cubes we extract spatially integrated on-source spectra from an 8$\times$8~arcmin$^{2}$ aperture centred on Cas~A. For the LBA 55-78 MHz range we used a 14$\times$14~arcmin$^{2}$ aperture centred on Cas~A. The larger aperture for the 55-78 MHz data is necessary given the lower spatial resolution of this observation. The CRRL $\alpha$ ($\Delta$n=1) lines are clearly visible in the individual spectra for both LOFAR and WSRT. We investigated the overlapping subbands containing CRRLs between the two LOFAR observations and we find that the line profile parameters agree within errors.

We removed the edge channels from the spectra and fitted a low order polynomial to the line free channels to estimate the continuum. We then convert the spectra to optical depth units following \citet{Oo14}. The typical spectral RMS per channel, in optical depth units, are 5$\times$10$^{-4}$, 6$\times$10$^{-4}$ and 4$\times$10$^{-4}$ for LOFAR 55-78 MHz, 33-57 MHz and WSRT respectively. The peak signal to noise for individual $\alpha$ lines in the LOFAR spectra is typically 7--9 for the -47 component and 2--4 for the -38~km~s$^{-1}$ component. Similarly for WSRT the typical peak signal to noise is about 5 for the -47 and 1.5 for the -38 component.

We perform line spectra stacking to the obtain higher signal to noise line profiles necessary to measure the line optical depth and line full width at half maximum (FWHM) of each of the velocity components. The initial stacking of line spectra was performed as described in \citet{Oo14}. These stacked spectra contain on average 6 $\alpha$ lines in the WSRT range and 10 to 20 stacked lines in the LBA range (Table~\ref{t_alln}). Stacking these lines over small changes in n is allowed as we expect the RRLs to change slowly and smoothly in the observed frequency ranges.

The stacked WSRT spectra are shown in Fig.~\ref{f_app_wsrt_substack} and the line profiles are found to be Gaussian. The three velocity components at -47, -38 and 0~km~s$^{-1}$ are narrow enough and sufficiently separated that we can fit them well with individual Gaussians. There is an additional line feature at -55~km~s$^{-1}$, likely due to RRLs from sulphur, that we blank prior to fitting the CRRLs. The results from the Gaussian fits are summarized in Table~\ref{t_wsrt_cia}. A stacked spectrum containing all $\alpha$ lines in the WSRT range is shown in Fig.~\ref{f_app_wsrt_full_stack}. For this stacked spectrum we also fit the -55~km~s$^{-1}$ feature after subtracting the CRRL fits from the spectrum. The results from the Gaussian fits to this stacked spectrum are summarized in Table~\ref{t_wsrt_meas_intg}. This latter spectrum is only used for our investigation of the gas ionization using the hydrogen lines in Sect.~\ref{s_ionisation}.

For the LBA spectra there is strong line broadening with decreasing frequency, as expected from the Stark effect (e.g. Sect.~\ref{s_line_width}). This leads to significant line blending for the -47 and -38~km~s$^{-1}$ components. Furthermore, this causes the line profiles to have Voigt profiles instead of Gaussian profiles. Voigt profiles are characterized by broad line wings. These broad wings can be affected by residuals in the continuum as well as nearby lines, such as CRRL $\beta$ ($\Delta$n=2), $\gamma$ ($\Delta$n=3) and $\delta$ ($\Delta$n=4) lines. In order to obtain the best possible fit we performed a different stacking procedure to optimize the continuum baseline in the LBA line spectra. This procedure is described in detail in Salas et al. (in prep.) and is similar to the procedure used by \citet{St07} for their low-frequency CRRL spectra.

Here we shortly summarize the main aspects of this procedure. For each stack the spectra are first searched for CRRL $\alpha$ and $\beta$ lines that are unblended with other lines and unaffected by RFI and bandpass roll-off. These lines were then stacked and fitted with Voigt profiles to create template line profiles. These profiles are subtracted from each (unstacked) line spectrum and the residual spectra are stacked to search and fit for the CRRL $\gamma$ lines. These $\gamma$ lines are then also subtracted from the individual line spectra and one final stack is performed to search and fit for the $\delta$ lines and also remove those from the individual line spectra. The residual baseline in the individual line spectra, where all $\alpha$, $\beta$, $\gamma$ and $\delta$ lines have the been removed, are then baseline corrected by a polynomial of order 0. Using the baseline corrected spectra we repeated the stack of the lines. 

This procedure is repeated 5 times on the LBA spectra by increasing the polynomial order by one in each step. Finally, stacked spectra with only one kind of transition are obtained by removing the corresponding best fit Voigt profiles from
the individual spectra. The $\alpha$ line spectra resulting from this procedure are shown in Figs.~\ref{f_app_lba_substack_1} and \ref{f_app_lba_substack_2}. The baseline corrected, stacked line spectra are then fitted with Voigt profiles for each of the three velocity components. The results are summarized in Table~\ref{t_lba_cia}. The line broadening continues to increase towards lower frequencies and below 40~MHz (n=550) it is no longer possible to robustly disentangle the -47 and -38~km~s$^{-1}$ components.

In Salas et al. (in prep.) we have verified this baseline correction procedure with detailed simulated LOFAR spectra that have the same resolution and noise characteristics as our observations and are processed in the same manner. In the LBA range studied here this baseline correction procedure provides only a minor improvement in our recovery of the line profiles. However, this procedure becomes increasingly important at frequencies below 33~MHz. This spectral line stacking procedure with baseline correction processing, as described above, is only used for the CRRL stacks as it removes all unidentified line features. Upper limits to the undetected HRRLs, Table~\ref{t_lba_hrrl}, are obtained from stacked spectra without these corrections applied. In this paper we will only discuss the $\alpha$ lines for carbon and hydrogen. The $\beta$, $\gamma$ and $\delta$ lines will be discussed in a future paper.

\section[]{Results}\label{s_results}
The WSRT spectra clearly show that there are at least 3 CRRL velocity components in emission at -47, -38 and 0~km~s$^{-1}$ relative to the local standard of rest (LSR), see Figs.~\ref{f_app_wsrt_substack} and \ref{f_app_wsrt_full_stack}. This is consistent with previous measurements by e.g. \citet[][hereafter PAE89]{Pa89} and \citet[][hereafter KAP98]{Ka98}. In addition the WSRT spectra also show evidence for the presence of a weak line near -55~km~s$^{-1}$. It is most prominent at the highest frequency stack, but observed at the 3$\sigma$ level in all stacks (e.g. Fig.~\ref{f_app_wsrt_substack}). A similar feature is not seen in HI absorption or CO emission spectra \citep[e.g.][]{Bi91,Mo06,Ki14} which makes it unlikely that it is associated with CRRL from another cold cloud at this velocity. A more likely explanation is that this feature is associated with RRL emission from sulphur (SRRL) and/or other elements at higher atomic numbers (sometimes referred to as XRRL or ZRRL) from the -47~km~s$^{-1}$ cloud.

Here we will focus on the CRRL and HRRL emission from the -47 and -38~km~s$^{-1}$ velocity components that arise in clouds situated in the Perseus arm. The -38 feature is rather broad and the WSRT spectra show tentative evidence that this feature may in fact consist of more than one component. This can also be seen in the assymetric line profiles of CO emission \citep{Ki14,Mo06,Li99} and HI 21~cm absorption \citep{Bi91,Sc97}. In particular the CO (J=2-1) emission spectrum from \citet{Li99} and \citet{Mo06} shows two emission peaks, one at -40 and one at -36~km~s$^{-1}$. For our current analysis we will treat the -38 feature as a single component.

The WSRT data also show the presence of hydrogen RRLs (HRRL). These lines are shifted by +149.4~km~s$^{-1}$ in the stacked CRRL spectrum (Fig.~\ref{f_app_wsrt_full_stack}). This difference corresponds exactly to the difference in rest frequencies between the CRRL and HRRL lines. This is only the 2nd detection of HRRLs along this line of sight and our detection is at a lower frequency than the first detection at 420~MHz by \citet[][; hereafter SS10]{So10}. This is the first interferometric detection and the first time that also the weaker -38~km~s$^{-1}$ component is detected. For the even weaker Orion spur CRRL component we did not detect the corresponding HRRLs. If the hydrogen to carbon RRL ratio in the Orion spur is similar to that in the Perseus arm components then this non-detection reflects that even higher signal to noise measurements are necessary to detect the HRRLs for the Orion component.

The HRRL line for the -47~km~s$^{-1}$ component has the same width as the corresponding CRRL line, indicating that they both arise in the same gas. However, we notice that the HRRL line for the -38~km~s$^{-1}$ component has a significantly narrower width than the corresponding CRRL line. This may constitute additional evidence that the -38~km~s$^{-1}$ CRRL component consists of multiple velocity components and that the HRRLs only trace part of this.

The observed CRRL LBA spectra also show 3 CRRL velocity components, but they appear in absorption, see Figs.~\ref{f_app_lba_substack_1} and \ref{f_app_lba_substack_2}. The relative LBA line centroids at -47, -38 and 0~km~s$^{-1}$ (relative to LSR) are consistent with the CRRL emission from WSRT, however the linewidths in the LBA range are observed to strongly increase in width with decreasing frequency (i.e. increasing n). This is expected and in Sect.~\ref{s_line_width} we will model this with collisional and radiation broadening. For n$>$550 the increase in linewidth of the -47 and -38 component becomes so large that deblending these components becomes degenerate and as such we will only consider the n$<$550 measurements in our analysis. The good correspondence between the absorption in the LBA and the emission in WSRT (Fig.~\ref{f_wsrt_lbahgh}) indicates that all of the CRRL emitting gas is situated in front of CasA. HRRLs were not detected in the LBA spectra and 3$\sigma$ upper limits are presented in Table~\ref{t_lba_hrrl}.

\section[]{RRL modeling}\label{s_rrl_model}
The CRRL $\alpha$ ($\Delta$n=1) transition spectra for both WSRT and the LBA allow us to distinguish at least 3 velocity components at -47, -38 and 0~km~s$^{-1}$. The measurements for the 0~km~s$^{-1}$ component, associated with the Orion spur, will be treated in a forthcoming paper. Here we will focus on interpreting the CRRL $\alpha$ transitions ($\Delta$n=1) emission from the -47 and -38~km~s$^{-1}$ components that are known to arise from clouds in the Perseus spiral arm (e.g. PAE89). We will use the high signal to noise stacked line spectra for our analysis (Figs.~\ref{f_app_lba_substack_1}, \ref{f_app_lba_substack_2} and \ref{f_app_wsrt_substack}). These CRRL spectra provide us with two observables to be modeled, (i) the line width and (ii) the optical depth. Both depend on the physical conditions of the emitting gas. In following we will first model the line width and then the optical depth. We will use the new CRRL models from \citet[][hereafter S16a,b]{S16a,S16b}. We find that combining the constraints from both observables is useful to disentangle the degeneracy between electron temperature, electron density and radiation field (see Sect.~\ref{s_combine_constraint}).

We adopt a homogeneous slab with constant density (n$_{e}$), temperature (T$_{e}$) and size (L$_{\rm{CII}}$). Equivalently, we could have selected the emission measure instead of the size of the cloud. This slab is bathed in an isotropic radiation field characterized by T$_{\rm{R}}~\propto~\lambda^{\beta}$ with $\beta$=2.6 and normalized in terms of T$_{\rm{R,100}}$ the value of T$_{\rm{R}}$ at 100~MHz. The level populations are fully described by the atomic physics involved (S16a,b). Following \citet{Se59} and \citet{Br70}, we define the departure coefficient b$_{\rm{n}}$ of level n as the weighted sum of the b$_{\rm{nl}}$ values (S16a). Here b$_{\rm{nl}}$=N$_{\rm{nl}}$/N$_{\rm{nl}}$(LTE), with N$_{\rm{nl}}$(LTE) the level populations as given by the Saha-Boltzmann equation under local thermal equilibrium conditions. We also introduce $\beta_{\rm{n}}$ as the correction factor for stimulated emission following \citet{Br72} and S16a. The b$_{\rm{n}}$ and $\beta_{\rm{n}}$ fully determine the optical depth given by a set of physical conditions T$_{e}$, n$_{e}$ and T$_{\rm{R}}$. In principle, the intensity also depends on the temperature of the background source but, in our analysis, we will assume that the intensity scales directly with the optical depth. S16b have shown that this is in general the case for quantum levels above 200 and we verify this aposteriori in Sect.~\ref{s_combine_constraint}. In our analysis, we use the models developed by S16a,b which solve the statistical equilibrium equations for arbitrary n and $\ell$ levels in terms of the b$_{\rm{n}}$ and $\beta_{\rm{n}}$ as a function T$_{e}$, n$_{e}$ and T$_{\rm{R}}$ fully self consistently. The gas density and temperature, together with the radiation temperature, also set the radiation and pressure line broadening at high n (S16a,b). We assume a filling factor of 1 for the CRRL emitting gas and address this point further in Sect.\ref{s_compare_prev}.

Previous studies of CRRL lines have been analyzed following the models by \citet{Wa82} and \citet{Po92}. Because of the limited computer power available at that time, considerable approximations had to be made and these models are not appropriate for quantitative analysis (S16a). In particular, as compared to previous models, we note that the b$_{\rm{n}}$ values for the models by S16a approach 1 faster at high n, i.e. n$\gtrsim$500-600, and as such the corresponding b$_{\rm{n}}\times\beta_{\rm{n}}$ values are smaller and have significantly flatter behavior at these high n.

We have used the models by S16a to create a detailed (T$_{e}$, n$_{e}$, T$_{\rm{R}}$) model grid for fitting our measurements. This grid is sampled in steps of 5~K for T$_{e}$ in the range 10-150~K and in steps of 0.005~cm$^{-3}$ for n$_{e}$ in the range 0.01 to 0.11~cm$^{-3}$. In addition this grid is computed with a non-zero Galactic power law radiation field T$_{\rm{R}}$ that is specified as above. For all T$_{e}$ and n$_{e}$ combinations in our grid we computed the departure coefficients for 5 values of T$_{\rm{R,100}}$ i.e. T$_{\rm{R,100}}$=800, 1200, 1400, 1600 and 2000~K. This range in T$_{\rm{R,100}}$ covers the range in expected values for the radio continuum temperature from the Milky Way along the line of sight to Cas~A \citep[e.g.][]{Ha82,La70,Ro99}. We will fit our data using a chi-squared method on this grid while adjusting the size of the cloud. 

\subsection[]{Line width}\label{s_line_width}
The measured FWHM line width for CRRLs depends on the instrumental resolution and three physical broadening terms; (i) Doppler, (ii) collisional and (iii) radiation broadening \citep[e.g. S16b;][]{Sh75}. The Doppler term is independent of frequency and set by the turbulence of the gas. The Doppler broadening is determined from the WSRT data and literature data at higher frequencies. We find that our WSRT data in the 300-390~MHz range are consistent with previously measured linewidth at 560~MHz by KAP98 and show no evidence for line broadening. From this we conclude that the line width in this range is dominated by Doppler broadening and derive a Doppler line width of 3.4~km~s$^{-1}$ for the -47~km~s$^{-1}$ component and 6.8~km~s$^{-1}$ for the -38~km~s$^{-1}$ component (Table~\ref{t_wsrt_cia} and \ref{t_wsrt_meas_intg}).

Whereas the WSRT data show constant line widths, dominated by Doppler broadening, the LBA data shows a clear increase in the FWHM with increasing n, as expected from pressure and radiation broadening. Having determined the Doppler contribution, which is modeled as a Gaussian, to the line profile we proceed to analyze the remaining line broadening in terms of pressure and radiation broadening. Both of these terms are modeled as Lorentzians and in order to properly recover the Lorentzian line wings we use the high signal to noise line profiles obtained from our line stacking procedure (Sect.~\ref{s_spc_stack}). We find that the line widths for the highest frequency stack (n=438) in the LBA are still consistent with pure Doppler broadening, see also Fig.~\ref{f_wsrt_lbahgh}, after which the Lorentzian contribution is found to increase and dominates the overall line profile for n$>$540. The total Lorentzian contribution to the line profile as a function of n in the LBA range is presented in Table~\ref{t_lba_cia}.

Collisional and radiation broadening are manifestations of the Stark effect and depend on the physical conditions of the gas and its environment in terms of the electron temperature T$_{e}$, the electron density n$_{e}$ and the ambient radiation field T$_{\rm{R}}$ \citep[e.g. S16b;][]{Go09,Wa82,Br77,Sh75}. We use the formulation by S16b for both collisional and radiation broadening. Here we parameterize the radiation field in terms of a Galactic power law radiation field as defined in Sect.~\ref{s_rrl_model}.

We calculate the total required Lorentzian contribution to the line width in terms of T$_{\rm{R,100}}$ as function of T$_{e}$ and n$_{e}$ in the ranges T$_{e}$=10-310~K and n$_{e}$=0.005-0.5~cm$^{-3}$. To avoid uncertainties from severe line blending we use only the data below n=550. The allowed parameter space is presented in Fig.~\ref{f_crrl_fwhm_n}. Within the allowed region of parameter space we find that there is no strong preference for a particular set of physical conditions, i.e. all allowed combinations provide similarly good (i.e. reduced $\chi^{2}\sim$1) fits to the data. The non-allowed, i.e. blanked, area in Fig.~\ref{f_crrl_fwhm_n} shows the region of parameter space that would overestimate the observed line widths beyond the measurement errors. 

For both velocity components constant T$_{\rm{R,100}}$ values trace smooth curves in (T$_{e}$, n$_{e}$) space and curves of increasing T$_{\rm{R,100}}$ move the allowed set of physical conditions to lower T$_{e}$ and n$_{e}$ values. Both pressure and radiation broadening have a very similar dependence on quantum number and hence fitting the data is degenerate \citep[e.g. S16b;][]{Sh75}. As both give rise to Lorentzian profiles, their contributions to the line broadening are additive. For any given radiation field, we can then subtract the radiation broadening component and derive the contribution required from pressure broadening. That will leave us with a relationship between the density and temperature of the gas, which is n$_{e}\times$T$_{e}^{-0.5}$ (S16b). Figs.~\ref{f_crrl_fwhm_n}, \ref{f_47_crrl_comb} and \ref{f_38_crrl_comb} illustrate this for a number of different values for T$_{\rm{R,100}}$. With increasing radiation field temperature, this relationship shifts down. As these figures demonstrate, a large fraction of the parameter space is not allowed.

We see in Sect.~\ref{s_optical_depth}, that the opposite behavior is found upon modeling the integrated optical depth and therefore the constraints obtained from modeling the line width provide us with useful information that is able to break the degeneracy between the different physical parameters. Finally we note that not only do we find that the -38~km~s$^{-1}$ component has a broader Doppler contribution than the -47~km~s$^{-1}$ component, but also its Lorentzian contribution increases slightly faster with increasing n than the -47~km~s$^{-1}$ component. This may indicate that the physical conditions differ between the -47 and the -38 component, or alternatively that the -38 feature consists of multiple velocity components with potentially different physical conditions.

\subsection[]{Optical depth}\label{s_optical_depth}
The measured CRRL integrated optical depth depends on T$_{e}$, n$_{e}$, T$_{\rm{R}}$ and L$_{\rm{CII}}$ or equivalently, the emission measure EM$_{\rm{CII}}$=n$_{e}\times$n$_{\rm{CII}}\times$L$_{\rm{CII}}$ \citep[e.g. S16a; PAE89;][]{Sh75}. CRRLs at low frequencies arise from quantum levels n that are not in local thermal equilibrium (LTE) and as such we need to evaluate the departure coefficients b$_{\rm{n}}$ and $\beta_{\rm{n}}$. These departure coeficients also depend on T$_{e}$, n$_{e}$ and T$_{\rm{R}}$ (e.g. S16a and references therein). Here we have used the models by S16a to create a detailed (T$_{e}$, n$_{e}$, T$_{\rm{R}}$) model grid for fitting our measurements as described in Sect.~\ref{s_rrl_model}. 

In the following we perform a grid search to find the best (T$_{e}$, n$_{e}$) model describing the data, for each T$_{\rm{R,100}}$ value, by optimizing the value of L$_{\rm{CII}}$. In Sect.~\ref{s_results} we showed that the WSRT emission and LBA absorption spectra are consistent in terms of the observed absolute and relative velocity centroids of the different CRRL components (e.g. Fig.~\ref{f_wsrt_lbahgh}). In addition we found in Sect.~\ref{s_line_width} that the observed line widths can be modelled with single physical models across the entire range in n from 225 to 550. This indicates that it is likely that all of the emitting gas observed from the -47 and -38~km~s$^{-1}$ Perseus arm components is situated in front of Cas~A and hence can be modeled across the entire range in n with a single value of L$_{\rm{CII}}$ for emission and absorption.

We have selected the n=301 (-47~km~s$^{-1}$ component only) and n=309 CRRL measurements from PAE89 and the n=225 CRRL measurement from KAP98 to complement our WSRT and LOFAR measurements upon fitting the models. The other data presented by these authors overlaps with our measurements and is consistent with these. We have not added these other measurements as they are either unresolved in velocity or have much lower signal to noise as compared to our measurements. In addition we want to avoid systematic uncertainties by adding measurements obtained with very different observing parameters. Finally we will only consider measurements with n in the range 225 to 550 as for n$>$550 it is not possible to reliably decompose the -38 and -47~km~s$^{-1}$ components. We exclude n$<$225, because we calculate the integrated optical depths using equation (6) in S16b. This equation is identical to what has been used in previous studies (e.g. PAE89,KAP98). However, as pointed out by S16b this equation is not exact and in the case of a strong background source the exact radiative transfer equation should be solved; i.e., equation (1) in S16b should be used for sufficiently low n levels. In the case of Cas~A, we find that for n$<$225 the differences between the approximate and exact solution start to become significant, i.e. greater than 1~percent, and hence we only consider measurements below this n value (see Sect.~\ref{s_model_uncertainties}).

The results of our grid search, in terms of the 1, 2 and 3 sigma confidence limits are shown by the red, blue and green colored boxes in Fig.~\ref{f_47_crrl_comb} and \ref{f_38_crrl_comb} for the different values of T$_{\rm{R,100}}$. For the -47~km~s$^{-1}$ component we find no significant difference in the quality (i.e. reduced $\chi^{2}\sim$1-2) of the best fit for each of the 5 different T$_{\rm{R,100}}$ values, but there is a systematic trend in that higher T$_{\rm{R,100}}$ values require (slightly) higher values of T$_{e}$ and n$_{e}$, see Fig.~\ref{f_47_crrl_comb}. This trend is the opposite of what we observed for our line width modeling in Sect.~\ref{s_line_width} and we will discuss this in more detail in Sect.~\ref{s_combine_constraint}. Considering the entire parameter space probed by our model grid for the -47~km~s$^{-1}$ component we find that only a very limited region in parameter space is allowed and that we can constrain T$_{e}$ to be in the range 80-90~K and n$_{e}$ to be in the range 0.035-0.045~cm$^{-3}$. However, the T$_{\rm{R,100}}$ value is not well constrained by only considering the integrated optical depth.

For the -38 component we find similar trends as for the -47 component in that we obtain equally good fits for each of the 5 different T$_{\rm{R,100}}$ values and higher T$_{\rm{R,100}}$ values require (slightly) higher combinations of T$_{e}$ and n$_{e}$ to be allowed, see Fig.~\ref{f_38_crrl_comb}. Given the lower signal-to-noise of the -38 component the parameter space is slightly larger than for the -47 component. In particular we see that a larger range in both T$_{e}$ and n$_{e}$ is allowed. However, this allowed range opens up along a particular curve in (T$_{e}$, n$_{e}$)-space that traces an almost constant (electron) pressure p$_{e}$. We will discuss this curve in more detail in Sect.~\ref{s_pressure}. Considering only the integrated optical depth models we constrain T$_{e}$ to be in the range 70-85~K and n$_{e}$ in the range 0.030-0.045~cm$^{-3}$. The model fits for the -38 component are not as good as for the -47 component and have a reduced $\chi^{2}\sim$4-5. In particular we note that the increase in integrated optical depth in the LBA range increases faster with increasing n than expected from the best-fit model.

\subsection[]{Combining Line width and Optical depth}\label{s_combine_constraint}
In Figs.~\ref{f_47_crrl_comb} and  \ref{f_38_crrl_comb} we have shown the independent constraints from both the line width and the optical depth in a single plot of T$_{e}$ vs. n$_{e}$ as a function of T$_{\rm{R,100}}$. We note that the constraints from the integrated optical depth are much more stringent than those obtained from the line width. However, as stated above, the integrated optical depth does not constrain T$_{\rm{R,100}}$ well. The line width does not provide very good constraints on either T$_{e}$, n$_{e}$ or T$_{\rm{R,100}}$, but we find that the allowed models for the line width move in an opposite direction in (T$_{e}$, n$_{e}$)-space as compared to the models for the integrated optical depth upon changing T$_{\rm{R,100}}$. Therefore the combination of the integrated optical depth and line width does allow us to constrain T$_{\rm{R,100}}$ and thus constrain T$_{e}$ and n$_{e}$ better.

Considering both measurements we find that the electron temperature and density for both compents can be constrained to better than 15 percent at the 1$\sigma$ confidence level, see Table~\ref{t_crrl_results}. We find very similar conditions, T$_{e}\sim$85~K and n$_{e}$=0.04~cm$^{-3}$, for both components. The background Galactic radiation field is marginally higher for the -38, as compared to the -47, component, but both are well within the range measured along the line of sight to Cas~A \citep[e.g.][]{Ha82,La70,Ro99}. The line of sight path length L$_{\rm{CII}}$ for these physical conditions is found to be about 35 and 19~pc for the -47 and -38 components respectively. Here we have assumed that the singly ionized carbon n$_{\rm{CII}}$ density is equal to the free electron density n$_{e}$. The contribution of ionized hydrogen to n$_{e}$, based on the HRRL measurements, is found to be of the order of a few percent and will be discussed in Sect.~\ref{s_ionisation}. This path length implies CII column densities N$_{\rm{CII}}$ of 4$\times$10$^{18}$ and 2$\times$10$^{18}$~cm$^{-2}$ and CII emission measures EM$_{\rm{CII}}$ of 0.06 and 0.03~cm$^{-6}$~pc, respectively. The results from the combined line width and optical depth constraints are summarized in Table~\ref{t_crrl_results} and shown in Figs.~\ref{f_47_bf} and \ref{f_38_bf}. The CII emission measure of both components, in terms of the associated free-free absorption, does not violate the observed low-frequency radio continuum turnover of Cas~A \citep*[e.g.][]{Ka95}.

\subsection[]{Comparison to earlier studies}\label{s_compare_prev}
PAE89 previously performed a velocity resolved CRRL investigation of the -47 and -38~km~s$^{-1}$ clouds observed along the line of sight to Cas~A. As discussed in Sect.~\ref{s_results} our measurements and those of PAE89 are broadly consistent, albeit that PAE89 have considerably higher scatter and larger errors in their measurements as compared to our data set. 

For the line width modelling both PAE89 and we considered a purely Galactic radiation field with a dilution factor of 1. PAE89 considered T$_{\rm{R,100}}$=800~K and Doppler widths of 6.7 and 5.9~km~s$^{-1}$ for the -47 and -38~km~s$^{-1}$ respectively. The emission line measurements by PAE89 and this work show that the Doppler width for the -47~km~s$^{-1}$ component is overestimated in the modeling by PAE89. Another difference in computing the line width between PAE89 and our work is that PAE89 use the \citet{Sh75} formulation, whereas we use the updated formulation by S16b. Comparing these we find that S16b predicts lower linewidths for both collisional and radiation broadening for a given combination of T$_{e}$, n$_{e}$ and T$_{\rm{R}}$. For a Galactic radiation field with $\beta$=2.6 the line FWHM from radiation broadening predicted by S16b is 24\% lower and this difference is independent of quantum number n. The lower values obtained from S16b are due to a more accurate approximation of the oscillator strength by S16b as compared to \citet{Sh75}. In terms of pressure broadening the difference is largest at low n and decreases towards higher n. This results from a more detailed fitting to the collisional cross-sections in S16b as compared to \citet{Sh75}. For the physical conditions of interest here S16b predicts a FWHM which is about 15\% lower at n=200 and this decreases to 5\% lower at n=600, as compared to \citep{Sh75}. The total difference in the FWHM thus amounts to about 30\% upon considering both broadening terms and, in combination with the different Doppler width, explains the higher Galactic T$_{\rm{R,100}}$ values that we obtain here. 

PAE89 also performed an investigation of the integrated CRRL optical depths. Our WSRT optical depth measurements for both the -47 and -38~km~s$^{-1}$ component at n$\sim$267 are about 20 percent larger than the measurements of PAE89. Given the lower signal to noise and narrow bandwidth of the spectra in PAE89 this difference can likely be attributed to uncertainties in the line fitting by PAE89 as well differences in observing parameters and their calibration. The LBA optical depths at n=438 to n=448 are slightly larger than the measurements of PAE89, but consistent within errors. At n$\sim$500 the sum of our measurements for the -38 and -47 components agrees with the sum of the PAE89 measurements, however the optical depths assigned to each of the two components by PAE89 differs from ours. We find that the lower signal to noise and the factor 2 lower velocity resolution of PAE89 likely makes their deblending more uncertain. At even higher n, i.e. n$\sim$575, we find that the sum of our -38 and -47 measurements agrees well with the high signal to noise measurement at 34~MHz by KAP98.

PAE89 performed a fit to their integrated CRRL optical depth measurements. They, and subsequent work by \citet{Pa94} and KAP98, used the models by \citet{Wa82} as the basis for their fitting procedures. PAE89 concluded that these models did not provide a satisfactory fit to their data. In particular they noted that their models could not simultaneously fit both the low n emission and high n absorption measurements and as such they could not discriminate between cold (T$_{e}\sim$20~K), high density (n$_{e}\sim$0.3~cm$^{-3}$) models and warm (T$_{e}\sim$100~K), lower density (n$_{e}\sim$0.05~cm$^{-3}$) models. As described in Sect.~\ref{s_rrl_model} the models by \citet{Wa82} are not appropriate for a quantitative analysis of the CRRL data. 

Most of the previous CRRL optical depth measurements agree well with the new high signal to noise measurements presented here. However, a few of the previous measurements are at variance with our data and given their lower signal to noise we deem those measurements to be unreliable and excluded them from our analysis. We find that the models by S16a,b are able to fit both the CRRL line width and the optical depth measurements well over the range n=225-550 for a single set of physical parameters (Sect.~\ref{s_combine_constraint} and Table~\ref{t_crrl_results}). 

In our above modeling of the optical depth, we have assumed a filling factor of 1 for the CRRL emitting gas. Previous studies by PAE89 and KAP98 also assumed this. Given that HI 21~cm absorption for both the -38 and -47~km~s$^{-1}$ components extends over the entire face of the remnant \citep{Bi91,Sc97} and the good spatial correlation of CRRL with HI 21~cm absorption \citep{An94} we believe that this is a reasonable choice. However, there is spatial structure within the gas on smaller scales and we will discuss this in more detail in Sect.~\ref{s_model_uncertainties}.

\section[]{Discussion}\label{s_discuss}
Since the study by PAE89 there have been two other detailed investigations of the CRRLs along the Cas~A line of sight by \citet{Pa94} and KAP98. Neither of these studies were able to fit both the integrated optical depths and the line widths of the CRRLs for a single set of physical parameters. This is likely due to the reasons outlined in Sect.~\ref{s_compare_prev}. With the more detailed models by S16a,b we have shown in Sect.~\ref{s_combine_constraint} that we are now able to obtain a satisfactory fit to both the linewidth and the optical depths.

The derived electron densities of $\sim$0.04~cm$^{-3}$ translate into a density of hydrogen nuclei of 286~cm$^{-3}$, adopting the gas phase carbon abundance of 1.4$\times$10$^{-4}$ \citep{Ca96,So97}. This density is high compared to the typical density of diffuse clouds traced by the 21~cm HI line \citep[n$_{\rm{H}}\sim$50~cm$^{-3}$; ][]{Wo03}. However, they are quite comparable to densities derived for the well studied diffuse sight-lines of $\zeta$~Oph, $\zeta$~Per, and $\omicron$~Per where simultaneous modeling of the observations of many atomic and molecular species result in densities in the range of 200--400~cm$^{-3}$ \citep*[e.g.][]{Di86,Vi88}.

The derived temperature of 85~K is well in the range of temperatures derived by the same studies as well as temperatures derived from HI 21~cm line studies \citep*[e.g.][]{Me75,Di82,Di90,He03}. The derived thermal pressure of 2$\times$10$^{4}$~K~cm$^{-3}$ agrees, of course, well with those measured towards $\zeta$~Oph, $\zeta$~Per, and $\omicron$~Per but they are an order of magnitude larger than the typical gas pressures derived from UV absorption lines measuring the CI fine structure line excitation ($\sim$4$\times$10$^{3}$~K~cm$^{-3}$; e.g. \citet{Je11}). 

Finally, our sizes are comparable to the sizes of typical HI diffuse clouds \citep[$\sim$10~pc, e.g.][]{Sp78} but the derived column densities are an order of magnitude higher. These differences may merely reflect that we are probing clouds in the Perseus and Orion spiral arms rather than diffuse clouds in the local Solar neighborhood. Specifically, the clouds probed by the CRRLs may be the atomic/CO-dark surfaces of molecular clouds. As our clouds are situated in spiral arms and molecular clouds have been detected along the same sight line at the same velocities \citep{Bi86,Li99,Mo06,Ki14}, this is quite reasonable. We note that it is difficult to keep CII ionized over the large path lengths inferred here, unless we are viewing the Cas~A clouds from a preferred angle. It is therefore likely that the clouds we are observing are sheet-like structures. Filamentary spurs springing off spiral arms are a common characteristic of a turbulent ISM and these are dominated by atomic and CO-dark molecular gas \citep{Sm14}.

The inferred pressures are also in line with measured pressures of molecular cloud surfaces \citep[e.g.][]{Bli80,He01}. Moreover, the very similar sight-lines towards $\zeta$~Per and $\omicron$~Per traverse the atomic HI surfaces associated with the B3/B4/B5 molecular clouds \citep*{An92}. On the other hand, the high HI 21~cm column densities and the implied high visual extinction of these clouds (N$_{HI}\sim$2$\times$10$^{22}$~cm$^{-2}$; A$_{V}\sim$10 magnitudes) are very high for atomic clouds (N$_{HI}\lesssim$~2$\times$10$^{21}$~cm$^{-2}$; A$_{V}\lesssim$~1; e.g. \citet{Di90}). Indeed, for visual extinctions in excess of 1 magnitude, much of the gas phase carbon is expected to be in CO (and to a lesser extent in CI) rather than CII \citep[e.g.][]{Di86,Vi88}. This may just be a matter of geometry as the clouds probed by the CRRL may be arranged into thin sheets as is common for large scale HI structures \citep{Sp75,Sa74,He84}. Future observations will be instrumental in settling the relationship between the CRRL gas and the molecular clouds in the direction of Cas~A.

\subsection[]{Gas pressure}\label{s_pressure}
In the previous sections we found that we can constrain the electron temperature and density for both components to better than 15 percent. If we consider T$_{e}$ and n$_{e}$ to be independent variables then this translates to an uncertainty of up to 20 percent for the electron pressure at the 1$\sigma$ confidence level. One would expect the uncertainty on the pressure to increase at the 2 and 3$\sigma$ levels, however this is not observed in Figs.~\ref{f_47_crrl_comb} and  \ref{f_38_crrl_comb}. These figures show that T$_{e}$ and n$_{e}$ are not independent and that the electron pressure remains to be constrained to better than 20 percent at the 3$\sigma$ confidence level. This tight relation ship between T$_{e}$ and n$_{e}$, along lines of almost constant pressure, is driven primarily by our constraints on the range in n where the CRRL emission to absorption transition takes place (Fig.~\ref{f_nt_emabsp}).

Although the electron pressure itself is well constrained by our measurements there is still an uncertainty in translating this to a thermal gas pressure due to the unknown abundance of carbon in these clouds. The typical gas phase carbon abundance in the interstellar medium has been found to be [C/H]$\sim$1.4$\times$10$^{-4}$ \citep[e.g.][]{Ca96}. It is possible to derive the carbon abundance [C/H] by comparing CRRL measurements with HI 21~cm absorption measurements, under the assumption that the lines arise from the same gas and that within this gas singly ionised carbon is the dominant state of carbon, i.e. N(CII)/N(HI)$\approx$N(C)/N(H)=[C/H] \citep[e.g.][PAE89]{Oo15}. 

HI 21~cm absorption measurements have been carried out by e.g., \citet{Me75,Bi91,Sc97}. These studies find three main HI absorbing components at -47, -38 and 0~km~s$^{-1}$. These HI 21~cm components, in terms of velocity and FWHM, are in good agreement with our CRRL measurements and indicate that it is likely that the HI 21~cm absorption and our CRRL measurements spatially trace the same gas structures. A similar conclusion was reached by KAP98 (and references therein). However, these studies also find that the HI 21~cm absorption measurements are heavily saturated for the -47~km~s$^{-1}$ component and mildly saturated for the -38~km~s$^{-1}$ component. This means that from these measurements we can only obtain a lower limit to true cold HI column density. Following \citet{Sc97} find N(HI)~$>$~4$\times$10$^{21}$~cm$^{-2}$ and N(HI)~$>$~3$\times$10$^{21}$~cm$^{-2}$ for the -47 and -38 component respectively. An upperlimit to the HI column density can be obtained by considering the total hydrogen column N(H) from X-ray observations. The maximum total N(H) reported by \citet{Hw12} is about 3.5$\times$10$^{22}$~cm$^{-2}$. From these measurements we can only constrain the carbon abundance to be in the range [C/H]=1.3-11$\times$10$^{-4}$ for the -47 component and [C/H]=0.7-7.7$\times$10$^{-4}$ for the -38 component.

This range in [C/H] allowed by the RRL and HI measurements is large and especially the upper limits of this range are likely not realistic. We therefore adopt the gas phase abundance by \citet{Ca96} here and then we find a thermal pressure p$_{thermal}$/k=2.4$\times$10$^{4}$~K~cm$^{-3}$. This is consistent with the model prediction of p$_{thermal}$/k$\sim$1$\times$10$^{4}$~K~cm$^{-3}$ by \citet[][their Fig.~7]{Wo03} for densities n$_{H}\sim$286~cm$^{-3}$ at a Galactocentric radius of about 10.5~kpc. It is also consistent with measurements and simulations of the external pressure for molecular clouds in the Galactic midplane \citep[e.g.][]{Bli80,He01,Gi16}.

The turbulent pressure in the gas can be obtained from the observed FWHM of the turbulent (Doppler) line broadening. We calculate the turbulent velocity dispersion as $\sigma_{turbulent}$=3$^{0.5}\times$FWHM/2.355 and find p$_{turbulent}$/k=1.9$\times$10$^{5}$~K~cm$^{-3}$ for the -47 component and p$_{turbulent}$/k=7.6$\times$10$^{5}$~K~cm$^{-3}$ for the -38 component. We thus find that the turbulent pressure dominates over the thermal pressure in both clouds as is typical in the interstellar medium of the Milky Way \citep[e.g.][ and references therein]{Wo03}. 

Another contribution to the pressure in the clouds are magnetic fields. Using OH measurements \citet{He86} infer an average magnetic field strength B$\sim$8~$\mu$G. HI 21~cm Zeeman splitting measurements by \citet{Sc86} indicate an average parallel component of the magnetic field B$_{\rm{||}}\sim$20~$\mu$G. The reason for the difference between these measurements is not clear. This range in measurements indicates a magnetic field pressure p$_{magnetic}$/k=(1.8-4.5)$\times$10$^{4}$~K~cm$^{-3}$. This shows that the magnetic field pressure is of the same order as the thermal pressure and less than the turbulent pressure, although the HI measurements do allow for higher magnetic field pressures that may rival the turbulent pressure.

\subsection[]{Hydrogen RRLs}\label{s_ionisation}
We have detected two HRRL emission lines in our stacked WSRT spectrum, see Fig.~\ref{f_app_wsrt_full_stack}. This is the second detection of HRRLs along this line of sight. The first detection of HRRL emission at n=250 and associated with the -47 component was made by SS10 at 420 MHz. Our n=267 detection of this component at 344~MHz agrees well with theirs. For the -38 component, SS10 do not claim a detection, but they do see a feature in their spectrum. We confirm this feature here at the 4$\sigma$ level. The main peak of our -38 HRRL line agrees well with the feature seen in the spectrum by SS10, however we do find that our line is narrower than theirs. Our -47 HRRL line is also narrower than the detection reported by SS10, but not as much as for the -38 component. Line broadening of RRLs typically increases with increasing n, as discussed above, and this therefore does not explain the difference. The difference between our spectrum and SS10 is close to the noise level of the SS10 spectrum and their broader feature may be caused by a noise peak. Deeper observations are necessary to investigate this further.

The HRRL velocity centroids in our WSRT spectrum, in the rest-frame for hydrogen, are at -47.4 and -38.6~km~s$^{-1}$. This agrees very well with the corresponding CRRL lines and shows that both the carbon and hydrogen lines likely originate in the same clouds, see Fig.~\ref{f_wsrt_hcrrl_overlay}. The line width of the CRRL and HRRL lines agree well for the -47 component, but the same is not true for the -38 component where our HRRL feature is significantly narrower than the corresponding CRRL feature (Fig.~\ref{f_wsrt_hcrrl_overlay}). This could indicate that only part of the -38 CRRL emitting gas is traced by the corresponding HRRL line. 

\subsubsection{Gas ionization}
Low frequency HRRLs can be used to trace the hydrogen ionization rates in the CRRL emitting clouds, if they trace the same gas. For the -47 component this is possible and two methods have been proposed to derive the total hydrogen ionization rate from HRRL measurements. The first method, proposed by Shaver (1976), uses the ratio between the HI 21~cm and the HRRL integrated optical depths. It is important to use the integrated and not the peak optical depths here, as the HRRL emission at sufficiently low frequencies can be affected by line broadening due to the Stark effect \citep[e.g. S16b; ][]{Sh75}. In the previous section we saw that the HI~21cm absorption measurements are saturated and thus underestimate the true HI optical depth. This method therefore only provides an upperlimit to the ionization rate $\zeta_{\rm{H}}$(-47)~$<$~5$\times$10$^{-17}$~s$^{-1}$. 

A second method to obtain $\zeta_{\rm{H}}$ was first proposed by \citet{So87}, later modified by \citet{So90} and SS10, and uses the ratio between the CRRL and the HRRL integrated optical depths. For convenience we repeat the equation here;

\begin{eqnarray}
\begin{aligned}
\zeta_{\rm{H}}&=&\alpha_{\rm{H}^(2)}[C/H]\left(\frac{\tau_{\rm{H_{n}}}\Delta\nu_{\rm{H_{n}}}}{\tau_{\rm{C_{n}}}\Delta\nu_{\rm{C_{n}}}}\right)\left(\frac{\Phi_{\rm{2}}n_{e}}{T_{e}^{0.5}}\right)\left(\frac{(b_{\rm{n}}\beta_{\rm{n}})_{\rm{C}}}{(b_{\rm{n}}\beta_{\rm{n}})_{\rm{H}}}\right)
\end{aligned}
\end{eqnarray}

Here we have reformulated it in terms of the hydrogen (gas phase) recombination coefficient $\alpha^{(2)}$=$\alpha_{\rm{H}}^{(2)}\Phi_{\rm{2}}T_{\rm{e}}^{-0.5}$, where captures to the n=1 level are excluded \citep{Sp82}. $\alpha_{\rm{H}}^{(2)}$=2.06$\times$10$^{-11}$ and $\Phi_{\rm{2}}$ is an integral over gaunt factors that is tabulated in table 5.2 in \citet{Sp82}. We use the b$_{\rm{n}}$ and $\beta_{\rm{n}}$ values from S16a. For the gas phase abundance of carbon we find that $\zeta_{\rm{H}}$(-47)=3$\times$10$^{-18}$~s$^{-1}$. Following PAE89 we can also use the ratio of the HRRL to CRRL optical depth to estimate the volume density ratio of ionised carbon to protons (n$_{\rm{CII}}$/n$_{p}$) and electrons (n$_{\rm{CII}}$/n$_{e}$). Using our WSRT measurements at n=267 and the b$_{\rm{n}}$, $\beta_{\rm{n}}$ values from S16a we find n$_{\rm{CII}}$/n$_{p}$~=~15.5 and n$_{\rm{CII}}$/n$_{e}$~=~0.94 This shows that 94 percent of the free electrons are donated by carbon and that the proton to electron ratio is n$_{\rm{p}}$/n$_{e}$=0.06. 

The HRRL to CRRL hydrogen ionization rate we derive for the -47 component is an order of magnitude lower than reported by SS10. We find that this is entirely due to a difference in models used to compute the departure coefficients. We have also computed $\zeta_{\rm{H}}$(-47) from their n=250 measurement using our models and find that it is consistent with our measurement at n=267. 

Our hydrogen ionization rate for the -47 component is a factor of a few lower than the modeled cosmic ray ionization rate ($\zeta_{\rm{CR,10~kpc}}\sim$1$\times$10$^{-17}$~s$^{-1}$) and the EUV plus X-ray ionization rate ($\zeta_{\rm{XR,10~kpc}}\sim$8$\times$10$^{-18}$~s$^{-1}$) at a Galactocentric radius R$_{c}$=10~kpc \citep{Wo03}. Recent measurements of the cosmic ray ionization rate in diffuse clouds, through H$_{3}^{+}$ observations, by \citet{In12} have shown that the cosmic ray ionization rate is higher by almost an order of magnitude (i.e. $\zeta_{\rm{C}R}\sim$10$^{-16}$-10$^{-15}$~s$^{-1}$) for total hydrogen column densities N$_{\rm{H}}\leq$10$^{22}$~cm$^{-2}$. This would be inconsistent with our measurement. However, Indriolo et al. made only a few measurements in the range l=90-130~deg, most of which are upper limits, and for N$_{\rm{H}}>$2$\times$10$^{22}$~cm$^{-2}$ they find a steep drop to about 2$\times$10$^{-17}$~s$^{-1}$ for the cosmic ray ionization rate. 

\subsubsection {Grain neutralization}
\citet{Li03} has pointed out that interactions with small grains may lower the RRL derived ionization rates. In this case our RRL measurement would provide a firm lower limit to the actual ionization rate. Atomic ions can not just be neutralized through gas phase recombination, but gas-grain interactions may also play a role through a process best called mutual neutralization \citep[e.g.][]{Dr87,LD88,Le88,Li03,Wo08}. In the literature this process is sometimes also referred to as grain neutralization. In this process large molecules and very small grains, or polycyclic aromatic hydrocarbons (PAH), can become negative charged by acquiring (free) electrons \citep{Ba94}. The models by \citet{Li03} show that mutual neutralization becomes increasingly important for regions that are more heavily shielded. At the gas density observed here these models show that, for the case of depleted gas phase abundances ([C/H]=1.4$\times$10$^{-4}$), mutual neutralization can decrease the proton density by up to a factor of $\sim$100 relative to gas phase recombination. While this process will affect the neutral atomic carbon abundance \citep{Ba98}, C$^{+}$ remains the main reservoir of carbon and the dominant free electron donor and \citet{Li03} show that at these gas densities the free electron density is very close to the singly ionized carbon density.

In equation 1 only gas phase recombination is taken in to account \citep[e.g.][]{So87}. Below we have modified this equation by specifically including a factor $\eta_{GN}$ which is the ratio of the total (gas and grain) recombination rate to the gas phase recombination rate (i.e. this is the inverse of value plotted in Fig.~1 in \citet{Li03});

\begin{eqnarray}
\begin{aligned}
\zeta_{\rm{H}}&=&\eta_{GN}\alpha_{\rm{H}^(2)}[C/H]\left(\frac{\tau_{\rm{H_{n}}}\Delta\nu_{\rm{H_{n}}}}{\tau_{\rm{C_{n}}}\Delta\nu_{\rm{C_{n}}}}\right)\left(\frac{\Phi_{\rm{2}}n_{e}}{T_{e}^{0.5}}\right)\left(\frac{(b_{\rm{n}}\beta_{\rm{n}})_{\rm{C}}}{(b_{\rm{n}}\beta_{\rm{n}})_{\rm{H}}}\right)
\end{aligned}
\end{eqnarray}

We find that upon including grain neutralization that the above hydrogen ionization rate, as derived from the HRRL over CRRL ratio, may need to be corrected upwards to a maximum of $\zeta_{\rm{H}}$(-47)$\sim$3$\times$10$^{-16}$~s$^{-1}$, if this process dominates within the environment in which the RRL emission arises. Similarly the upperlimit on $\zeta_{\rm{H}}$(-47) derived from the HRRL over HI 21~cm ratio would also have to be corrected upwards. Currently there are no direct observational constraints on the importance of mutual neutralization for the clouds studied here.

\subsubsection{HRRL vs. CRRL gas conditions}
In the above calculations for the hydrogen ionization rate we have implicitly assumed that the CRRLs and HRRLs trace gas with the same physical conditions and the same geometry. To test this we plot in Fig.~\ref{f_wsrt_hcrrl spectra} the HRRL optical depth as a function of n with the HRRL models from S16a for the best-fit CRRL conditions overplotted. In order to fit the model to the data we allow the ionized hydrogen column density N$_{\rm{HII}}$=n$_{\rm{HII}}$$\times$L$_{\rm{HII}}$ to differ from the singly ionized carbon column density N$_{\rm{CII}}$=n$_{\rm{CII}}$$\times$L$_{\rm{CII}}$. If we demand HII and CII to follow the same geometry, i.e. L$_{HII}$=L$_{\rm{CII}}$, then the HRRL to CRRL optical depth ratio directly traces n$_{HII}$/n$_{\rm{CII}}$ i.e. the relative fraction of free electrons donated by hydrogen and carbon. In agreement with above we find n$_{\rm{HII}}$/n$_{\rm{CII}}$=0.06.

Fig.~\ref{f_wsrt_hcrrl spectra} shows that for the -47 component the HRRL model with the best-fit CRRL gas conditions can match the high frequency detections by us, \citet{Oo15} and SS10, but not our low frequency 3$\sigma$ LBA HRRL limits. This indicates that for the -47 component the CRRL and HRRL do not trace the same gas. To determine whether any other HRRL model can fit the HRRL measurements we ran a full HRRL grid search with the same parameters as done for the CRRLs above. We constrain the model fits by demanding that the model must be able reproduce all the detections within their 1$\sigma$ errors and provide low frequency values that fall within the 3$\sigma$ LBA limits. For the -47 component we find that only some significantly colder and higher density models, i.e. T$_{e}$=30-50~K, n$_{e}$=0.065-0.11~cm$^{-3}$ and EM$_{H}$=0.0007-0.002, are able to fit the measurements. 

This comparison of the HRRL and CRRL models suggests that the HRRL and CRRL emission may not originate in the same region with the same physical conditions. Alternatively, it is possible that the CRRLs probe CO-dark molecular gas (Sect.~\ref{s_discuss}). In such gas carbon is ionized but hydrogen is in molecular form, a large contribution to the CRRL emission from CO-dark gas may then be expected. Spatially resolved investigations at higher frequencies (1-8~GHz) of dense PDRs have shown that in some cases narrow HRRL and CRRL lines do not trace the same gas \citep[e.g.][ and references therein]{Ro89}. Deeper and higher spatial resolution measurements are necessary to investigate this in cold clouds at low frequencies (Salas et al. in prep.).

\subsection{CRRL modeling uncertainties}\label{s_model_uncertainties}
The modeling approach that we used here has a number of uncertainties that we will discuss below.

\subsubsection{Spatial structure}
We have used a homogenous slab model with a constant temperature and density and a filling factor of unity to model the clouds in front of Cas~A. This approach is likely a simplification of the true situation for these clouds. There have been a number of studies performed to explore the physical conditions and spatial structure in the cool atomic and molecular gas of the Cas~A clouds \citep[e.g.][]{Ki14,Mo06,Li99,Sc97,An94,Bi91,Ba84,Ba83,dJ78}. These studies show: (i) that the cool gas extends over most of the face of the remnant, (ii) that the cool gas has spatial structure on arcmin-scales and peaks in a filamentary-like structure over the southern part of the remnant running from the western hotspot to the south-east of the remnant, and (iii) the -38 component is more localized and shifted towards the western part of the remnant, as compared to -47 component, although this could be a sensitivity issue. The observed CRRL emission is consistent with this arcmin-scale picture and \citet{An94} argue that on 2.7$\times$2.4 arcmin$^{2}$ scales the CRRL emission shows a better spatial and velocity correlation with HI 21~cm absorption than with $^{12}$CO emission, although they also note that there are differences between all three tracers. 

Observations of $^{12}$CO on larger scales show that the southern filamentary structure seen in absorption against Cas~A is likely the edge of a larger molecular cloud complex that is located to the south of the remnant \citep[e.g.][]{Ki14,Ba84}. Higher spatial resolution studies have been performed in HI 21~cm absorption and in several molecular tracers \citep[e.g.][]{Bi91,Ba83,Ba84}. The HI 21~cm study by \citet{Bi91}, with a spatial resolution of 7 arcsec, shows that the HI absorption extends over the entire face remnant and that there is great complexity within it. They find several morphological structures identified based on velocity cuts and describe these structures as filaments, arcs, loops and irregular with typical sizes varying from about 0.5 to 3 arcmin. Molecular absorption line studies are consistent with these sizes. Additional substructure may be present and \citet{Ba84} based on the low filling factor (f=0.04) they derive for one of their ammonia clouds find that the molecular clouds may contain structure as small as 8 arcsec. They interprete this as evidence for dense, high pressure clumps with physical sizes of 0.12~pc and pressures p$_{thermal}$/k$\sim$7$\times$10$^{4}$~K~cm$^{-3}$.

Given the available data we recognize that our CRRL measurements likely measure an average which is weighted towards the southern part of the remnant. By assuming a unity filling factor we average over a range in physical conditions that may be in present in the gas. On arcmin-scales the total, i.e. the sum of the -38 and -47 component, CRRL optical depth does not vary by more than about a factor of 2 across the remnant \citep{An94}. \citet{As13} shows that, except for an overall increase in optical depth, there are no significant differences for the total CRRL integrated optical depth as function of n (albeit over a very limited range in n) observed against the western hotspot as compared to the sum over the entire remnant.

The low-frequency CRRLs observed here likely trace gas intermediate between the atomic and molecular phase (e.g. Sect~\ref{s_discuss}) and a substantial filling factor seems reasonable. Low filling factors for the CRRL emitting gas would imply even larger, and likely unrealistic, CII column densities given the required beam dilution corrections. The physical conditions derived here may therefore represent a reasonable average of the true physical conditions. If higher spatial resolution observations find evindence for low filling factors of the CRRL emitting gas then this can effect the observed optical depths as a function of n as the underlying continuum emission from Cas~A is also highly structured in terms of the surface brightness and spectral index \citep[e.g.][]{Ka95}. In particular this may effect the measurements for the potentially more localized -38 component. A more detailed analysis of the spatial structure for the CRRLs along the line of sight to Cas~A will be presented elsewhere (Salas et al. in prep.).

\subsubsection{Radiation field \& radiative transfer}
In this work we have assumed that the clouds are embedded within an isotropic radiation field. At the distance of of Cas A an angular scale of 1~arcmin corresponds to 1~pc. Over these small spatial scales we feel it is reasonable to expect that the low-frequency Milky Way synchrotron radiation field is isotropic. Comparison of our LOFAR interferometric data with existing single dish radio surveys \citep[e.g.][]{Ha82,La70,Ro99} shows that the bulk ($\gtrsim$80\%) of the low-frequency Milky Way synchrotron emission is emitted on scales larger than about 10 wavelengths, or equivalently on angular scales larger than of about 6 degrees. The existing low-frequency Milky Way single dish measurements do not allow for a diffuse Milky Way contribution that is less than about 800~K at 100~MHz \citep[e.g.][]{Ha82,La70,Ro99} and our measured linewidths do not allow for a total contribution greater than about 1600~K at 100~MHz. Given the above constraints we have investigated lowering the Milky Way contribution and adding a contribution from Cas~A to the radiation field. For the data presented here we find that this does not provide a better fit.

As discussed in S16b the equations used here to calculate the line widths and optical depths are approximations. These approximations are valid at sufficiently high n levels. This is especially true in the case of a bright background source such as Cas~A. Following the prescription in S16b we have, for a few representative cases, calculated the exact solutions to the radiative transfer equation for the CRRLs along the line of sight to Cas~A. We find that the differences between the approximate and exact solutions in the case of Cas~A are less than one percent for n$>$225 and as such do not affect our results.

\subsubsection{Collision rates}
Our CRRL models depend on the ratio of the $^{2}$P$_{3/2}$ over the $^{2}$P$_{1/2}$ population, also referred to as the R value \citep{Wa82,Po92}. We have used the formulation by \citet{S16a} who give R in terms of collisions with electrons and hydrogen atoms using the rates given in \citet*{Ti85} and \citet*{La77}. The \citet{S16a} model does not include collisions with molecular hydrogen. We have investigated adding these collisions, using the rate given in \citet{Ti85}, and we find a small systematic effect on the computed b$_{n}\beta_{n}$ values that are shifted by about 5\%, for n$_{\rm{H_{2}}}$=n$_{\rm{HI}}$, towards lower values upon including these collisions, see Fig.~\ref{f_cmp_bbn_h2}.

A more important effect on the CRRL models, in terms of b$_{n}\beta_{n}$, is the choice in $\ell$-changing collisions. S16a uses rates by \citet{Vr12}. These rates were the most up to date rates for the $\ell$-changing collisions at the time of writing. S16a also investigated rates by \citet{Pe64}. These rates give qualitatively similar b$_{n}\beta_{n}$ behavior, but quantitatively the results can differ substantially. The \citet{Pe64} rates are not suitable for this work as they diverge for densities n$_{e}<$0.05~cm$^{-3}$. Very recently there has been renewed interest in these rates by \citet*{Gu16a} and \citet*{Gu16b}. In a future paper we will address the influence of the $\ell$-changing rates on the Cas~A clouds in more detail.

\section[]{Conclusions}\label{s_conclus}
In this paper we present the first results from our ongoing LCASS radio survey of the cold ISM along the line of sight to Cas~A. We have obtained high signal to noise observations with LOFAR in 33-78 MHz range and complemented these with WSRT observations in the range 304-386 MHz. The high signal to noise and high spectral resolution CRRL spectra allow us to carry out a detailed velocity resolved study of the foreground cold clouds in Perseus arm at -47 and -38~km~s$^{-1}$. We have used our new CRRL models (S16a,b) to interprete the dependence of CRRL line width and integrated optical depth on quantum number n. For the first time we have found a set of physical parameters that can describe both the line width and integrated optical depth.

\begin{itemize}
\item The CRRL line width and optical depths for the -47 and the -38 components can be modeled as two uniform clouds. The best-fit model for the -47~km~s$^{-1}$ cloud has T$_{e}$(-47)=85$\pm$5~K, n$_{e}$(-47)=0.040$\pm$0.005~cm$^{-3}$, L$_{\rm{CII}}$=35.3$\pm$1.2~pc and T$_{\rm{R,100}}$=1351$\pm$81~K. The best-fit model for the -38~km~s$^{-1}$ cloud has T$_{e}$(-38)=85$\pm$10~K and n$_{e}$(-38)=0.040$\pm$0.005~cm$^{-3}$, L$_{\rm{CII}}$=18.6$\pm$1.6~pc and T$_{\rm{R,100}}$=1507$\pm$128~K. The -38~km~s$^{-1}$ cloud has a steeper increase in optical depth at low frequencies, as compared to the -47~km~s$^{-1}$ cloud, and is fitted less well by our models. Together with the broad line profile this may indicate that the -38~km~s$^{-1}$ cloud consists of multiple velocity components with potentially different physical conditions.
\\
\item The derived CII column density is (4.4$\pm$0.6)$\times$10$^{18}$~cm$^{-2}$ and (2.3$\pm$0.3)$\times$10$^{18}$~cm$^{-2}$ for the -~47 and -~38~km~s$^{-1}$ clouds respectively. This is higher than in previous investigations as our b$_{n}\times\beta_{n}$ values for carbon from the S16a models are significantly lower than in previous models \citep[e.g.][]{Wa82,Po92}. The large pathlengths inferred from these CII column densities indicate that likely we are viewing sheet-like structures.
\\
\item The electron pressure, even for the lower signal to noise measurements of the -38~km~s$^{-1}$ cloud, is determined to better than 20\% uncertainty out to the 3$\sigma$ confidence level. The resulting thermal hydrogen pressure is (2.4$\pm$0.5)$\times$10$^{4}$~K~cm$^{-3}$, if we assume the gas phase abundance of carbon. This pressure is high, but consistent with our expectations of dense clouds in the Galactic midplane at a Galactocentric radius of about 10.5~kpc.
\\
\item The detection of HRRLs with the WSRT allows us to assess that the ionized hydrogen to carbon ratio is about 0.06. This means that 94 percent of the free electrons are donated by carbon. The corresponding hydrogen ionization rate would be (3$\pm$0.05)$\times$10$^{-18}$~s$^{-1}$ if the HRRL and CRRL trace the same gas. Our models indicate that this is not likely and much of the CRRL emission may be associated with CO-dark molecular gas. Deeper and higher spatial resolution measurements are necessary to investigate this further.
\end{itemize}

From our measurements and model we find a consistent picture for the -47 and -38 clouds in the Perseus arm towards Cas~A, in that they are dense, cool clouds with very low hydrogen ionization levels. The saturated HI 21~cm absorption measurements prevent us from determining the carbon to hydrogen abundance for these clouds. We have here used the gas phase abundance of carbon (i.e. [C/H]=1.4$\times$10$^{-4}$) to convert our carbon measurements to hydrogen measurements. We find high thermal pressures and high hydrogen column densities (N$_{\rm{H}}\sim$10$^{22}$~cm$^{-2}$). 

For these conditions we may be tracing the atomic to molecular hydrogen interface, and the gas cooling of such a cloud will be dominated by the [CII] 158~micron line \citep[][their Fig.~10]{Wo03}. S16b shows that the [CII] 158 micron line can provide an independent constraint on the temperature if it traces the same gas as the CRRLs.

The work presented here shows that we can use low-frequency CRRLs to accurately determine the physical conditions of cold neutral clouds in the ISM. Future observations with LOFAR using LBA and also the High Band Antennas (HBA) will allow us to determine this for many other cold clouds making up the cold neutral medium and thus assess their role within the ISM and their relationship with molecular gas. The CRRL models used here do contain some systematic uncertainties, in particular concerning the $\ell$-changing collisions. We will investigate the effects of some very recent updates of these rates by \citet{Gu16a} and \citet{Gu16b} in a future paper.

\section*{Acknowledgments}
LOFAR, the Low Frequency Array designed and constructed by ASTRON, has facilities in several countries, that are owned by various parties (each with their own funding sources), and that are collectively operated by the International LOFAR Telescope (ILT) foundation under a joint scientific policy

The authors would like to thank the LOFAR and WSRT observatory staff for their assistance in obtaining and handling of this large data set. The LOFAR obervations presented here are taken as part of LCASS survey. We gratefully acknowledge that LCASS is carried out using Directors discretionary time under project DDT001. We are grateful to the referee, Malcolm Walmsley for his very helpful and constructive review.

JBRO, AGGMT, HJAR and PS acknowledge acknowledge financial support from NWO Top LOFAR-CRRL project, project No. 614.001.351. R.J.W. is supported by a Clay Fellowship awarded by the Harvard-Smithsonian Center for Astrophysics. AGGMT acknowledges support through the Spinoza premie of the Dutch Science Organization (NWO).

\clearpage

\newpage


\begin{table*}
 \centering
  \begin{tabular}{|l|l|l|l|} \hline
  Parameter                & LOFAR LBA (1)      & LOFAR LBA (2)   & WSRT P-band \\ \hline
  Data ID                  & L40787             & L184343         & S12A/002 \\
  Field center RA (J2000)  & 23h23m22.8s        & 23h23m22.8s     & 23h23m27.9s \\
  Field center DEC (J2000) & +58d50m16s         & +58d50m16s      & +58d48m42s \\
  Observing date           & 2011 December 27   & 2013 October 31 & 2012 January 28 \\
  Total on-source time     & 10.5 h             & 10 h            & 12 h \\
  Frequency range          & 33-57 MHz          & 55-78 MHz       & 300-390 MHz \\
  Number of sub-bands      & 122                & 122             & 6 \\
  Width of a sub-band      & 0.195 MHz          & 0.195 MHz       & 1.25 MHz \\
  Channels per subband     & 512                & 512             & 2048 \\
  Channel width            & 2.0-3.5~km~s$^{-1}$ & 1.5-2.1~km~s$^{-1}$ & 0.5-1.0~km~s$^{-1}$ \\ \hline
  \end{tabular}
 \caption[]{Details of the Observations. Note that for WSRT we cycle through 6x8 spectral setups in time so that the on-source time per line amounts to about 1.5~h.}\label{t_obs}
\end{table*}

\begin{table*}
 \centering
  \begin{tabular}{|l|l|l|l|} \hline
  Stack & Species  & Individual $\alpha$ line transitions & Observation \\ \hline
          n  &      & n  \\
         260 & C~\&~H & 257,258,260,261,262 & WSRT \\
         266 & C~\&~H & 263,264,265,267,268 & WSRT \\
         271 & C~\&~H & 270,271,272,273 & WSRT \\
         276 & C~\&~H & 274,275,276,277,278 & WSRT \\
             &      & \\
         438 & C    & 435,436,437,438,439,440,442 & LBA (2) \\
         448 & C    & 443,444,445,447,449,452,454 & LBA (2) \\
         459 & C    & 456,457,459,460,461,463 & LBA (2) \\
         467 & C    & 464,465,466,467,468,472 & LBA (2) \\
         477 & C    & 473,474,475,479,480,481 & LBA (2) \\
         485 & C    & 482,483,484,486,487,488,489 & LBA (2) \\
         496 & C    & 491,492,493,495,496,497,498,499,500,501,503,504,505 & LBA (1) \\
         510 & C    & 506,507,508,509,510,511,512,513,514,515,517,518,519 & LBA (1) \\
         527 & C    & 522,523,525,526,528,529,530,531,532,533,534,535 & LBA (1) \\
         542 & C    & 536,537,538,540,541,542,543,544,547,548,549,550 & LBA (1) \\
         559 & C    & 551,552,553,554,558,559,560,561,562,563,565,566 & LBA (1) \\
         575 & C    & 567,568,569,573,574,575,576,577,578,579,580,581,584 & LBA (1) \\
             &      & \\
         439 & H    & 435,436,437,438,439,440,441,442 & LBA (2) \\
         447 & H    & 443,445,446,447,448,449,450,451 & LBA (2) \\
         458 & H    & 454,455,456,457,458,459,460,461 & LBA (2) \\
         466 & H    & 462,463,464,465,466,469,470,471 & LBA (2) \\
         475 & H    & 472,473,474,475,476,477,478 & LBA (2) \\
         485 & H    & 481,482,483,484,485,487,488,489 & LBA (2) \\
         496 & H    & 491,492,493,495,496,498,499,501,502,503,504,505 & LBA (1) \\
         510 & H    & 506,507,508,509,510,511,512,513,514,515,516,517 & LBA (1) \\
         527 & H    & 519,522,523,525,526,528,529,531,532,533,535,536 & LBA (1) \\
         542 & H    & 537,538,540,541,542,543,544,546,547,548,549,550 & LBA (1) \\
         559 & H    & 551,552,556,557,558,559,560,561,562,563,565,566 & LBA (1) \\
         575 & H    & 567,568,569,574,575,576,577,578,579,580,581,582,584 & LBA (1) \\ \hline
  \end{tabular}
 \caption[]{Individual $\alpha$ line transitions included in each CRRL and HRRL line stack.}\label{t_alln}
\end{table*}

\begin{table*}
 \centering
  \begin{tabular}{|l|l|r|r|r|} \hline
  transition & freq  & $\int \tau$~d$\nu$ & v$_{LSR}$          & FWHM$_{T}$            \\ \hline
          n  & [MHz] & [Hz]             & [km~s$^{-1}$]            & [km~s$^{-1}$]          \\
         260 & 372.2 & -8.54 $\pm$ 0.29 & -47.71 $\pm$ 0.06 & 3.43 $\pm$ 0.13 \\
             &       & -4.85 $\pm$ 0.41 & -38.81 $\pm$ 0.28 & 6.89 $\pm$ 0.69 \\
         266 & 347.6 & -8.27 $\pm$ 0.28 & -47.63 $\pm$ 0.05 & 3.24 $\pm$ 0.13 \\
             &       & -4.06 $\pm$ 0.42 & -37.70 $\pm$ 0.36 & 7.10 $\pm$ 0.86 \\
         271 & 328.8 & -8.48 $\pm$ 0.32 & -47.76 $\pm$ 0.07 & 3.65 $\pm$ 0.16 \\
             &       & -4.23 $\pm$ 0.41 & -38.24 $\pm$ 0.30 & 6.32 $\pm$ 0.72 \\
         276 & 311.2 & -7.60 $\pm$ 0.24 & -47.60 $\pm$ 0.05 & 3.24 $\pm$ 0.12 \\
             &       & -3.68 $\pm$ 0.34 & -38.05 $\pm$ 0.31 & 6.93 $\pm$ 0.76 \\ \hline
  \end{tabular}
 \caption[]{WSRT P-band measured line properties for Cn$\alpha$ recombination line stacks. $\int \tau$~d$\nu$ is the integrated optical depth, v$_{LSR}$ is the velocity relative to the local standard of rest and FWHM$_{T}$ is total full width at half maximum. The values are obtained from a Gaussian fit to the spectrum.}\label{t_wsrt_cia}
\end{table*}

\begin{table*}
 \centering
  \begin{tabular}{|l|r|r|r|r|} \hline
  RRL & v$_{LSR}$          & FWHM$_{T}$     & $\int \tau$~d$\nu$               & $\tau_{\rm{peak}}$ \\ \hline
      & [km~s$^{-1}$]      & [km~s$^{-1}$]  & [Hz]                             &  \\
    S & -55.70 $\pm$ 0.30 & 2.40 $\pm$ 0.70 & -0.57 $\pm$ 0.13 & ( 2.1 $\pm$ 0.5) $\times$ 10$^{-4}$ \\
    C & -47.67 $\pm$ 0.03 & 3.39 $\pm$ 0.08 & -8.26 $\pm$ 0.16 & (21.5 $\pm$ 0.5) $\times$ 10$^{-4}$ \\
    H & 101.99 $\pm$ 0.14 & 3.81 $\pm$ 0.34 & -1.96 $\pm$ 0.15 & ( 4.5 $\pm$ 0.5) $\times$ 10$^{-4}$ \\
    C & -38.24 $\pm$ 0.18 & 6.78 $\pm$ 0.44 & -4.19 $\pm$ 0.23 & ( 5.4 $\pm$ 0.5) $\times$ 10$^{-4}$ \\
    H & 110.80 $\pm$ 0.27 & 2.20 $\pm$ 0.63 & -0.46 $\pm$ 0.11 & ( 1.8 $\pm$ 0.5) $\times$ 10$^{-4}$ \\ \hline
  \end{tabular}
 \caption[]{WSRT stacked, over the full band, CRRL, HRRL and SRRL $\alpha$ line profile properties for the line of sight to Cas~A. The average frequency is 343.7~MHz which corresponds to n=267. v$_{LSR}$ is the velocity relative to the local standard of rest, FWHM$_{T}$ is total full width at half maximum and $\int \tau$~d$\nu$ is the integrated optical depth. The values are obtained from a Gaussian fit to the spectrum. The peak optical depth $\tau_{\rm{peak}}$ is determined directly from the stacked spectrum that has 0.5~km~s$^{-1}$ channels, see Fig.~\ref{f_app_wsrt_full_stack}. The 1$\sigma$ spectral RMS per 0.5~km~s$^{-1}$ channel is 0.5$\times$10$^{-4}$ in units of optical depth.}\label{t_wsrt_meas_intg}
\end{table*}

\begin{table*}
 \centering
  \begin{tabular}{|l|l|r|r|r|r|r|} \hline
  transition & freq  & $\int \tau$~d$\nu$ & v$_{LSR}$         & FWHM$_{T}$         & FWHM$_{L}$        & Observation\\ \hline
          n  & [MHz] & [Hz]               & [km~s$^{-1}$]     & [km~s$^{-1}$]      & [km~s$^{-1}$]     & \\
         438 & 78.03 &  5.63 $\pm$ 0.50   & -47.39 $\pm$ 1.46 &   5.60 $\pm$ 0.55  & ---               & 2 \\
             &       &  1.75 $\pm$ 0.50   & [-37.99]          &   7.41 $\pm$ 0.54  & ---               & 2 \\
         448 & 72.93 &  5.58 $\pm$ 0.27   & -47.69 $\pm$ 1.57 &   5.56 $\pm$ 0.57  &  1.64 $\pm$ 0.30  & 2 \\
             &       &  1.97 $\pm$ 0.21   & [-38.29]          &   7.49 $\pm$ 0.54  & ---               & 2 \\
         459 & 67.82 &  6.15 $\pm$ 0.26   & -47.66 $\pm$ 1.69 &   5.80 $\pm$ 0.57  &  1.75 $\pm$ 0.26  & 2 \\
             &       &  2.89 $\pm$ 0.30   & [-38.26]          &   8.10 $\pm$ 0.56  &  0.92 $\pm$ 0.94  & 2 \\
         467 & 64.39 &  6.61 $\pm$ 0.24   & -47.69 $\pm$ 1.78 &   6.27 $\pm$ 0.57  &  2.31 $\pm$ 0.23  & 2 \\
             &       &  2.96 $\pm$ 0.28   & [-38.29]          &   8.61 $\pm$ 0.58  &  1.68 $\pm$ 0.88  & 2 \\
         477 & 60.43 &  6.81 $\pm$ 0.21   & -47.86 $\pm$ 1.89 &   6.50 $\pm$ 0.57  &  2.42 $\pm$ 0.20  & 2 \\
             &       &  3.79 $\pm$ 0.25   & [-38.46]          &   9.11 $\pm$ 0.58  &  2.35 $\pm$ 0.65  & 2 \\
         485 & 57.49 &  7.30 $\pm$ 0.25   & -47.95 $\pm$ 1.99 &   6.99 $\pm$ 0.57  &  2.95 $\pm$ 0.23  & 2 \\
             &       & (4.71 $\pm$ 0.34)  & [-38.55]          & (11.61 $\pm$ 0.67) & (6.06 $\pm$ 0.87) & 2 \\
         496 & 53.76 &  7.47 $\pm$ 0.47   & -46.10 $\pm$ 2.13 &   7.29 $\pm$ 0.60  &  3.10 $\pm$ 0.45  & 1 \\
             &       &  3.82 $\pm$ 0.54   & [-36.70]          &   9.47 $\pm$ 0.63  &  2.54 $\pm$ 1.33  & 1 \\
         510 & 49.45 &  7.97 $\pm$ 0.54   & -45.89 $\pm$ 2.31 &   8.48 $\pm$ 0.63  &  4.44 $\pm$ 0.54  & 1 \\
             &       &  4.41 $\pm$ 0.61   & [-36.49]          &  10.57 $\pm$ 0.67  &  3.99 $\pm$ 1.30  & 1 \\
         527 & 44.82 &  8.82 $\pm$ 0.60   & -46.02 $\pm$ 2.55 &  10.01 $\pm$ 0.66  &  6.09 $\pm$ 0.58  & 1 \\
             &       &  5.13 $\pm$ 0.63   & [-36.62]          &  12.01 $\pm$ 0.69  &  5.79 $\pm$ 1.11  & 1 \\
         542 & 41.21 &  8.74 $\pm$ 0.91   & -45.95 $\pm$ 2.78 &  11.92 $\pm$ 0.75  &  8.19 $\pm$ 0.90  & 1 \\
             &       &  6.16 $\pm$ 0.90   & [-36.55]          &  14.52 $\pm$ 0.78  &  8.97 $\pm$ 1.21  & 1 \\
         559 & 37.57 &  (7.95 $\pm$ 1.00) & (-46.03 $\pm$ 3.04) & (12.76 $\pm$ 0.77) &  (8.75 $\pm$ 0.99) & 1 \\
             &       &  (7.43 $\pm$ 1.00) & [-36.63]            & (15.82 $\pm$ 0.75) & (10.30 $\pm$ 0.99) & 1 \\
         575 & 34.52 &  (9.05 $\pm$ 1.34) & (-45.76 $\pm$ 3.31) & (15.17 $\pm$ 0.87) & (11.31 $\pm$ 1.22) & 1 \\
             &       &  (7.61 $\pm$ 1.29) & [-36.36]            & (18.67 $\pm$ 0.82) & (13.58 $\pm$ 1.11) & 1 \\ \hline
  \end{tabular}
 \caption[]{LOFAR LBA: Measured line properties for Cn$\alpha$ recombination line stacks. In our line profile fitting procedure we have fixed the velocity offset between the -47 and the -38~km~s$^{-1}$ component to 9.4~km~s$^{-1}$. For n$>$550 the line blending of the two Perseus arm components is so severe that fitting two components, although necessary to describe the total line profile, is very sensitive to the local spectral noise and bandpass features. We therefore do not use decomposed optical depth and linewidth values for the individual component above n=500. Note that there is a small constant offset in velocity of 1-2~km~s$^{-1}$ between the second LBA (n=438-485) and the first LBA (n=496-575) observations. This is due to the inaccuracies in our offline Doppler correction. We have not attempted to correct this as it does not influence the results for the integrated optical depth or the line width. FWHM$_{T}$ is total full width at half maximum and FWHM$_{L}$ is the Lorentzian contribution.}\label{t_lba_cia}
\end{table*}

\begin{table*}
 \centering
  \begin{tabular}{|l|l|r|r|} \hline
  transition & freq  & $\int \tau$~d$\nu$ (3$\sigma$) & v$_{LSR}$ \\ \hline
          n  & [MHz] & [Hz]             &  [km~s$^{-1}$]   \\
         439 & 77.46 & 0.410            & -47              \\
             &       & 0.472            & -38              \\
         447 & 73.38 & 0.212            & -47              \\
             &       & 0.246            & -38              \\
         458 & 68.23 & 0.151            & -47              \\
             &       & 0.178            & -38              \\
         466 & 64.78 & 0.140            & -47              \\
             &       & 0.163            & -38              \\
         475 & 61.17 & 0.136            & -47              \\
             &       & 0.162            & -38              \\
         485 & 57.46 & 0.111            & -47              \\
             &       & 0.143            & -38              \\
         496 & 53.73 & 0.165            & -47              \\
             &       & 0.189            & -38              \\
         510 & 49.43 & 0.163            & -47              \\
             &       & 0.182            & -38              \\ 
         527 & 44.80 & 0.130            & -47              \\
             &       & 0.143            & -38              \\ 
         542 & 41.19 & 0.108            & -47              \\
             &       & 0.119            & -38              \\ 
         559 & 37.55 & 0.087            & -47              \\
             &       & 0.097            & -38              \\ 
         575 & 34.50 & 0.110            & -47              \\
             &       & 0.122            & -38              \\ \hline
  \end{tabular}
 \caption[]{Integrated optical depth limits (3$\sigma$) for the non-detected Hn$\alpha$ lines for hydrogen (HRRL) in the LOFAR LBA range. For the HRRL we use the stacked LBA spectra without baseline correction processing. The upper limits for the integrated optical depth are calculated from $\tau_{rms,chn}$ and by assuming that the HRRLs are at the same velocity and have the same width as CRRLs.}\label{t_lba_hrrl}
\end{table*}

\begin{table*}
 \centering
  \begin{tabular}{|l|l|l|l|} \hline
  Parameter              & unit           & -47~km~s$^{-1}$                   & -38~km~s$^{-1}$                   \\ \hline
  T$_{\rm{R,100}}$       & [K]            & 1400 (1351~$\pm$~83)              & 1600 (1507~$\pm$~128)             \\ 
  T$_{e}$                & [K]            & 85~$\pm$~5                        & 85~$\pm$~10                       \\ 
  n$_{e}$                & [cm$^{-3}$]    & 0.040~$\pm$~0.005                 & 0.040~$\pm$~0.005                 \\ 
  L$_{\rm{CII}}$         & [pc]           & 35.3~$\pm$~1.2                    & 18.6~$\pm$~1.6                    \\ 
  EM$_{\rm{CII}}$        & [cm$^{-6}$~pc] & 0.056~$\pm$~0.014                 & 0.030~$\pm$~0.008                 \\ 
  N$_{\rm{CII}}$         & [cm$^{-2}$]    & (4.4~$\pm$~0.6)$\times$10$^{18}$  & (2.3~$\pm$~0.3)$\times$10$^{18}$  \\ 
  N$_{\rm{H}}$           & [cm$^{-2}$]    & (3.1~$\pm$~0.4)$\times$10$^{22}$  & (1.6~$\pm$~0.2)$\times$10$^{22}$  \\ 
  n$_{\rm{H}}$           & [cm$^{-3}$]    & 286~$\pm$~36                      & 286~$\pm$~36                      \\ 
  p$_{\rm{thermal}}$/k   & [K~cm$^{-3}$]  & (2.4~$\pm$~0.5)$\times$10$^{4}$   & (2.4~$\pm$~0.5)$\times$10$^{4}$   \\ 
  p$_{\rm{turbulent}}$/k & [K~cm$^{-3}$]  & (1.9~$\pm$~0.1)$\times$10$^{5}$   & (7.6~$\pm$~1.0)$\times$10$^{5}$   \\ 
  p$_{\rm{magnetic}}$/k  & [K~cm$^{-3}$]  & (1.8-4.5)$\times$10$^{4}$         & ---                               \\ 
  $\zeta_{\rm{H}}$            & [s$^{-1}$]     & (0.3~$\pm$~0.05)$\times$10$^{-17}$ & ---                               \\ \hline  
  \end{tabular}
  \caption[]{CRRL model results. Here we have adopted the gas phase abundance of carbon by \citet{Ca96} to convert our CRRL measurements to hydrogen column densities N$_{H}$, volume densities n$_{H}$ and thermal pressure. The range in magnetic pressures is taken from the measurements by \citet{He86} and \citet{Sc86}.}\label{t_crrl_results}
\end{table*}

\clearpage

\newpage



\clearpage

\begin{figure*}
    \includegraphics[width=0.44\textwidth, angle=90]{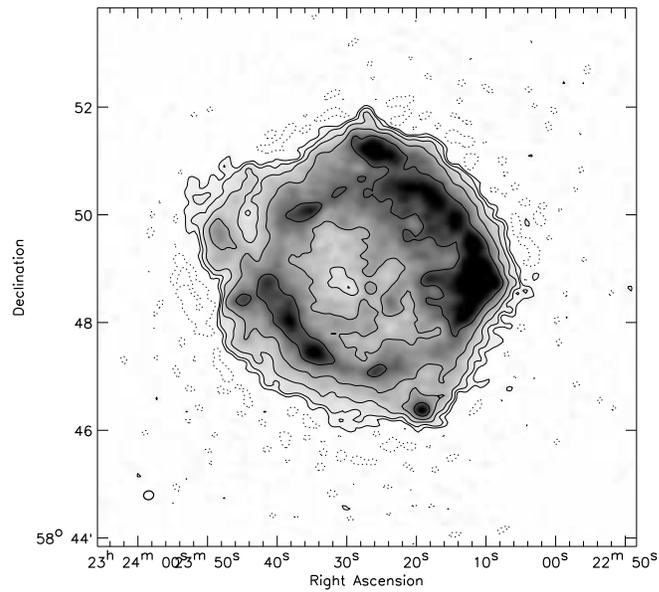}
  \vspace{0.5cm}
  \caption{Cas~A continuum image at 69~MHz obtained from a single 0.2~MHz sub-band. This image was made from a LOFAR LBA observation, taken on October 15th in 2011, using uniform weighting and has a resolution of $11.2\arcsec \times 9.8\arcsec$.}\label{f_cas_model}
\end{figure*}

\begin{figure*}\vspace{1cm}
\mbox{
    \includegraphics[width=0.33\textwidth, angle=90]{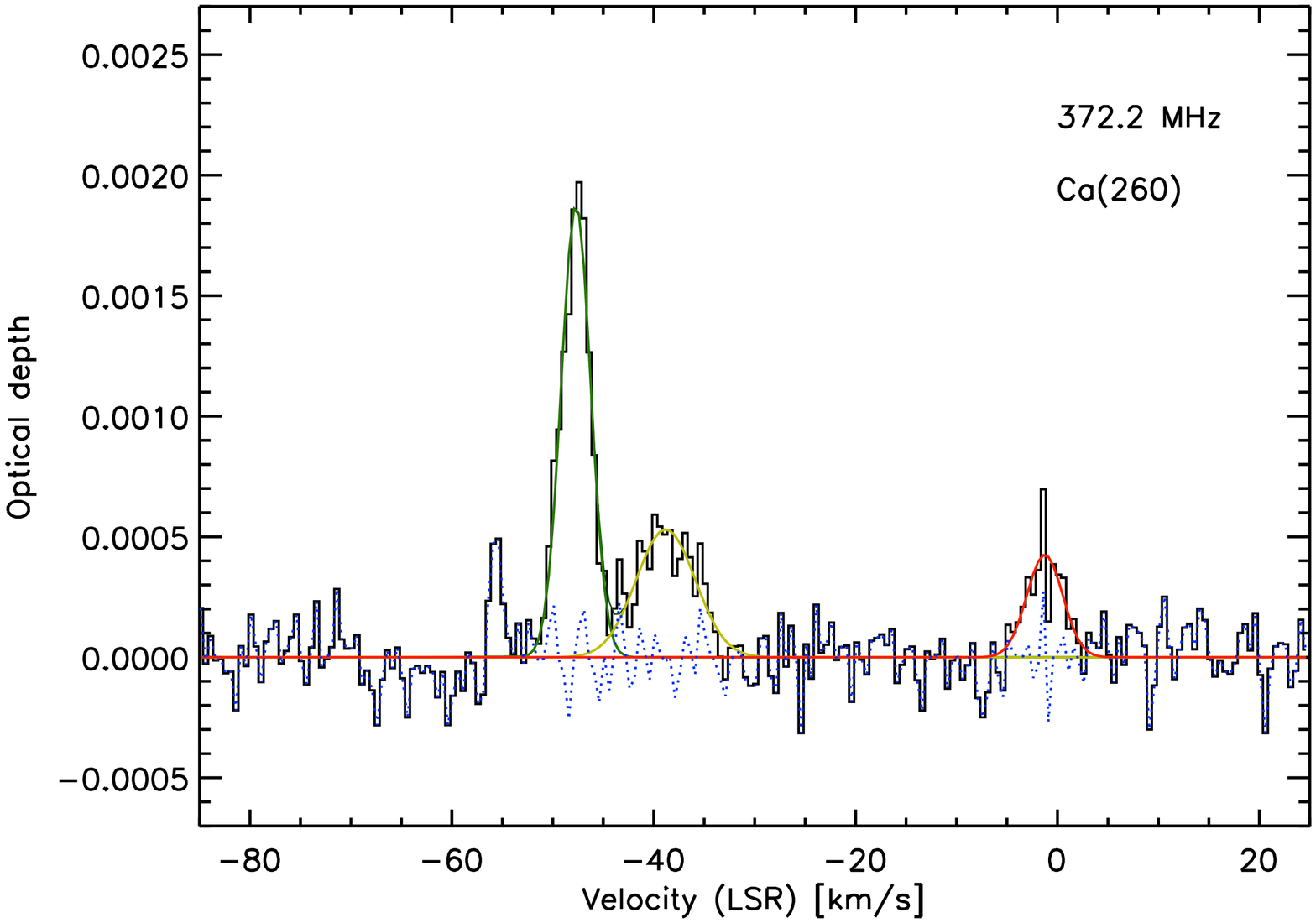}\hspace{1cm}
    \includegraphics[width=0.33\textwidth, angle=90]{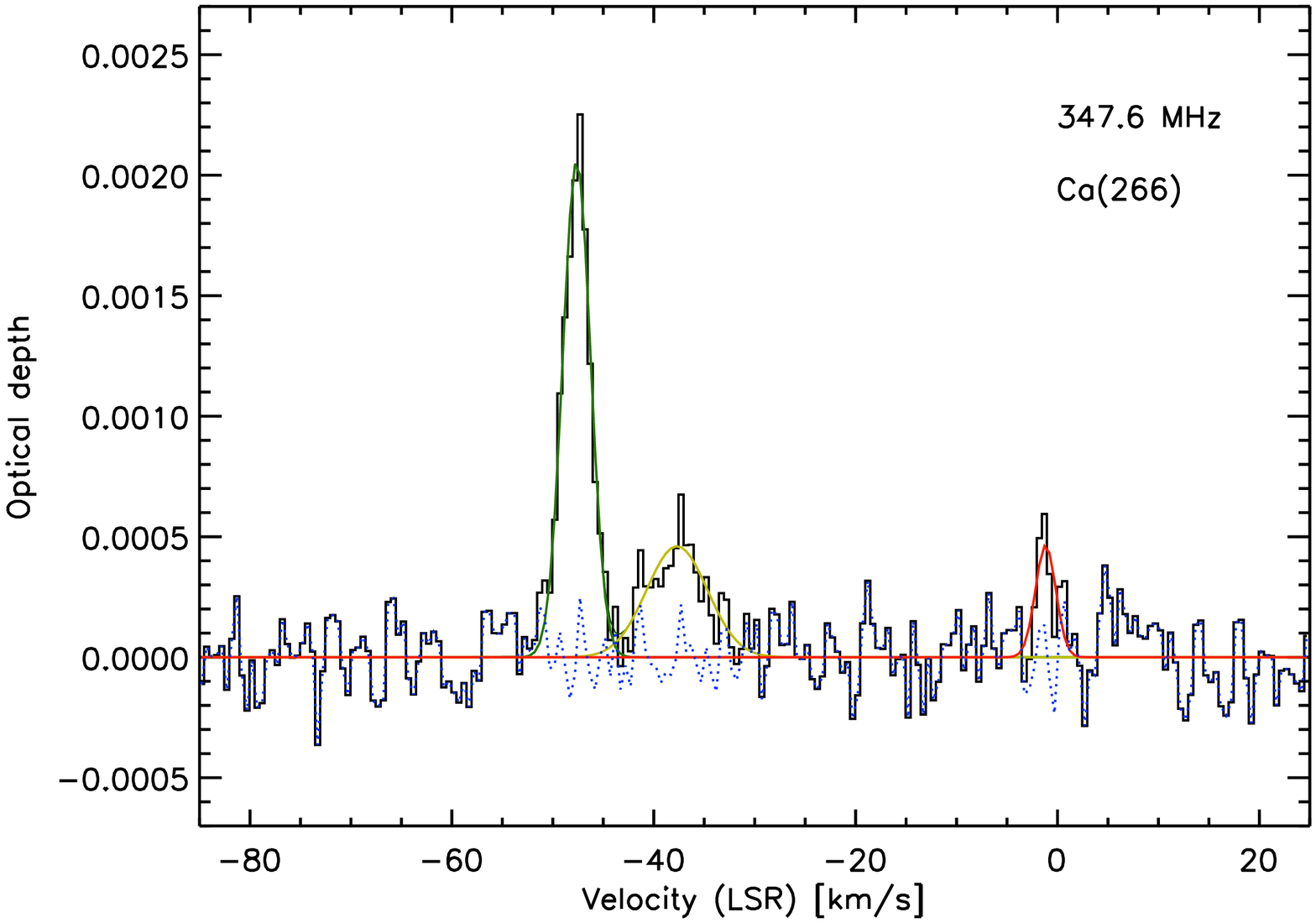}
}
\mbox{
    \includegraphics[width=0.33\textwidth, angle=90]{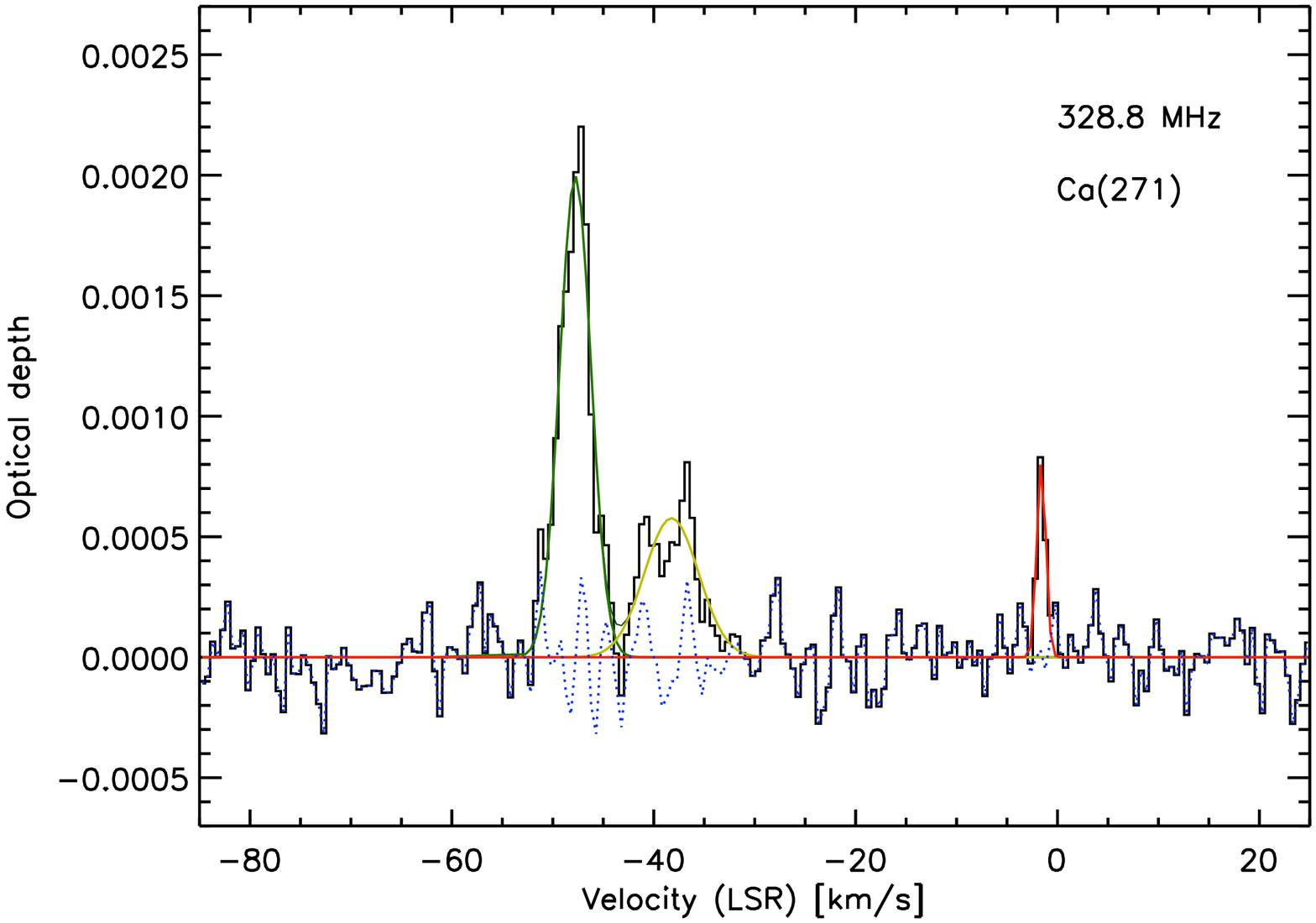}\hspace{1cm}
    \includegraphics[width=0.33\textwidth, angle=90]{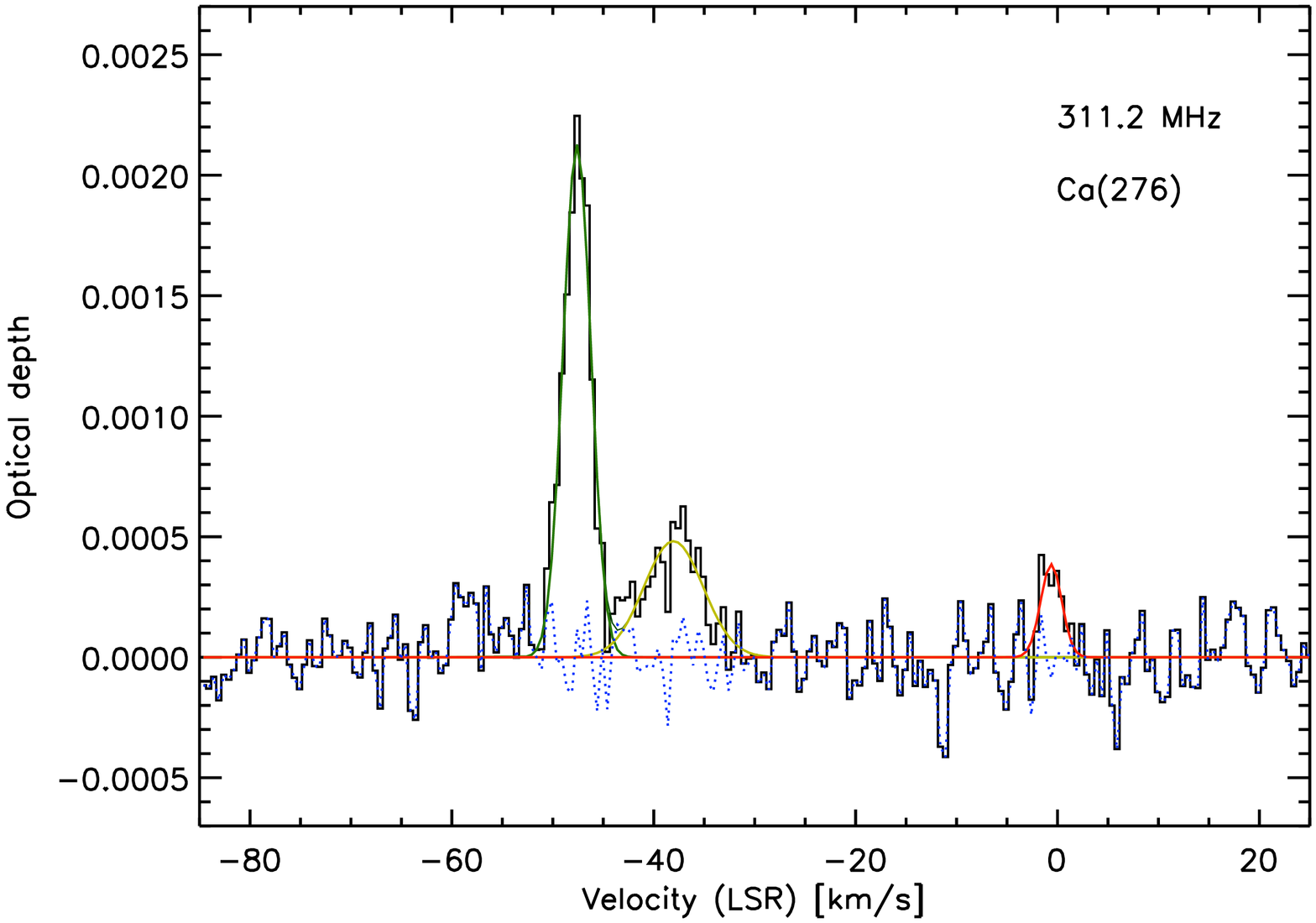}
}
  \caption{WSRT P-band 310-390 MHz: stacked CRRL spectra. The green, yellow and red lines show the decomposition into the -47, -38 and 0~km~s$^{-1}$ components. The blue dotted line shows the residuals after the subtracting the fitted line profiles.}\label{f_app_wsrt_substack}
\end{figure*}

\begin{figure*}\vspace{1cm}
\mbox{
    \includegraphics[width=0.6\textwidth, angle=90]{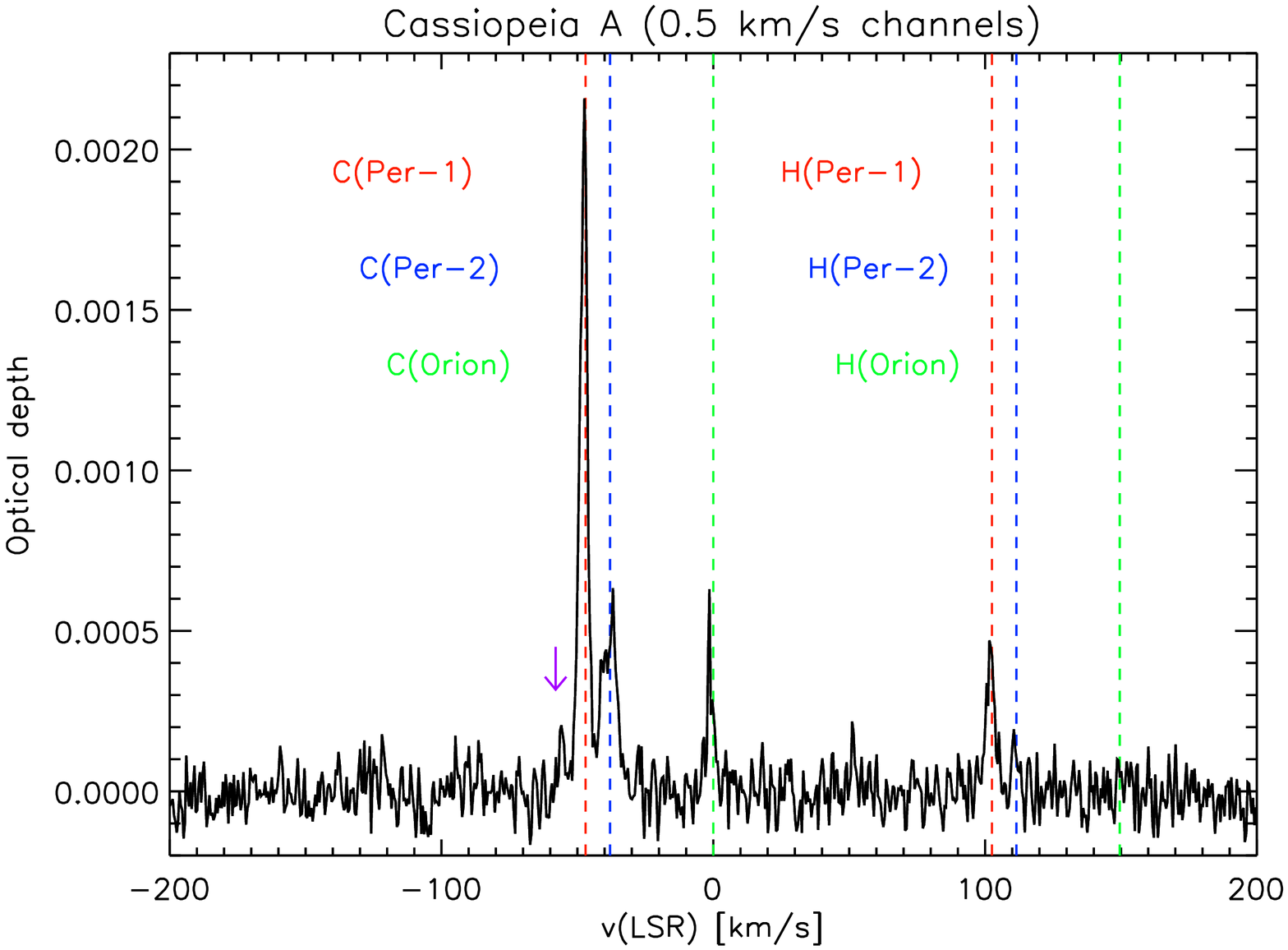}
}\vspace{1cm}
\mbox{
    \includegraphics[width=0.33\textwidth, angle=90]{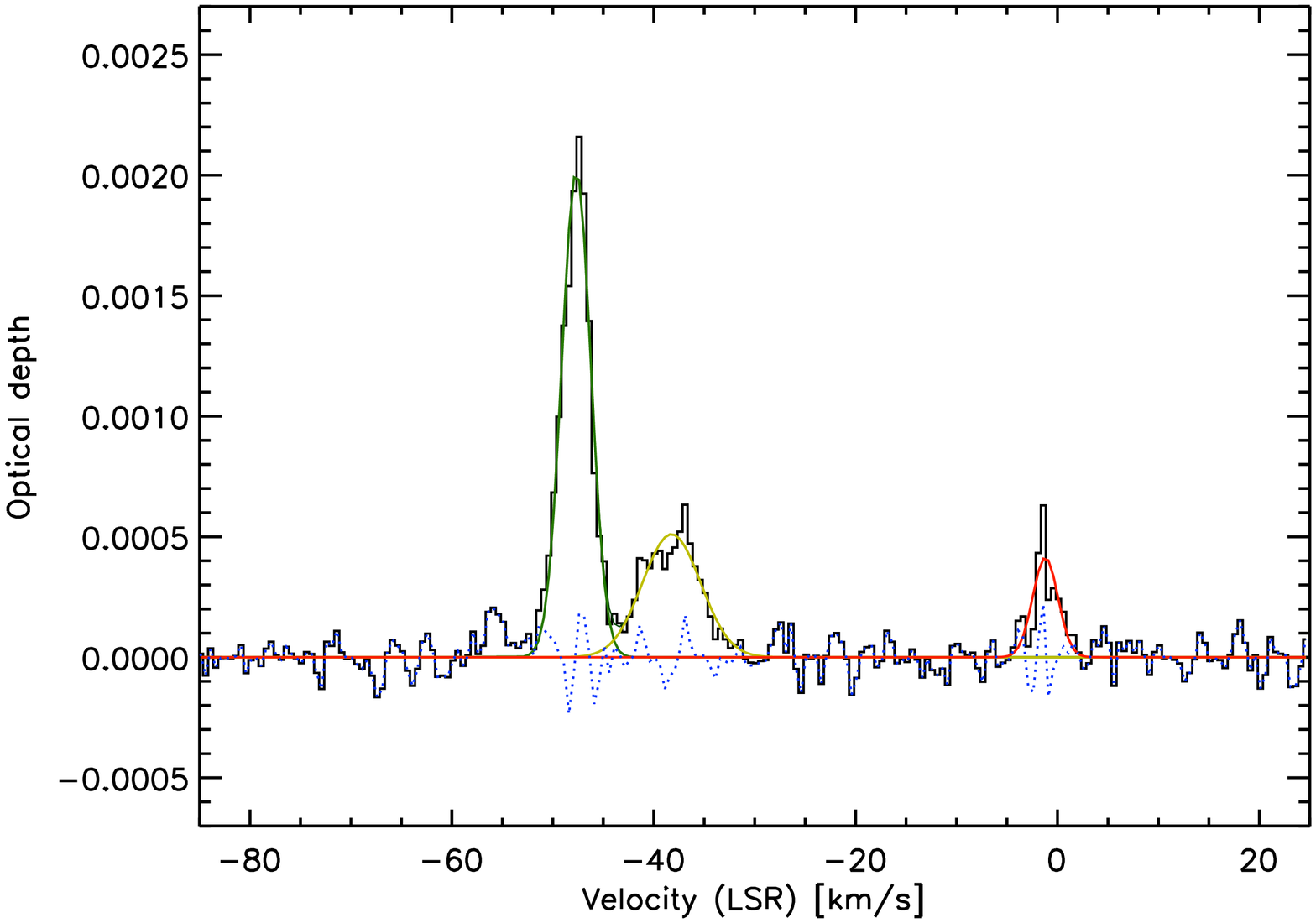}\hspace{1cm}
    \includegraphics[width=0.33\textwidth, angle=90]{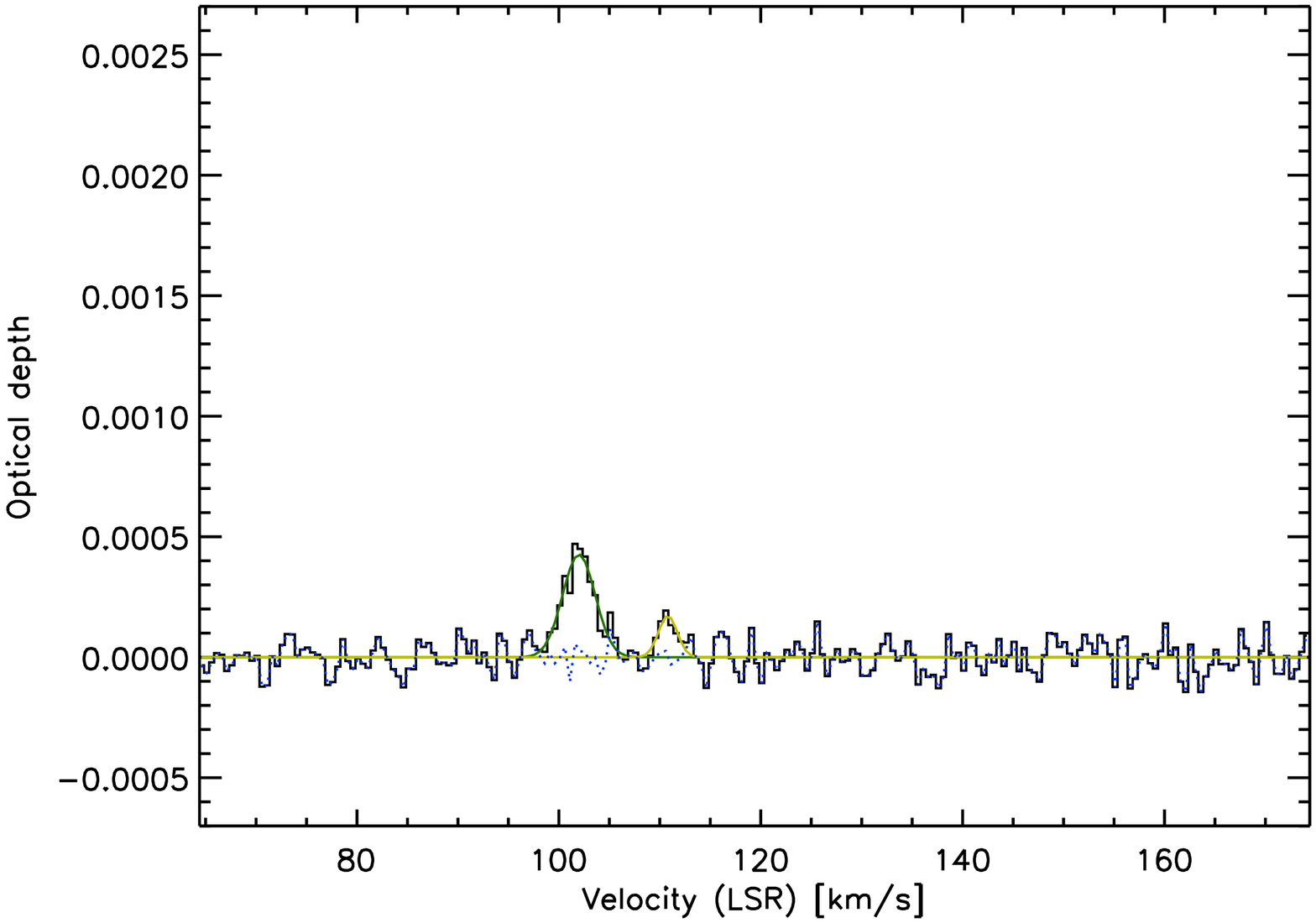}
}
  \vspace{0.5cm}
  \caption{WSRT P-band stacked, over the full band, RRL spectra. The spectra are centered for CRRLs on v(LSR)=0.0~km~s$^{-1}$ and spatially integrated over the remnant. Gaussian fits to the RRL lines are shown by the green (-47 component), yellow (-38 component) and red (0 component) solid lines in the bottom spectra. \textit{(Top)} Stacked WSRT RRL spectrum showing both the CRRLs and HRRLs. The purple arrow shows the location of the SRRL feature at -55~km~s$^{-1}$. \textit{(Bottom-left)} Zoom in on the CRRL components. \textit{(Bottom-right)} Zoom in on the HRRL components. In the bottom panels the green, yellow and red lines show the decomposition into the -47, -38 and 0~km~s$^{-1}$ components and the blue dotted line shows the residuals after the subtracting the fitted line profiles.}\label{f_app_wsrt_full_stack}
\end{figure*}

\clearpage

\newpage

\begin{figure*}\vspace{0.5cm}
\mbox{
    \hspace{0.8cm}\includegraphics[width=0.36\textwidth, angle=-90]{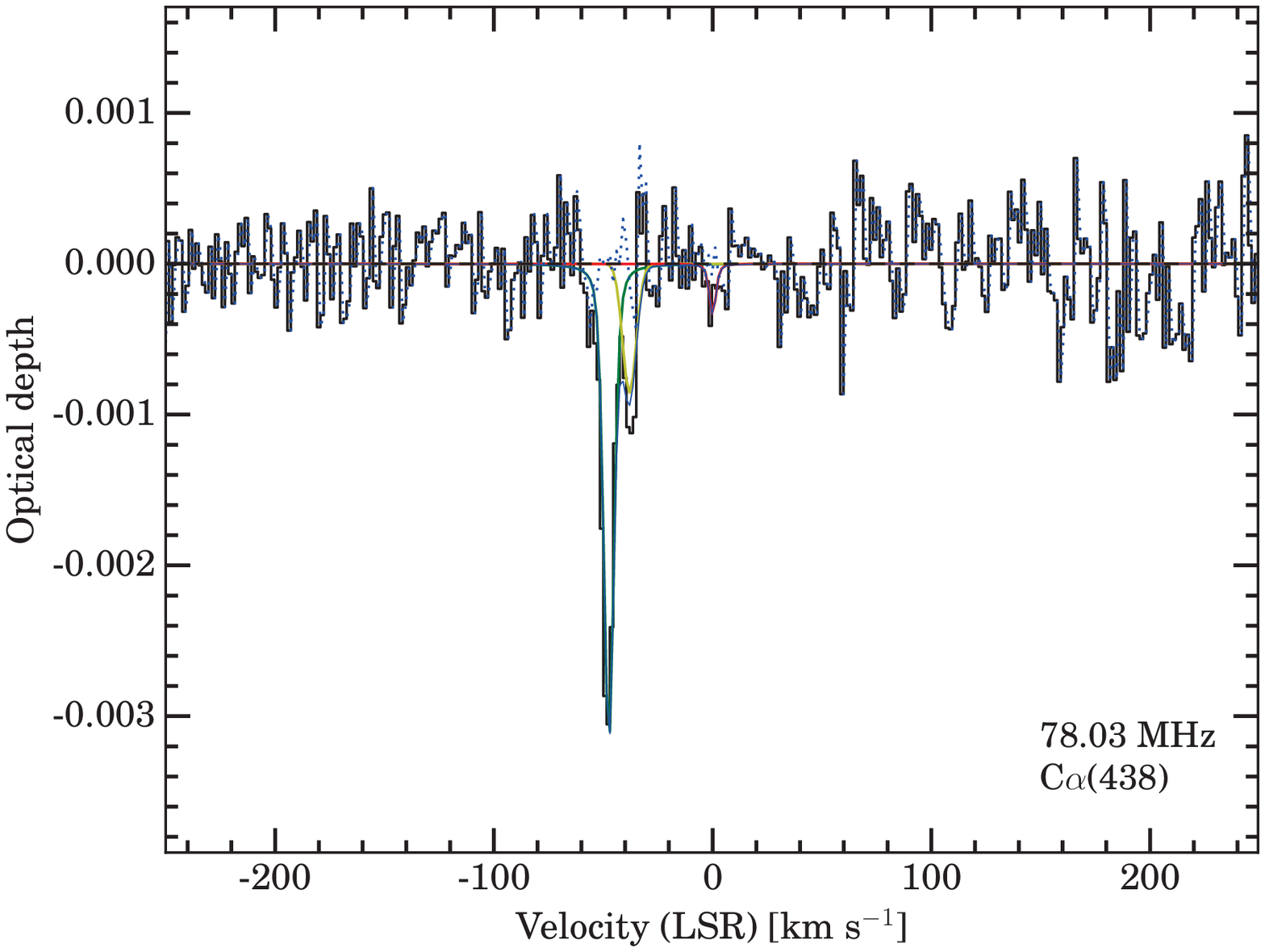}\hspace{0.3cm}
    \includegraphics[width=0.36\textwidth, angle=-90]{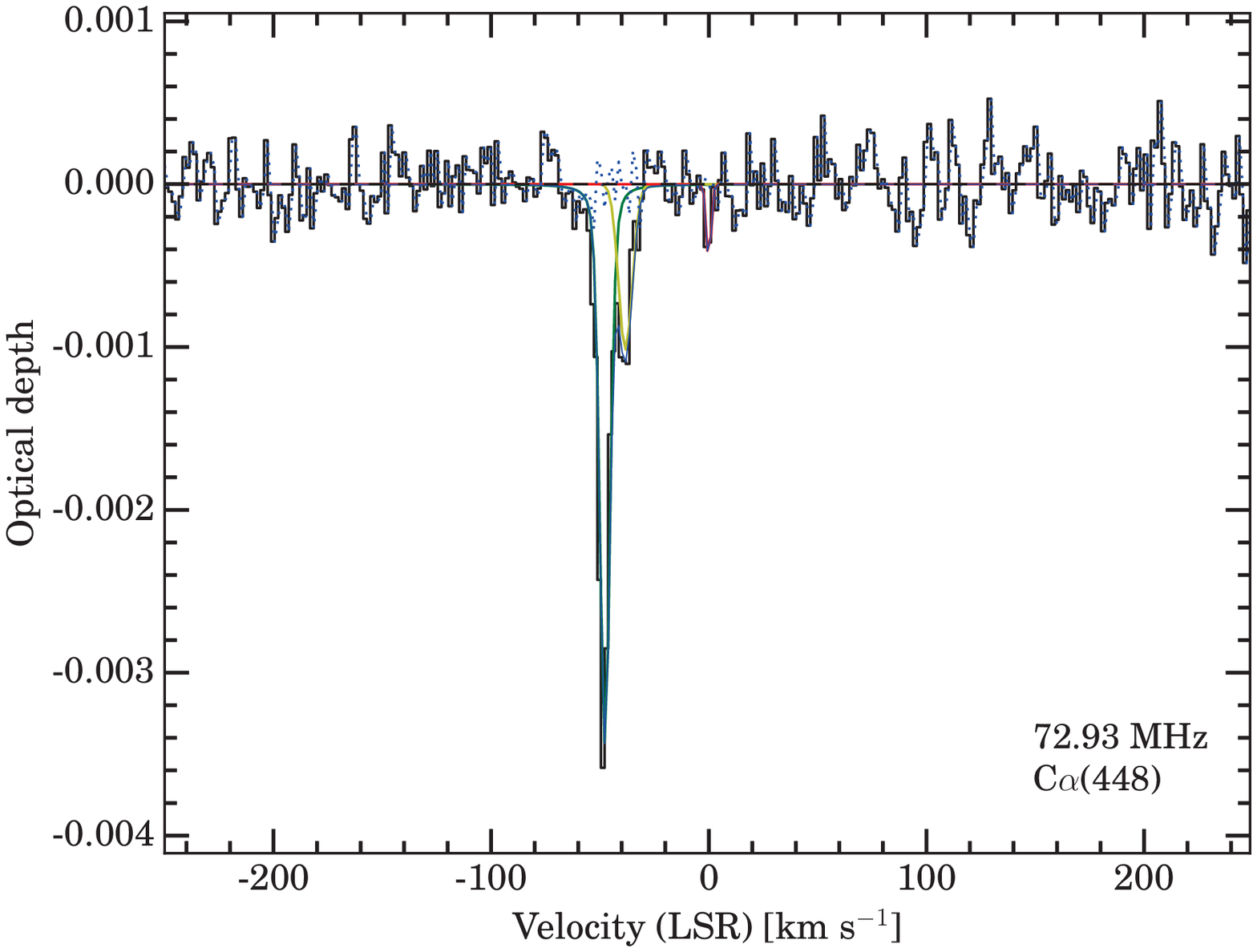}
}\vspace{0.8cm}
\mbox{
    \hspace{0.8cm}\includegraphics[width=0.36\textwidth, angle=-90]{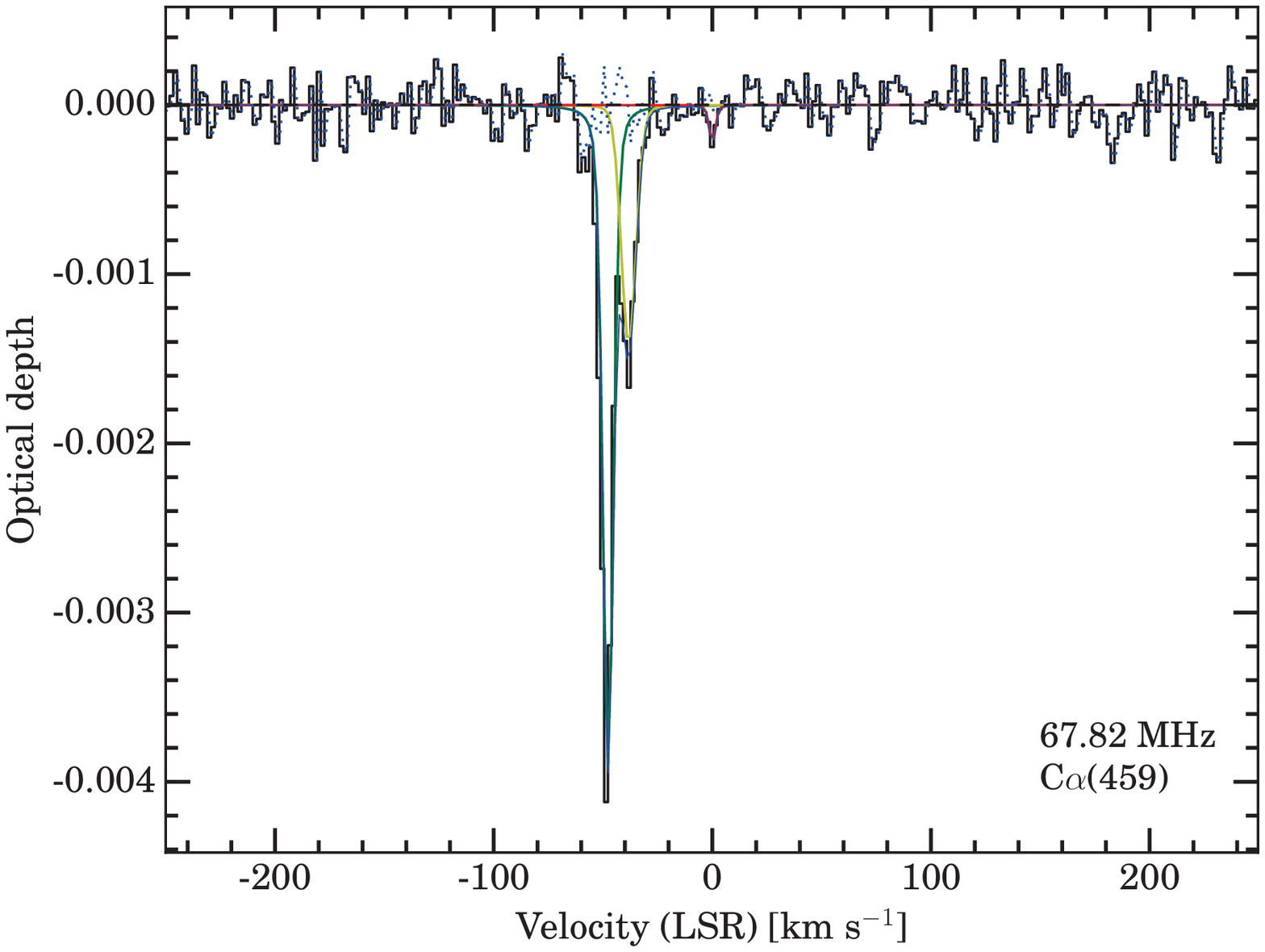}\hspace{0.3cm}
    \includegraphics[width=0.36\textwidth, angle=-90]{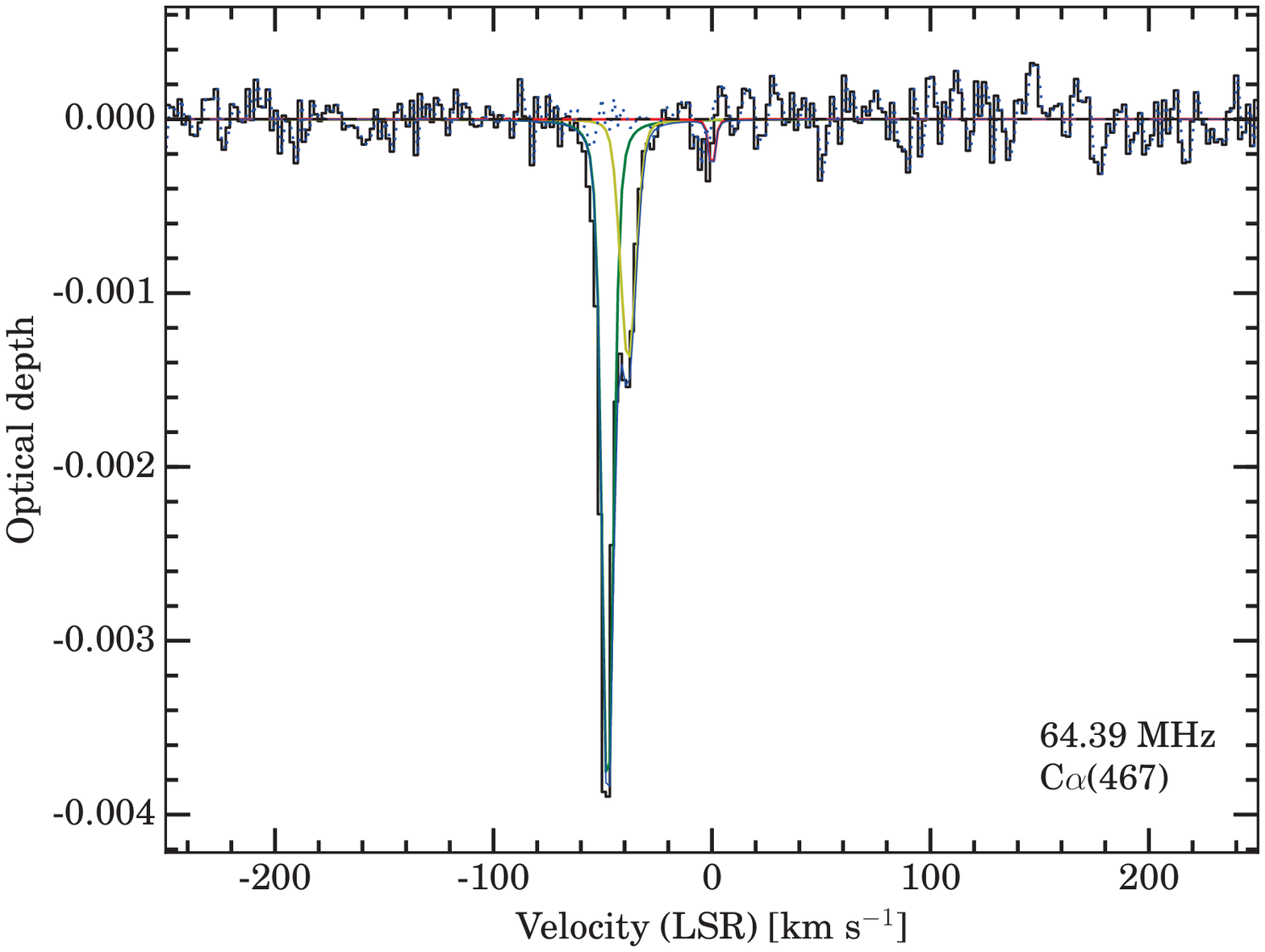}
}\vspace{0.8cm}
\mbox{
    \hspace{0.8cm}\includegraphics[width=0.36\textwidth, angle=-90]{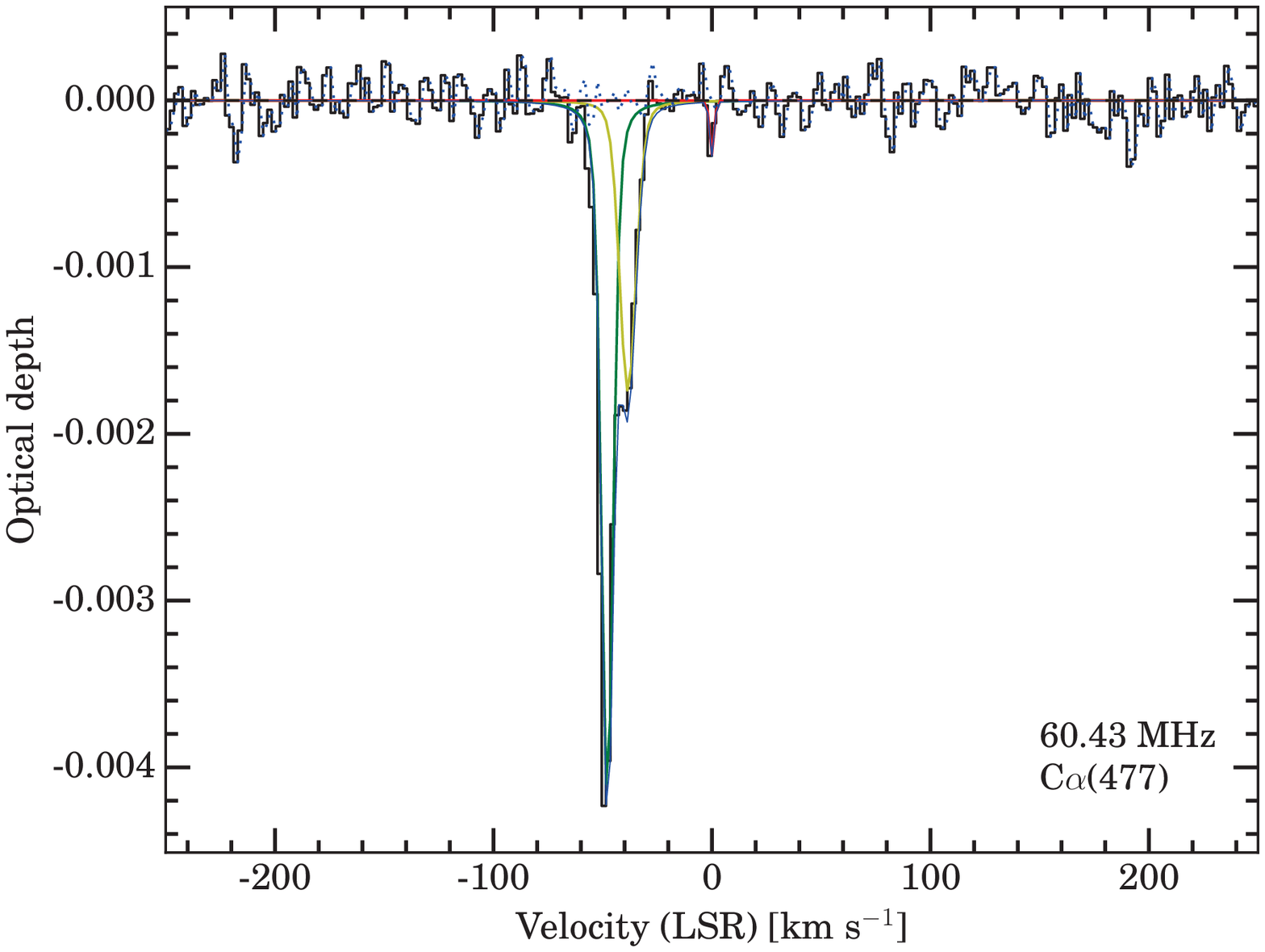}\hspace{0.3cm}
    \includegraphics[width=0.36\textwidth, angle=-90]{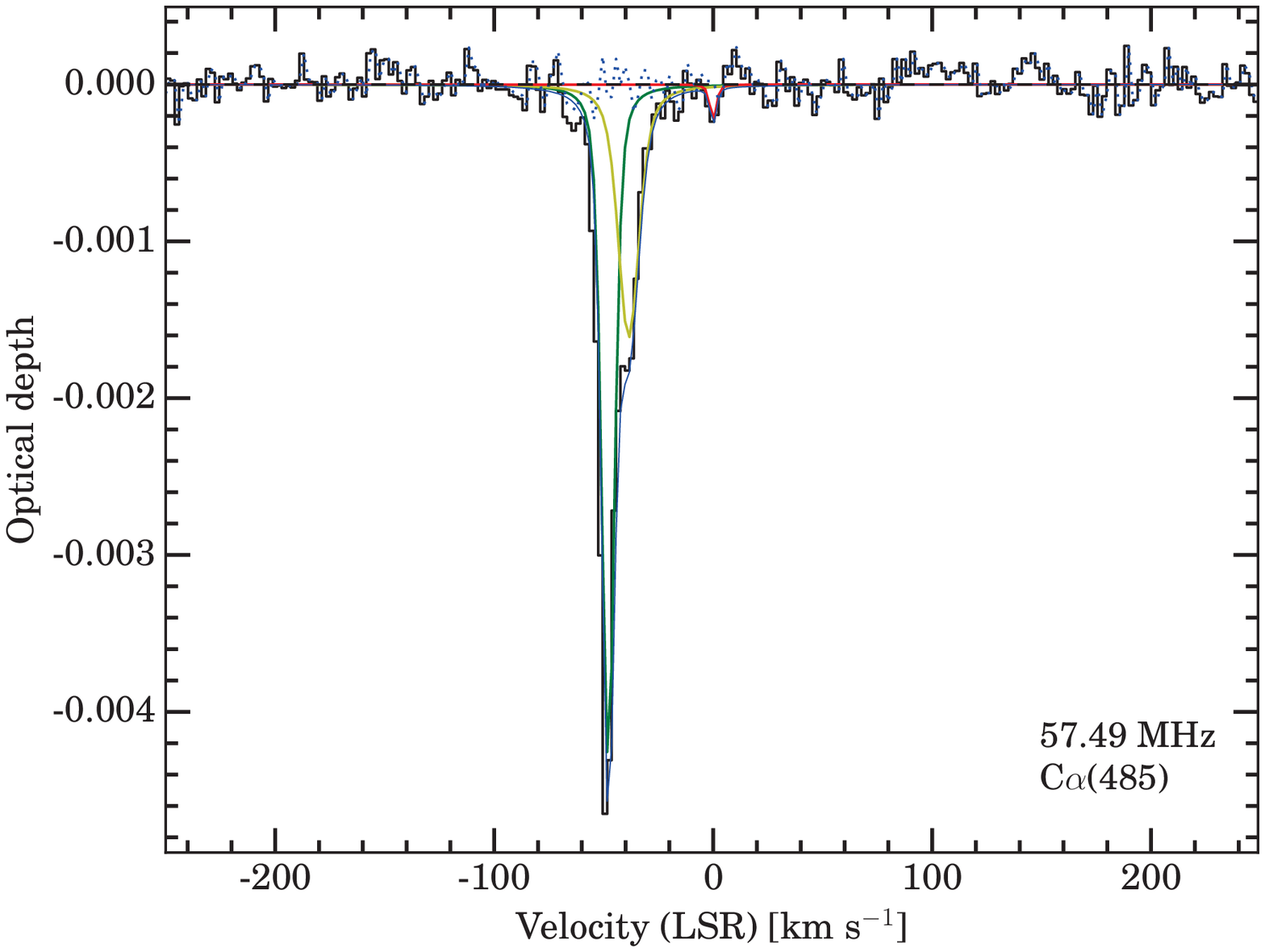}
}
  \vspace{0cm}
  \caption{LOFAR LBA 55-78 MHz: stacked CRRL spectra. The green, yellow and red lines show the decomposition into the -47, -38 and 0~km~s$^{-1}$ components. The blue dotted line shows the residuals after the subtracting the fitted line profiles.}\label{f_app_lba_substack_1}
\end{figure*}

\newpage

\begin{figure*}\vspace{0.5cm}
\mbox{
    \hspace{0.8cm}\includegraphics[width=0.36\textwidth, angle=-90]{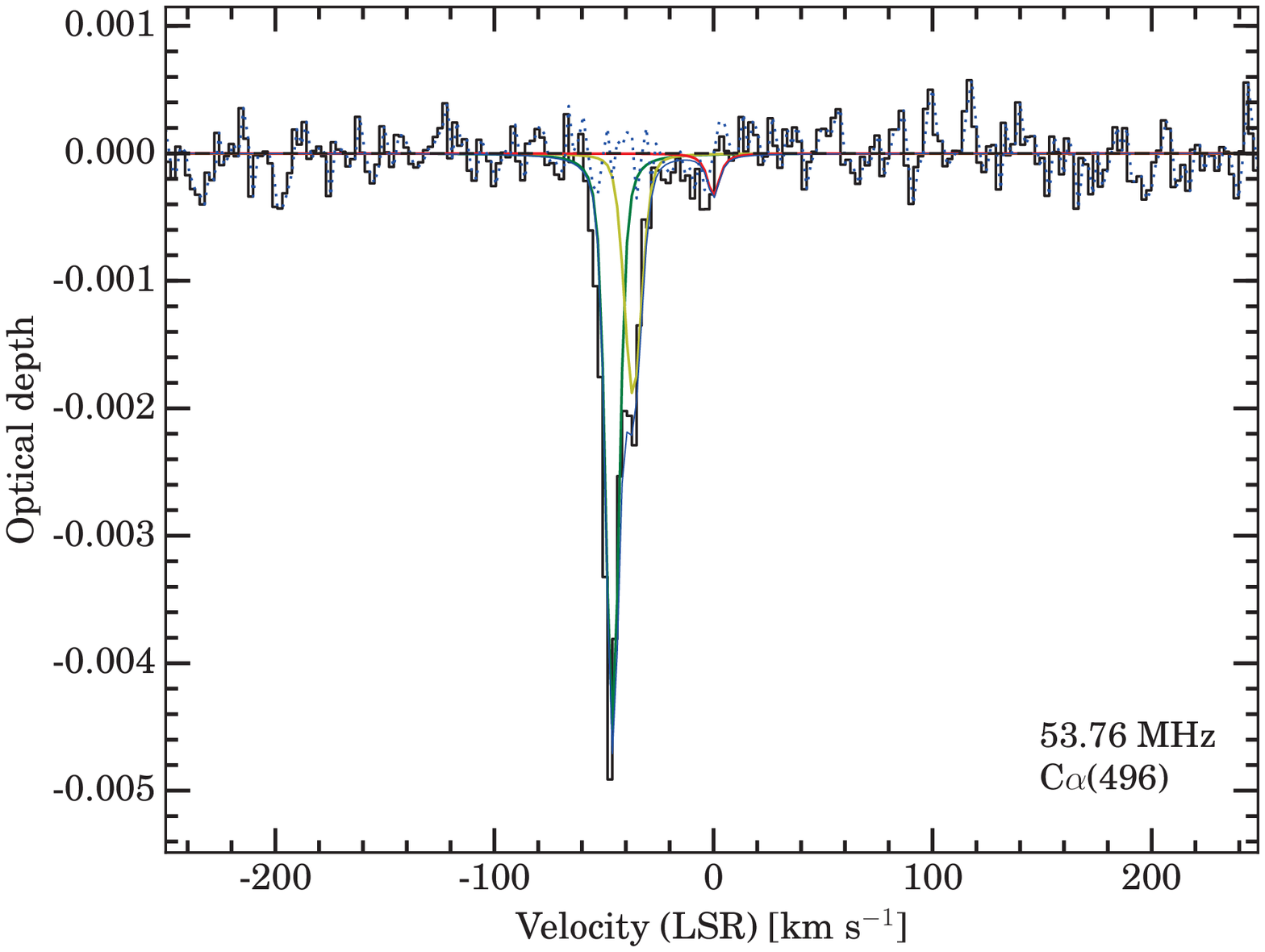}\hspace{0.3cm}
    \includegraphics[width=0.36\textwidth, angle=-90]{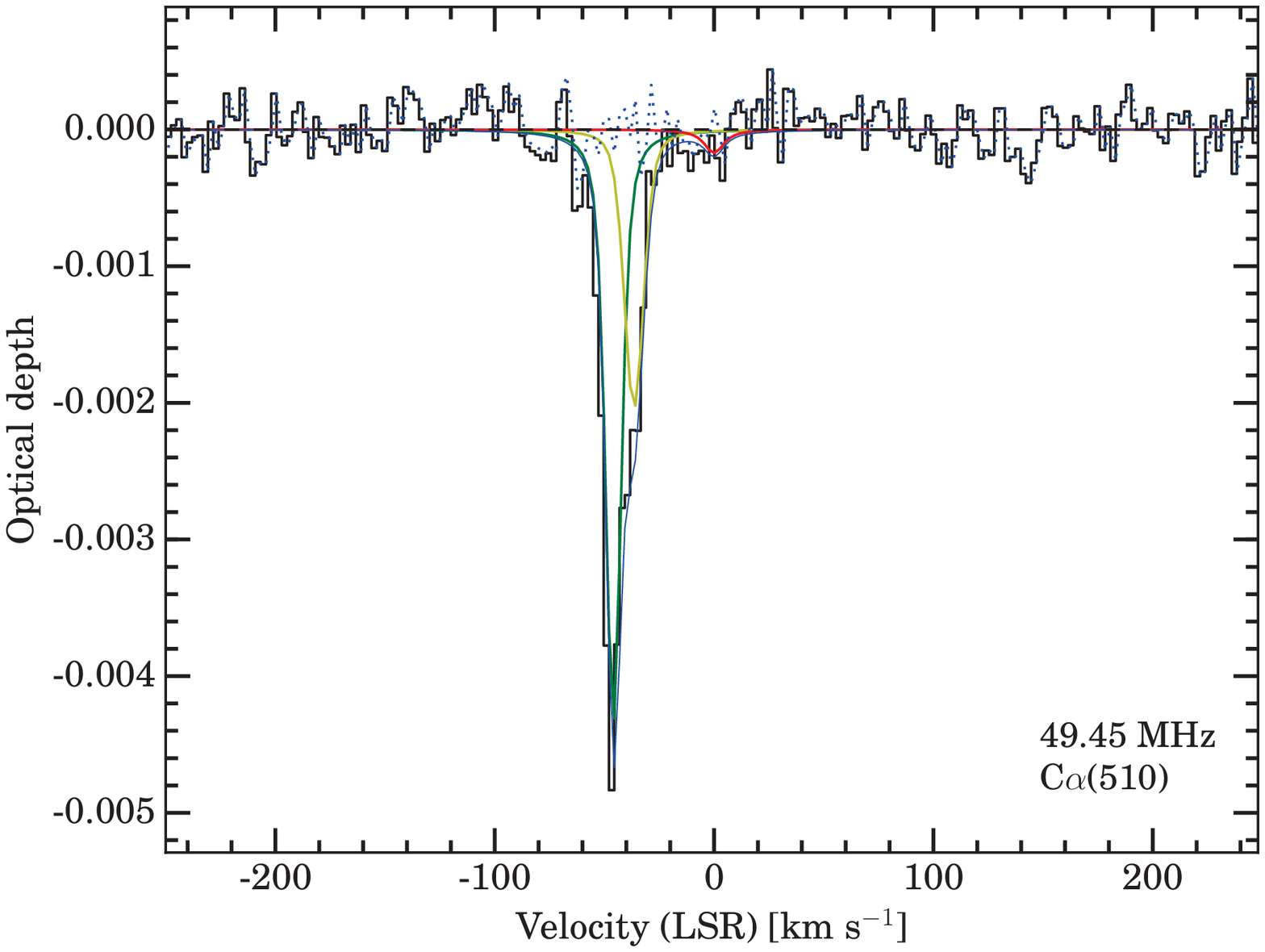}
}\vspace{0.8cm}
\mbox{
    \hspace{0.8cm}\includegraphics[width=0.36\textwidth, angle=-90]{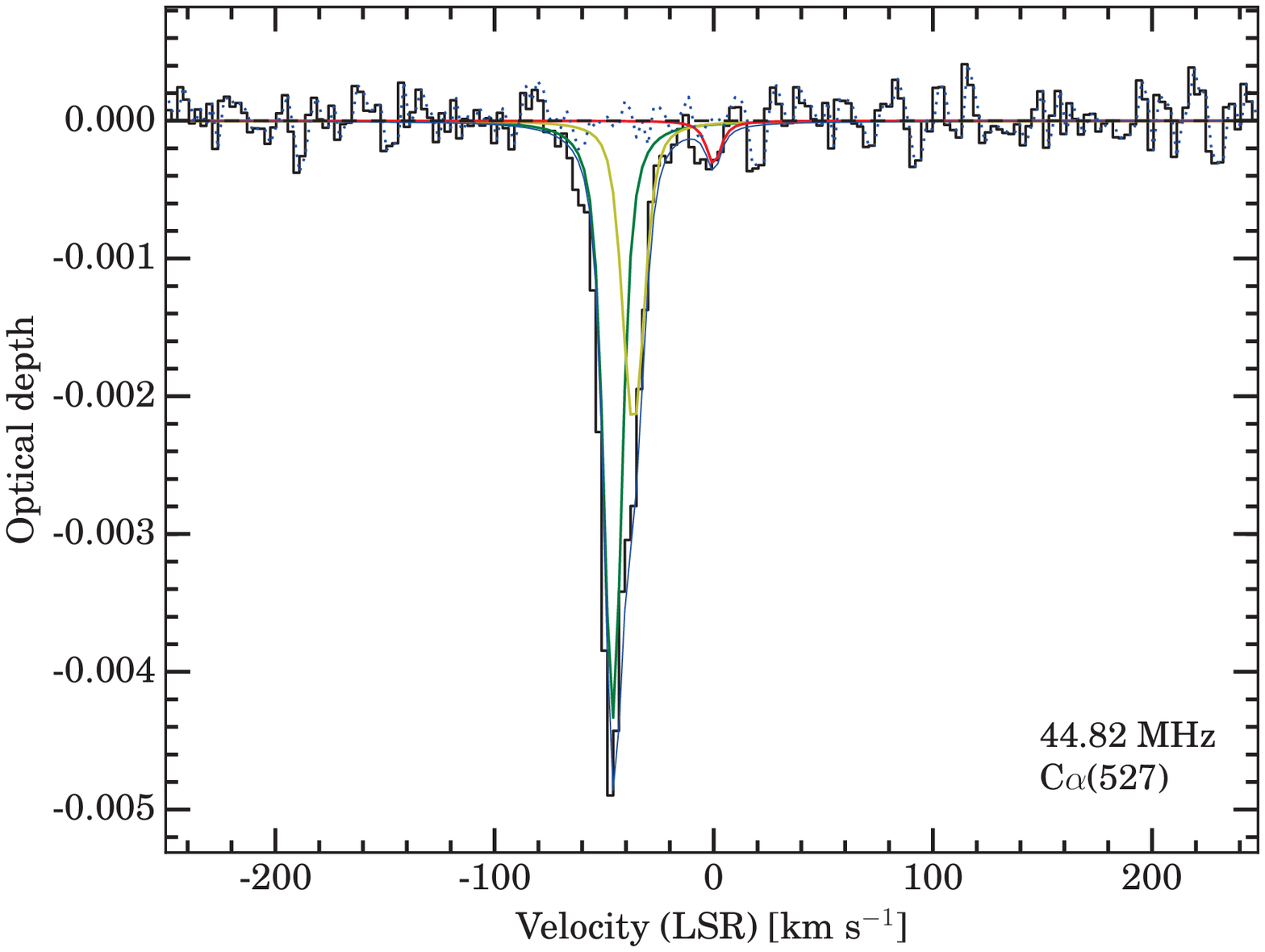}\hspace{0.3cm}
    \includegraphics[width=0.36\textwidth, angle=-90]{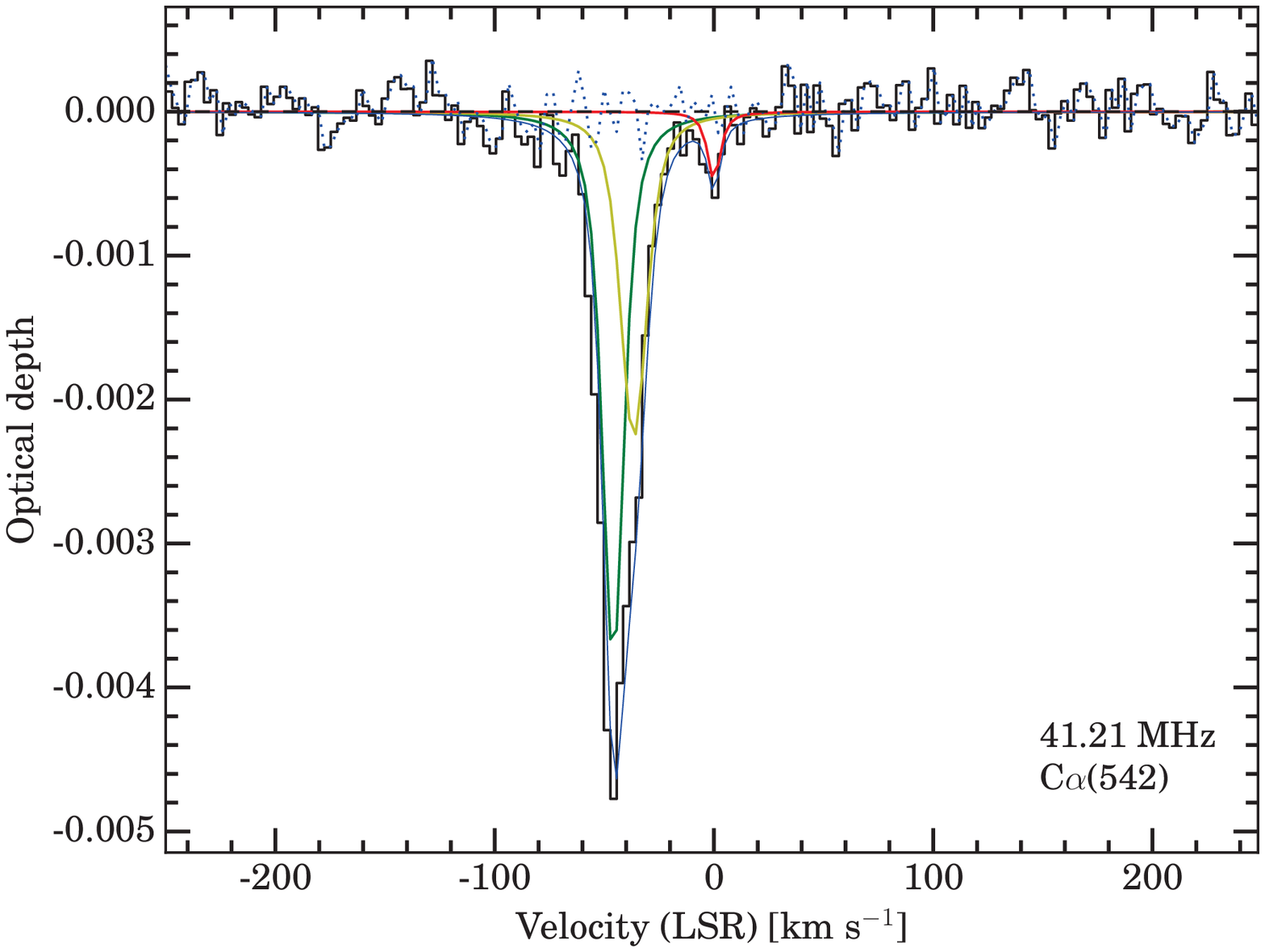}
}\vspace{0.8cm}
\mbox{
    \hspace{0.8cm}\includegraphics[width=0.36\textwidth, angle=-90]{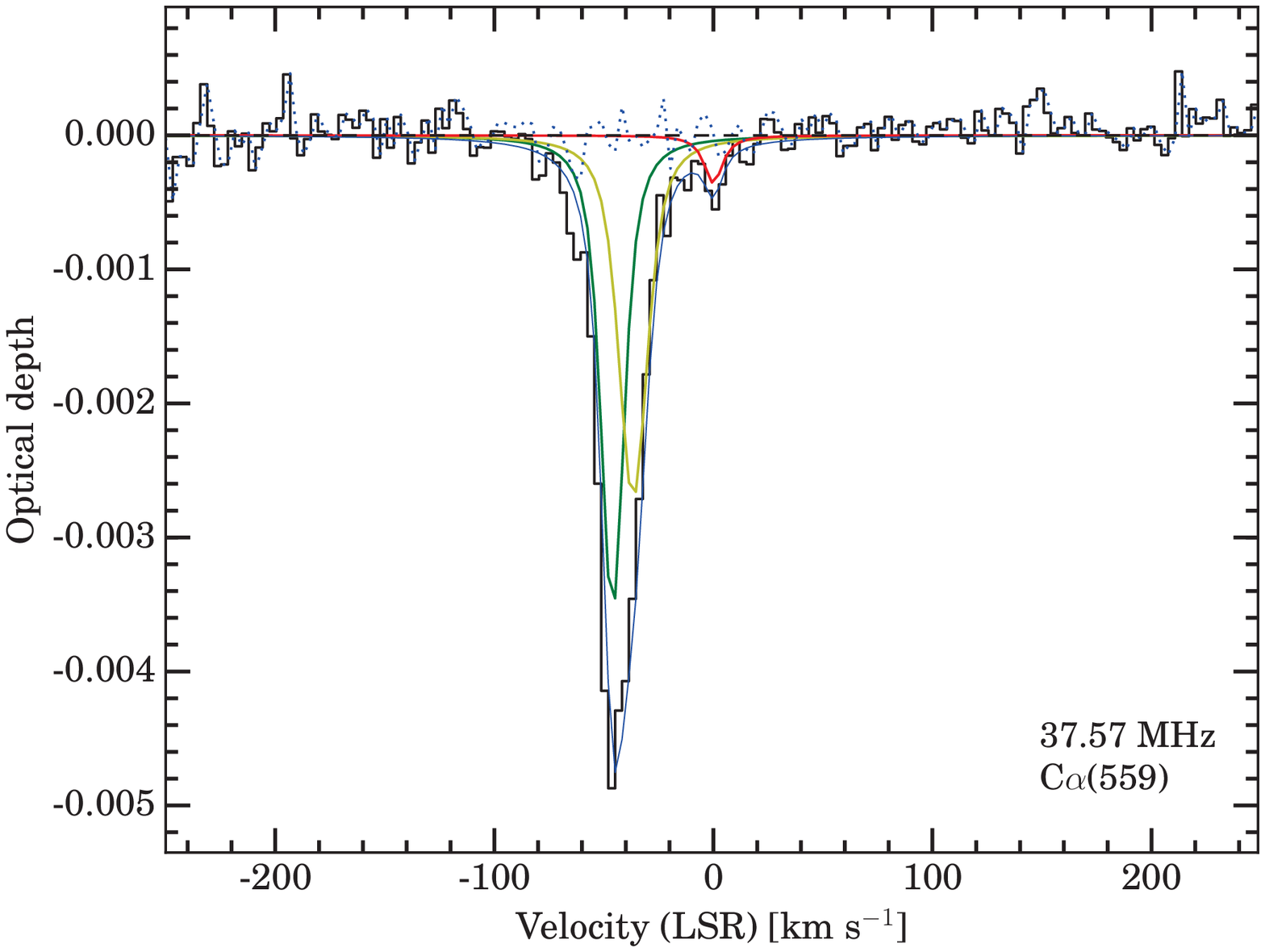}\hspace{0.3cm}
    \includegraphics[width=0.36\textwidth, angle=-90]{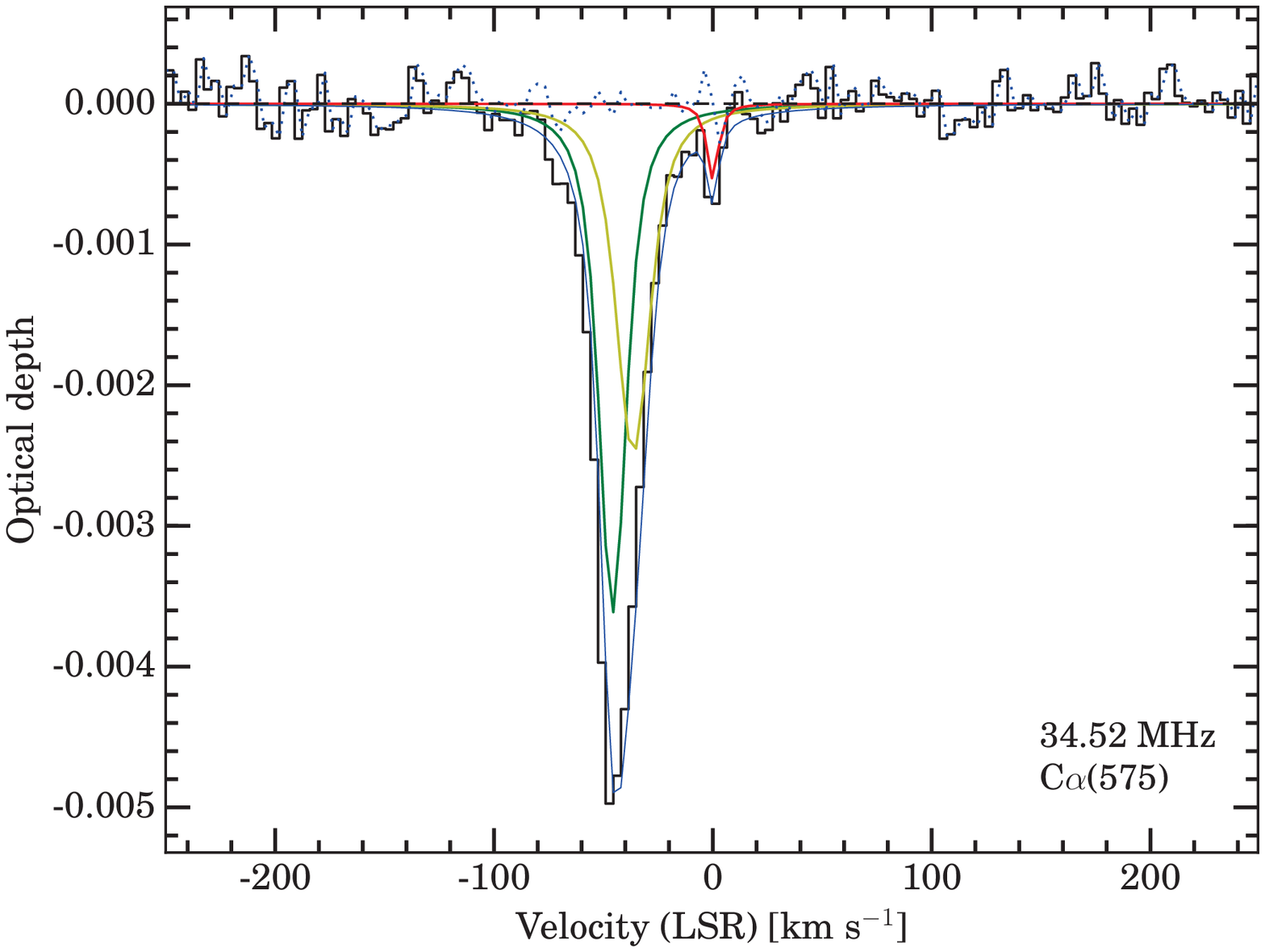}
}
  \vspace{0cm}
  \caption{LOFAR LBA 56-33 MHz:  stacked CRRL spectra. The green, yellow and red lines show the decomposition into the -47, -38 and 0~km~s$^{-1}$ components. The blue dotted line shows the residuals after the subtracting the fitted line profiles.}\label{f_app_lba_substack_2}
\end{figure*}

\begin{figure*}\vspace{1cm}
    \includegraphics[width=0.4\textwidth, angle=90]{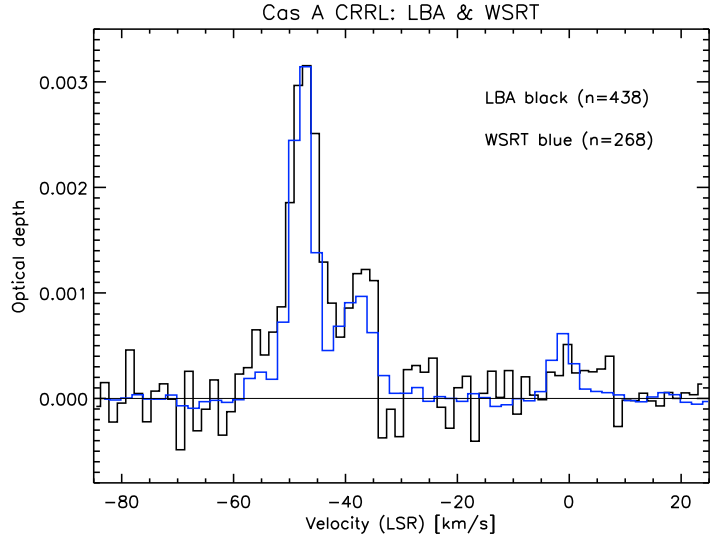}
  \vspace{0.5cm}
  \caption{Overlay of the WSRT P-band (n=268, blue) and LBA (n=438, black) stacked CRRL spectra for the line of sight to Cas~A. The LBA spectrum has been inverted for this comparison and the WSRT spectrum was re-scaled to match the peak of the -47~km~s$^{-1}$ component in the LBA spectrum. The good match of the line profile widths shows that at the highest LBA frequency the line profile is still dominated by Doppler broadening. Both the WSRT and the (unprocessed bandpass) LBA spectra show an excess at -55~km~s$^{-1}$ which could be associated with sulphur RRLs.}\label{f_wsrt_lbahgh}
\end{figure*}

\begin{figure*}\vspace{0cm}
\mbox{
    \includegraphics[width=0.48\textwidth, angle=0]{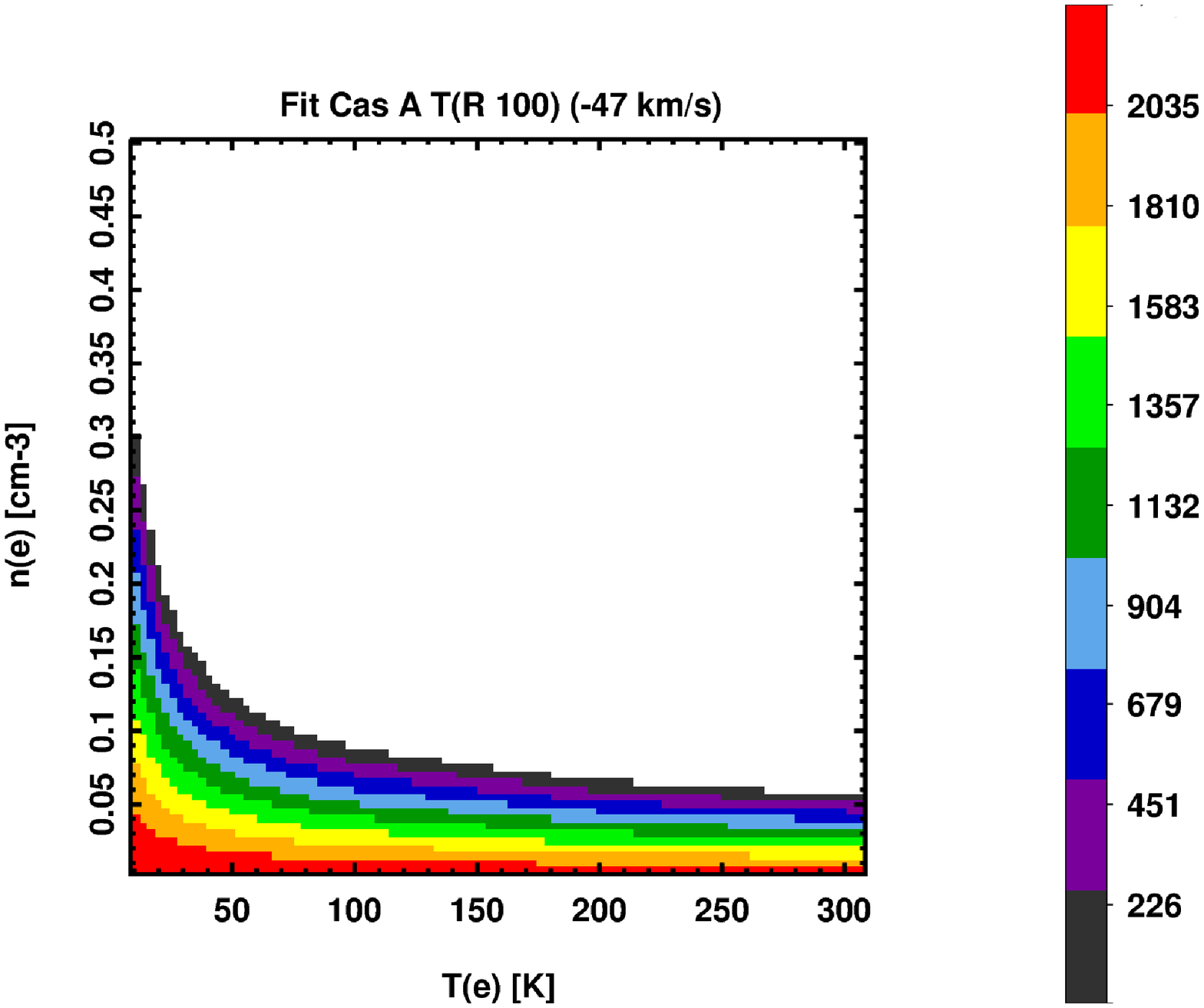}\hspace{0.5cm}
    \includegraphics[width=0.48\textwidth, angle=0]{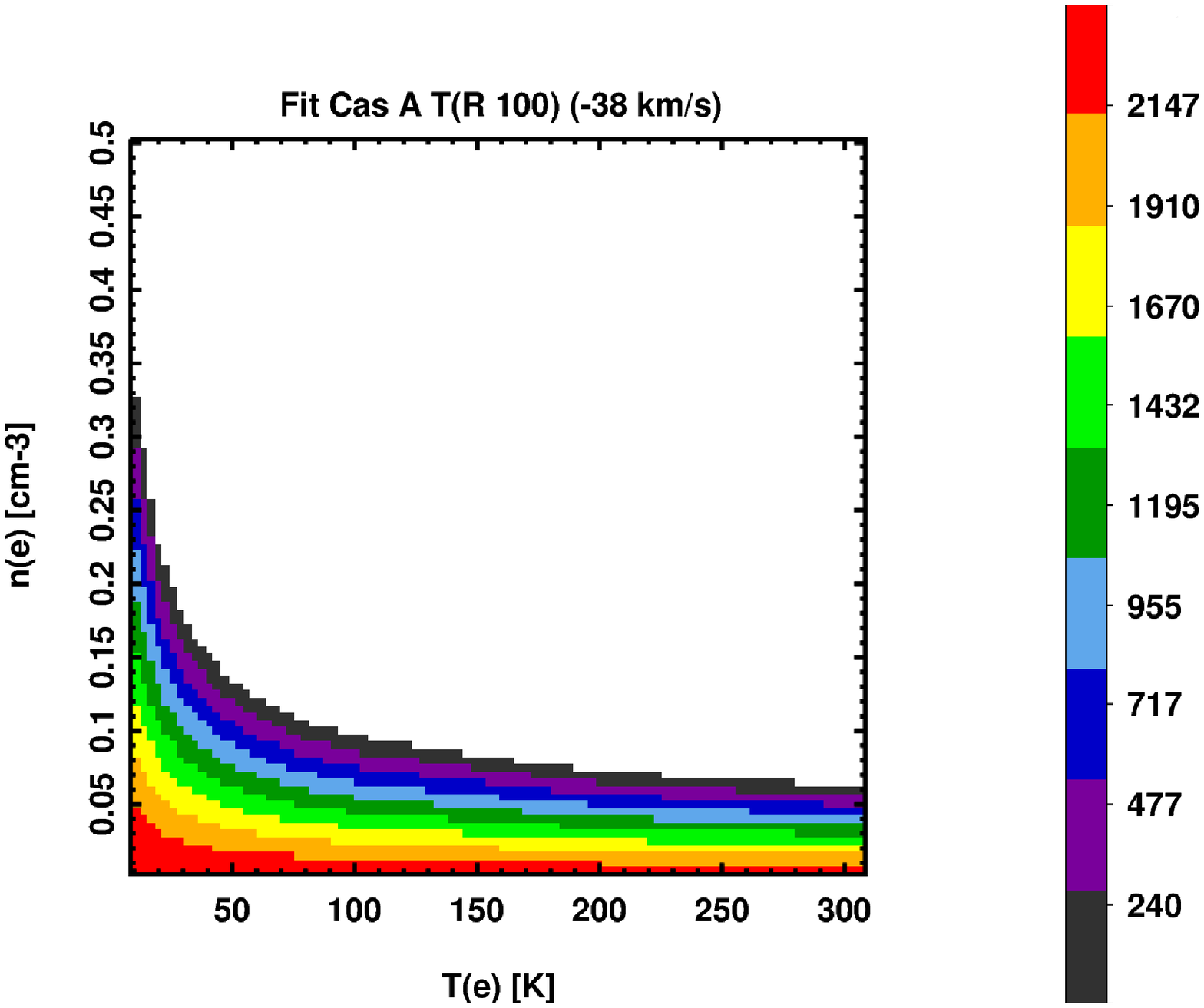}
}
  \vspace{0.5cm}
  \caption{T$_{\rm{R,100}}$ value as a function of T$_{e}$ and n$_{e}$ from our model fits to the CRRL line width (FWHM$_{L}$) vs. quantum number (n) for the Perseus arm components at -47~km~s$^{-1}$ (left) and -38~km~s$^{-1}$ (right). For the -47~km~s$^{-1}$ component we find that T$_{\rm{R,100}}$=1328~K for the best-fit (T$_{e}$, n$_{e}$) combination from the optical depths, see also Fig.~\ref{f_47_bf}. For the -38~km~s$^{-1}$ component we find that T$_{\rm{R,100}}$=1507~K for the best-fit (T$_{e}$, n$_{e}$) combination from the optical depths, see also Fig.~\ref{f_38_bf}. The reduced chi square values for the points shown for both the -47~km~s$^{-1}$ and the -38~km~s$^{-1}$ component are all around 1 showing the strong degeneracy between pressure and radiation broadening. The 1$\sigma$ errors associated with T$_{\rm{R,100}}$ are independent of (T$_{e}$, n$_{e}$) and found to be 83~K and 128~K for the -47~km~s$^{-1}$ and -38~km~s$^{-1}$ component respectively. Constant T$_{\rm{R,100}}$ values trace curves of the form n$_{e}\times$T$_{e}^{-0.5}$.}\label{f_crrl_fwhm_n}
\end{figure*}

\newpage

\begin{figure*}\vspace{0cm}
\mbox{
    \includegraphics[width=0.37\textwidth, angle=90]{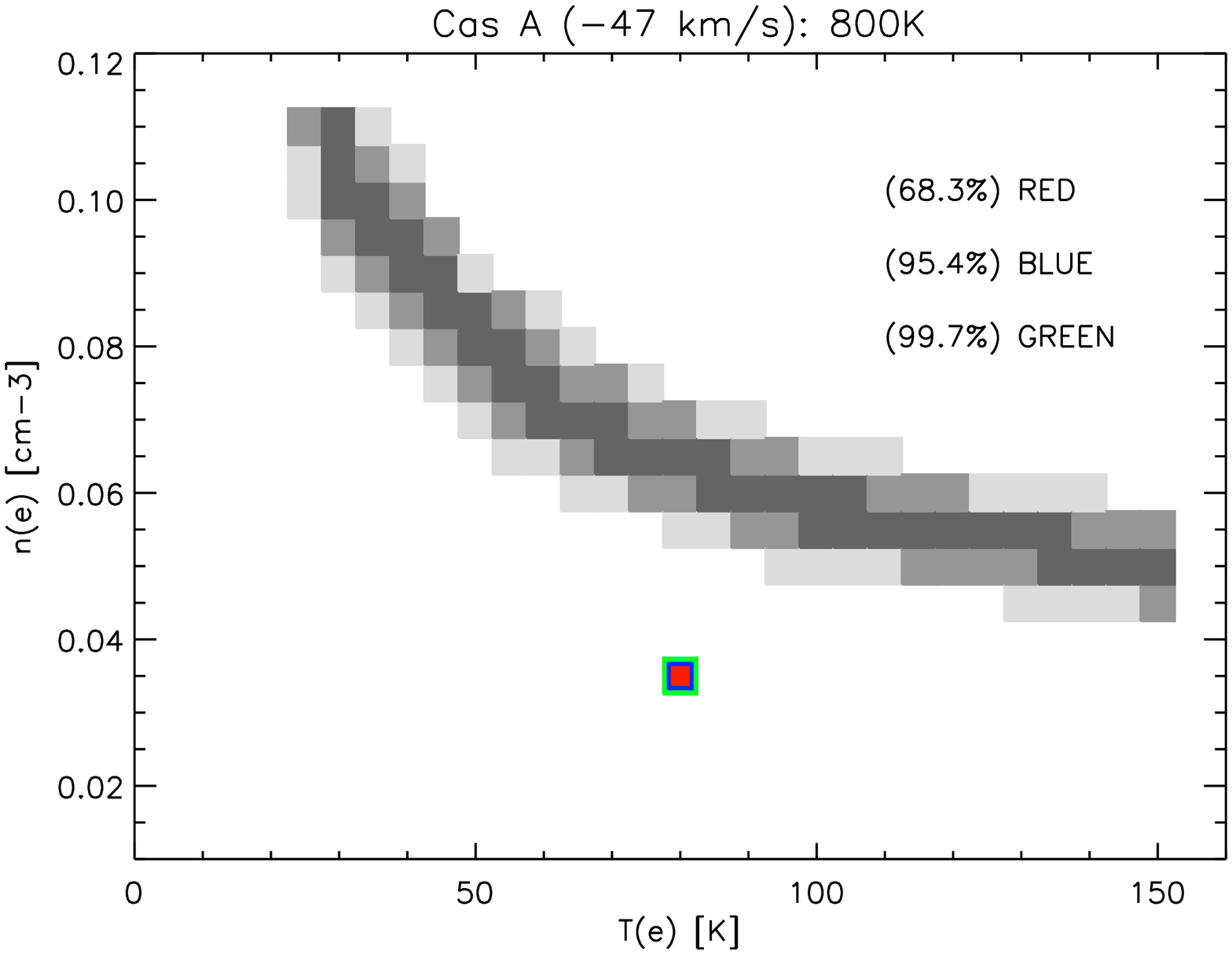}\hspace{0.5cm}
    \includegraphics[width=0.37\textwidth, angle=90]{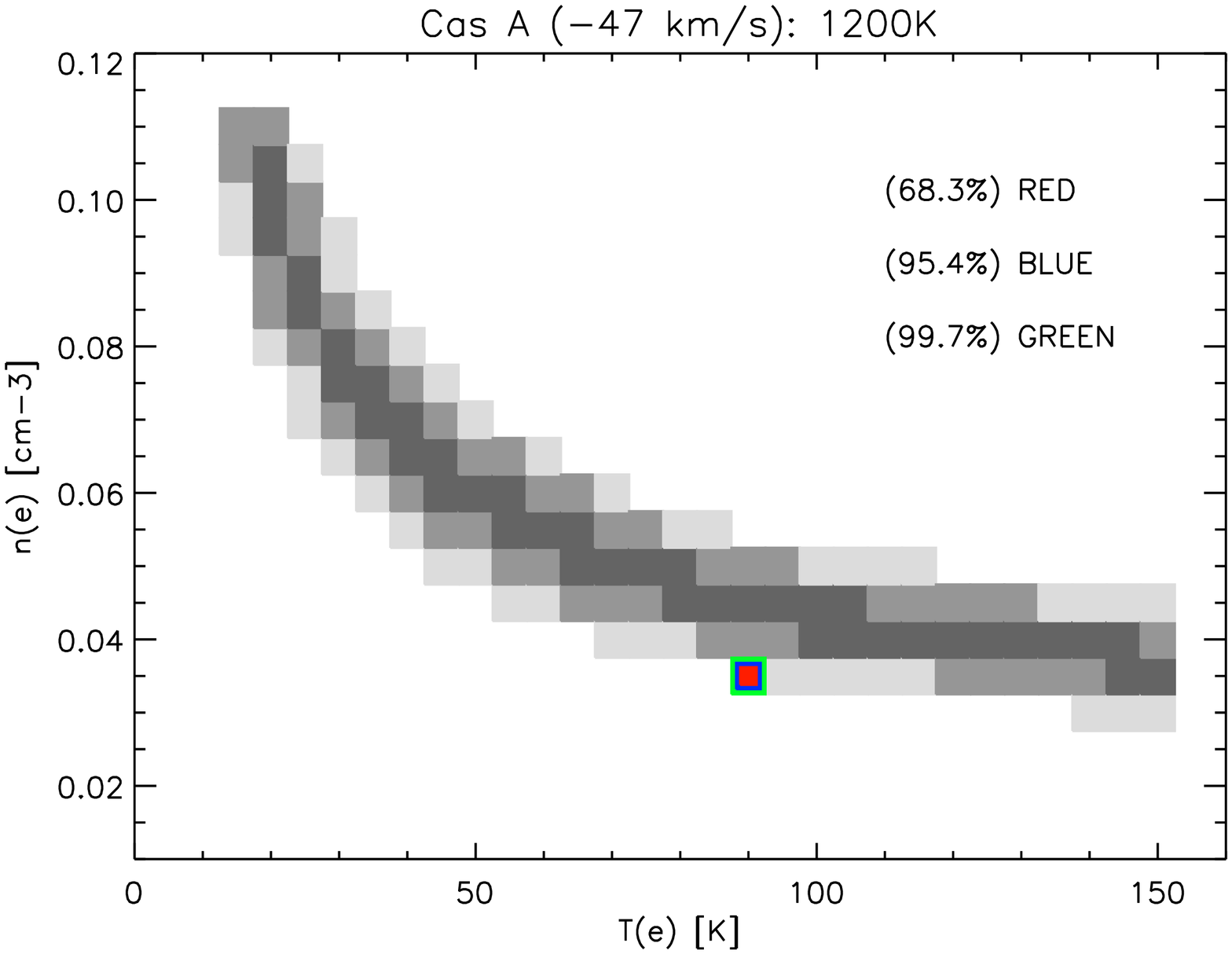}
}\vspace{0.3cm}
\mbox{
    \includegraphics[width=0.37\textwidth, angle=90]{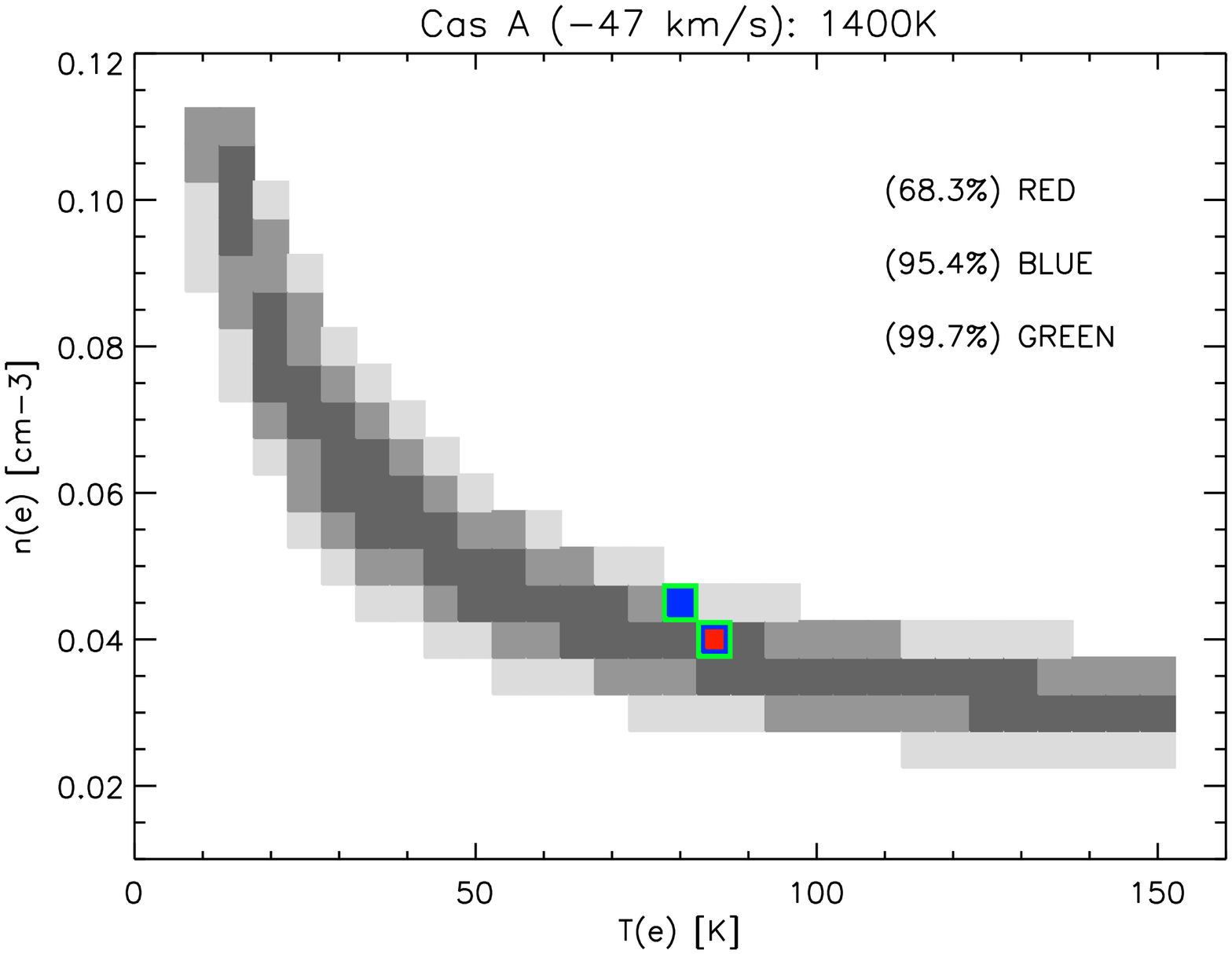}\hspace{0.5cm}
    \includegraphics[width=0.37\textwidth, angle=90]{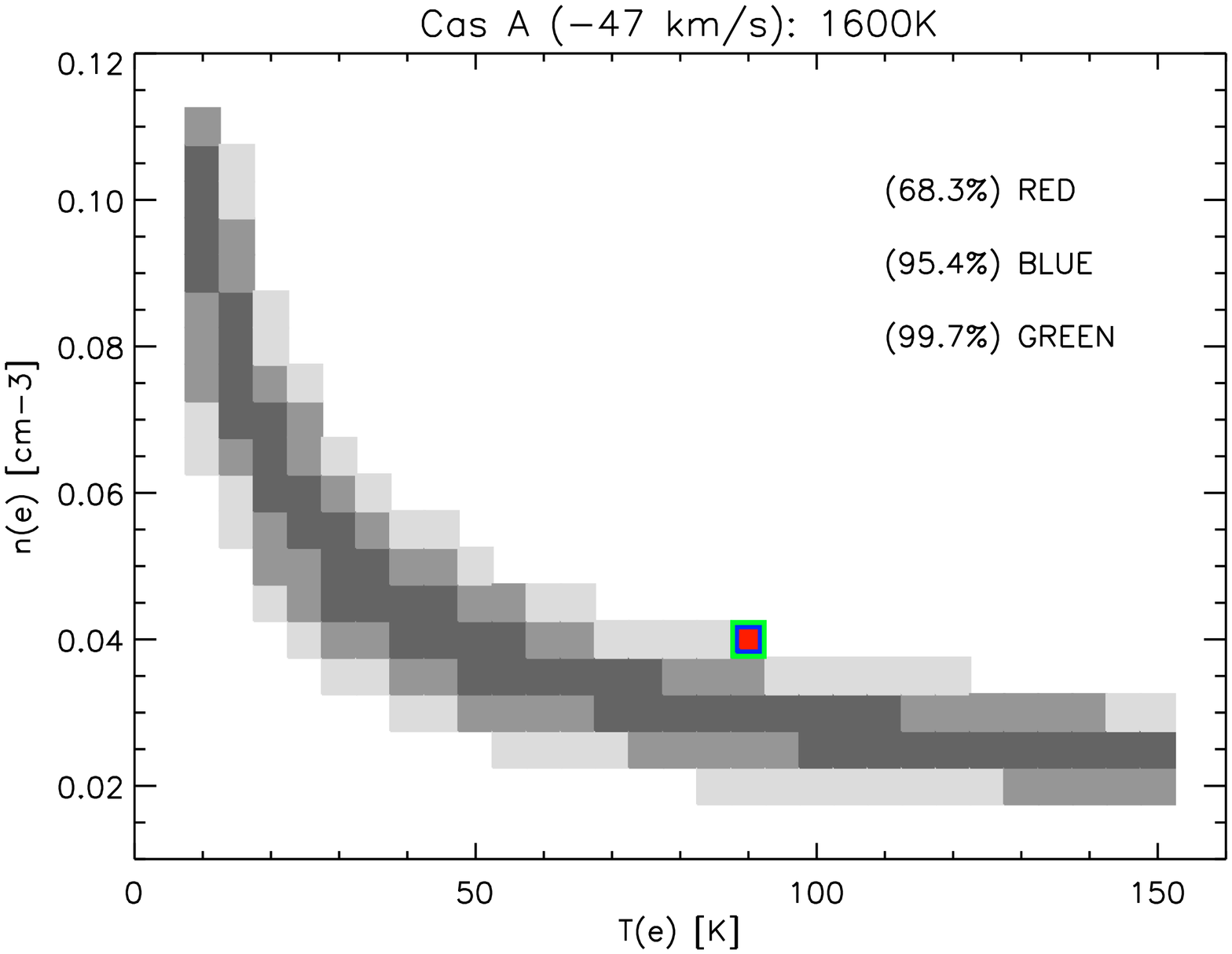}
}\vspace{0.3cm}
\mbox{
    \includegraphics[width=0.37\textwidth, angle=90]{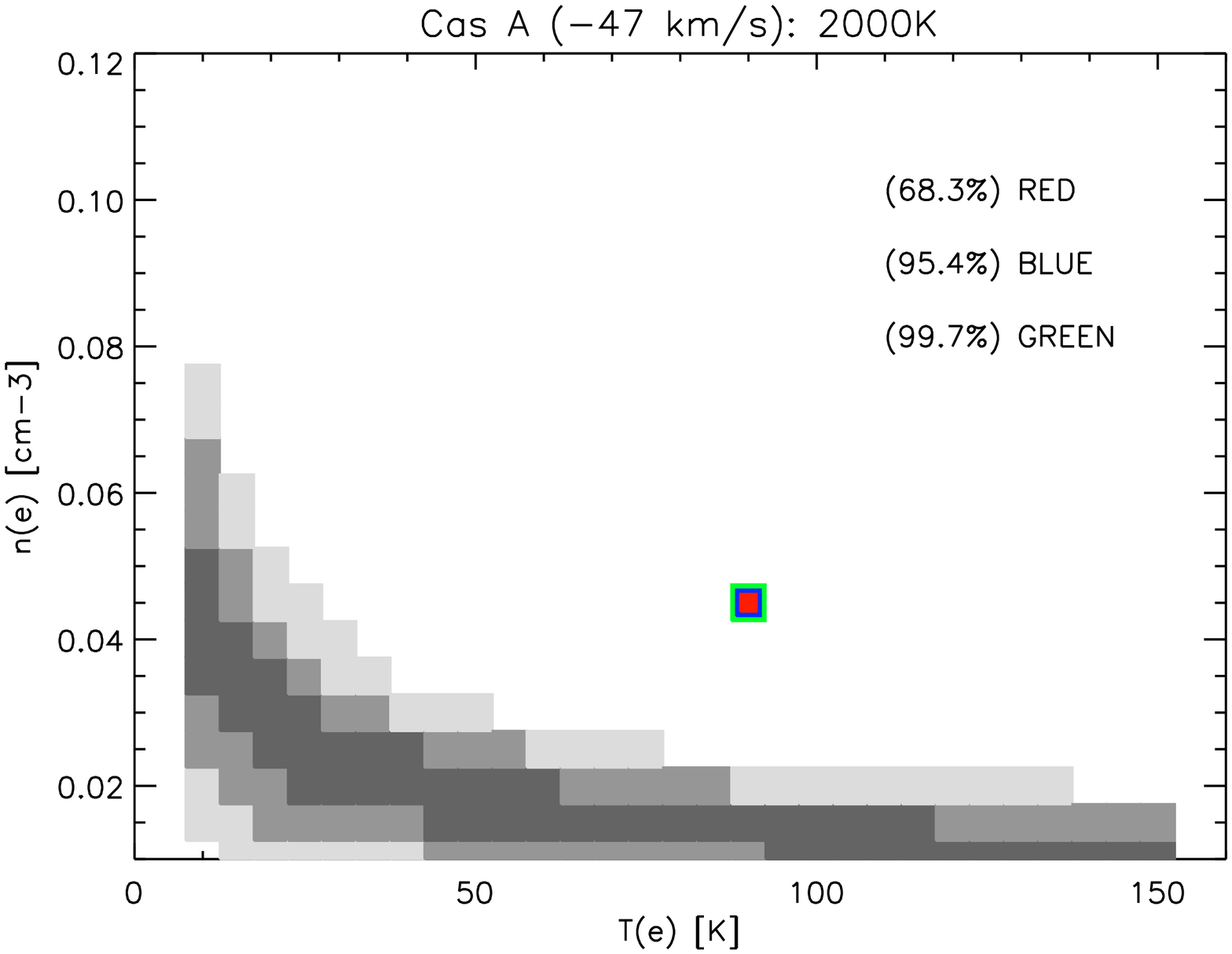}
}
  \vspace{0.5cm}
  \caption{Combined model constraints for the CRRL integrated optical depth ($\tau$) and line width (FWHM) for the Perseus arm component at -47~km~s$^{-1}$. The 1, 2, and 3 sigma confidence limits from the integrated optical depth fitting are shown by the red, blue and green boxes respectively. The red and blue boxes should have the same size as the green boxes, but they have been decreased in size for clarity. The 1, 2, and 3 sigma line width error limits are shown by the black, dark-grey and light grey boxes respectively. The model fits shown have been carried out for five different Tr,100 values of our (T$_{e}$, n$_{e}$) grid; (top-left) Tr,100~=~800~K. (top-right) Tr,100~=~1200~K. (middle-left) Tr,100~=~1400~K. (middle-right) Tr,100~=~1600~K. (bottom-left) Tr,100~=~2000~K.}\label{f_47_crrl_comb}
\end{figure*}

\newpage

\begin{figure*}\vspace{0cm}
\mbox{
    \includegraphics[width=0.37\textwidth, angle=90]{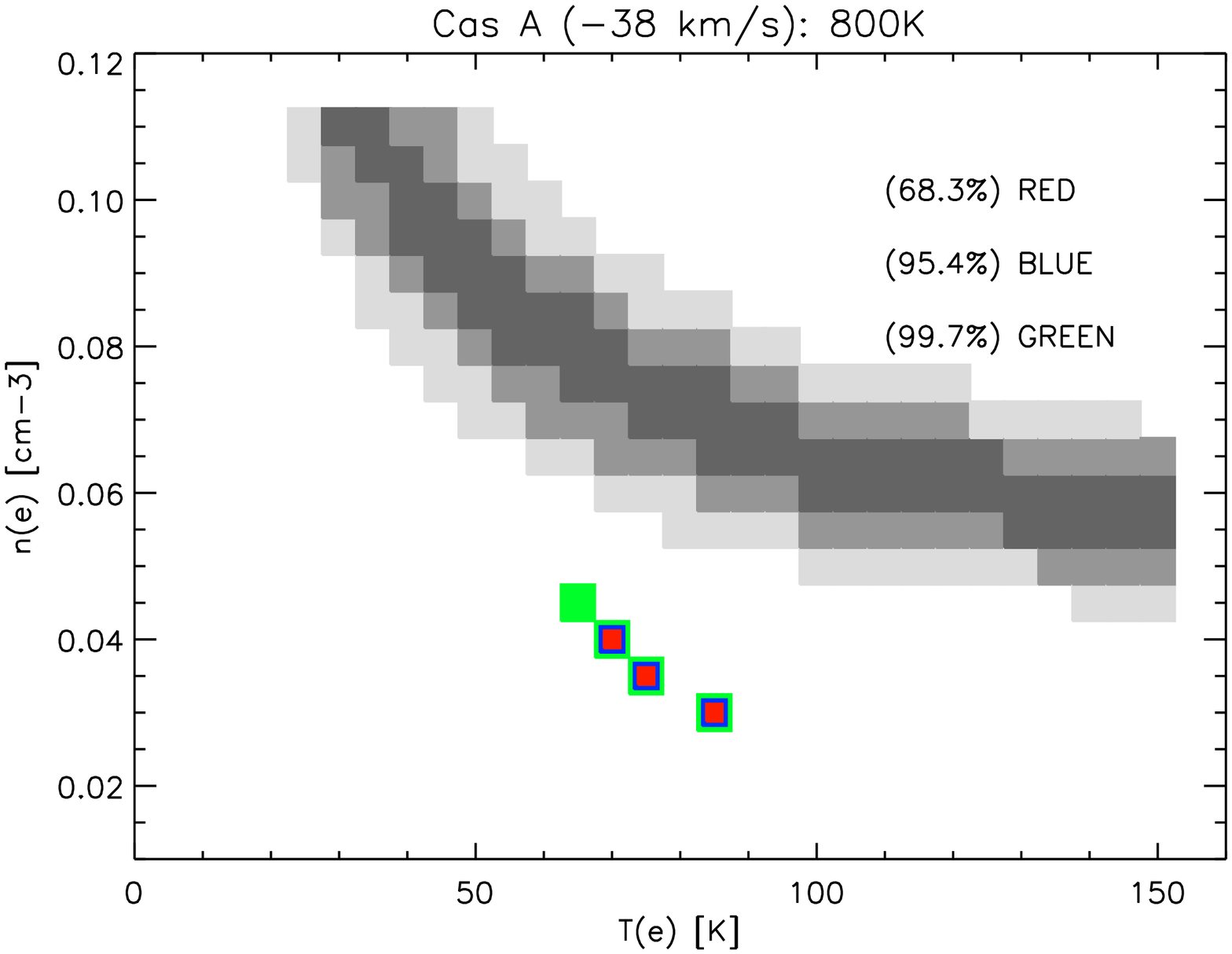}\hspace{0.5cm}
    \includegraphics[width=0.37\textwidth, angle=90]{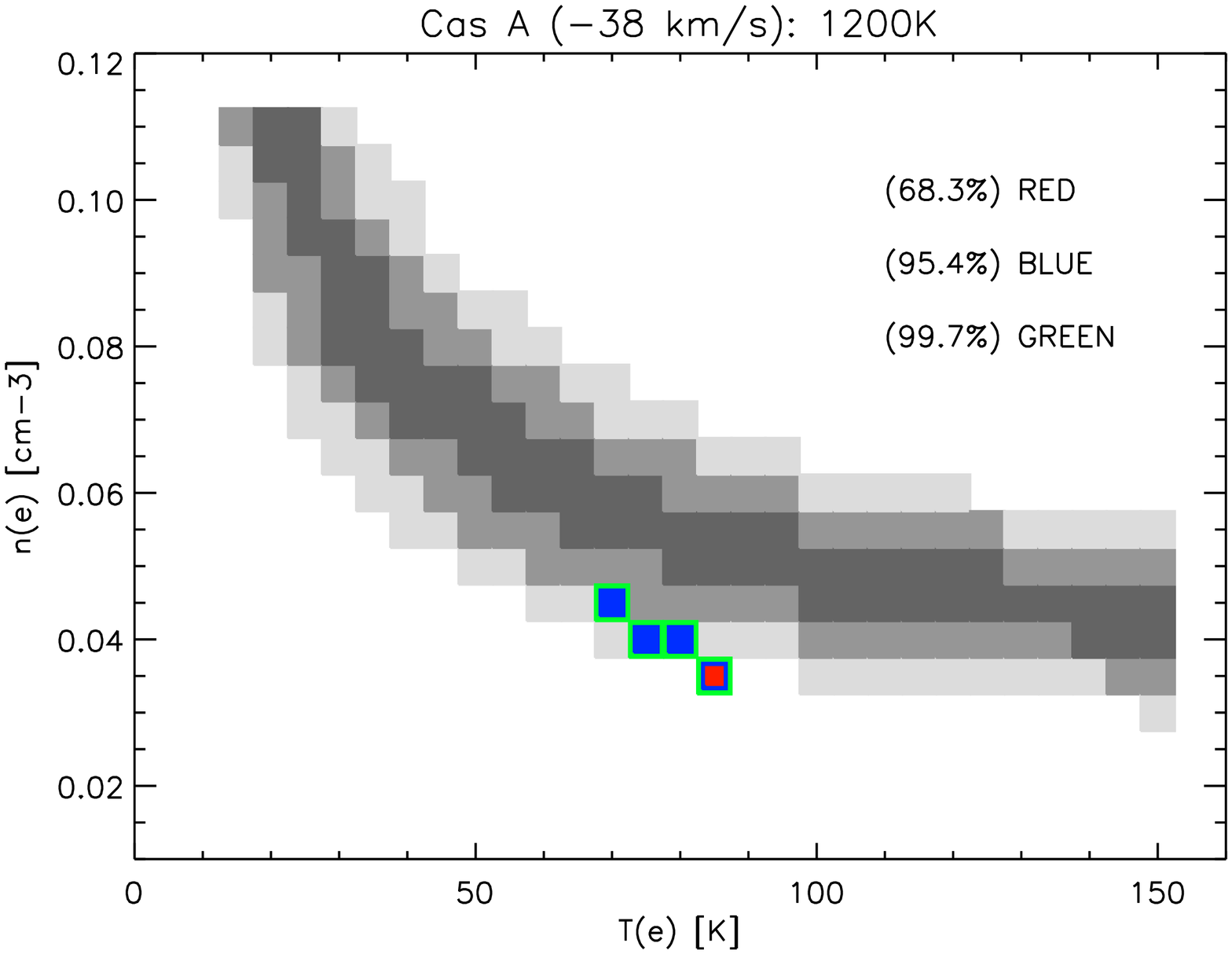}
}\vspace{0.3cm}
\mbox{
    \includegraphics[width=0.37\textwidth, angle=90]{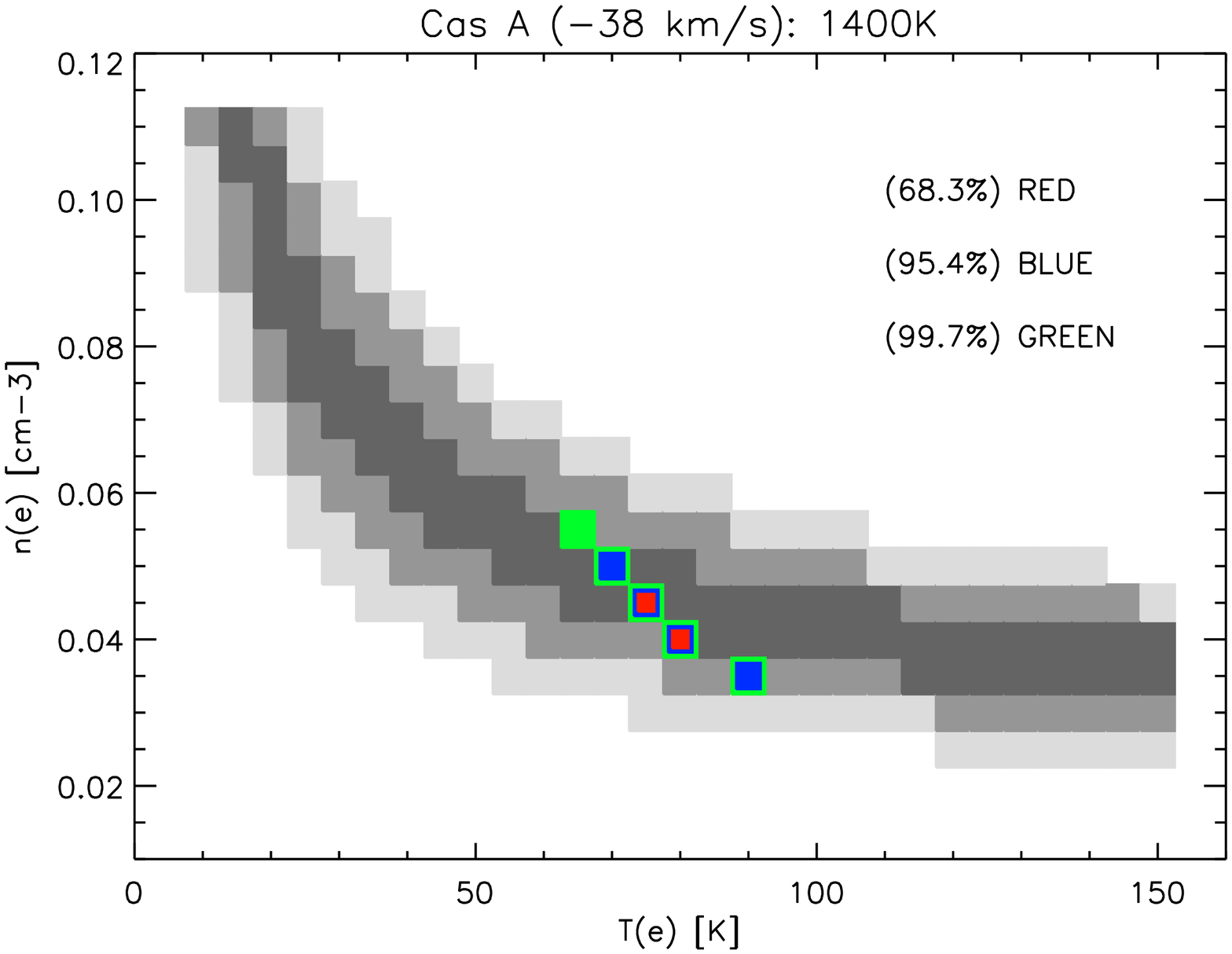}\hspace{0.5cm}
    \includegraphics[width=0.37\textwidth, angle=90]{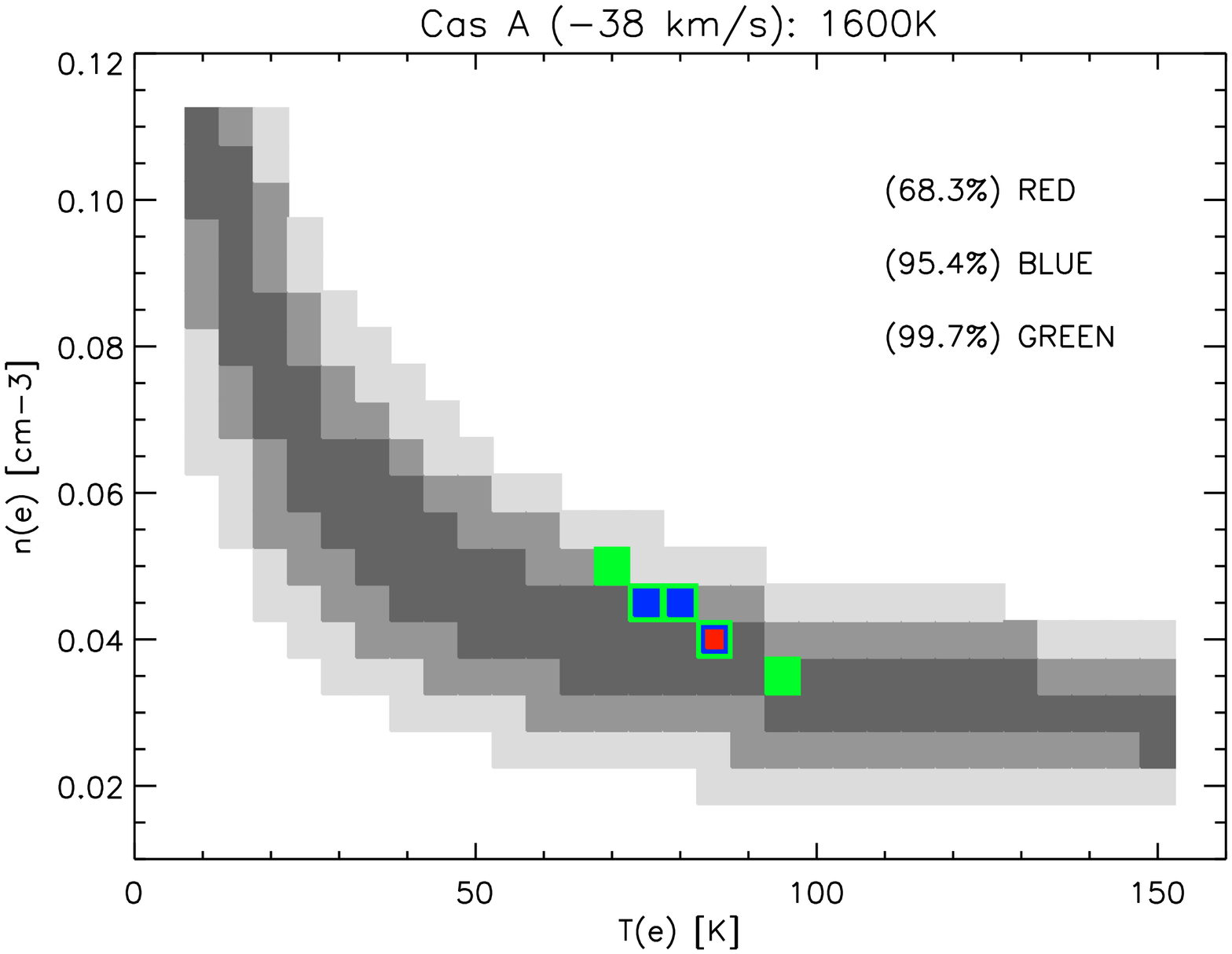}
}\vspace{0.3cm}
\mbox{
    \includegraphics[width=0.37\textwidth, angle=90]{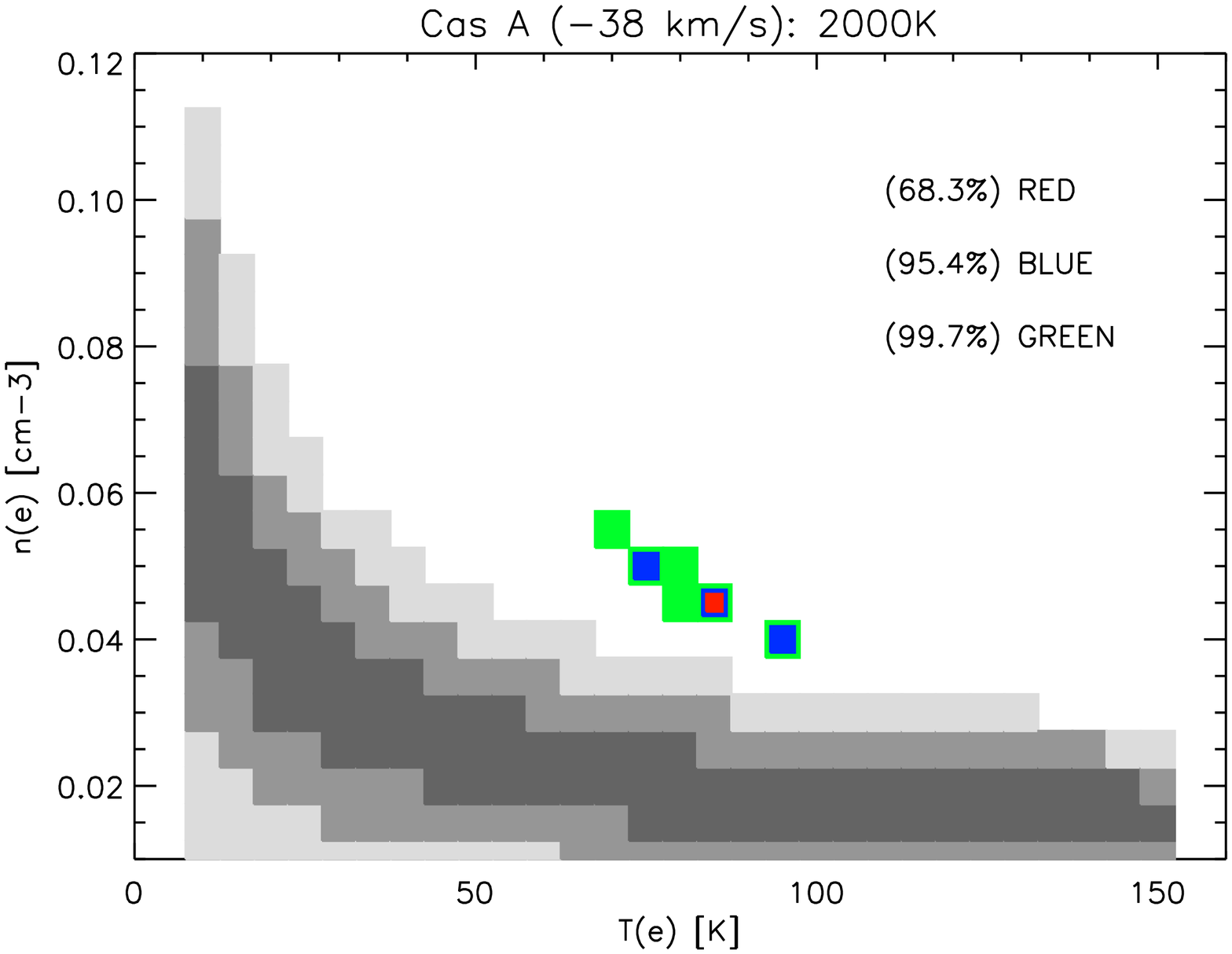}
}
  \vspace{0.5cm}
  \caption{Combined model constraints for the CRRL integrated optical depth ($\tau$) and line width (FWHM) for the Perseus arm component at -38~km~s$^{-1}$. The 1, 2, and 3 sigma confidence limits from the integrated optical depth fitting are shown by the red, blue and green boxes respectively. The red and blue boxes should have the same size as the green boxes, but they have been decreased in size for clarity. The 1, 2, and 3 sigma line width error limits are shown by the black, dark-grey and light grey boxes respectively. The model fits shown have been carried out for five different Tr,100 values of our (T$_{e}$, n$_{e}$) grid; (top-left) Tr,100~=~800~K. (top-right) Tr,100~=~1200~K. (middle-left) Tr,100~=~1400~K. (middle-right) Tr,100~=~1600~K. (bottom-left) Tr,100~=~2000~K. Our (T$_{e}$, n$_{e}$) grid is sampled in steps of 5~K for T$_{e}$ in the range 10-150~K and in steps of 0.005~cm$^{-3}$ for n$_{e}$ in the range 0.01 to 0.11~cm$^{-3}$.}\label{f_38_crrl_comb}
\end{figure*}

\begin{figure*}\vspace{0cm}
    \includegraphics[width=0.36\textwidth, angle=90]{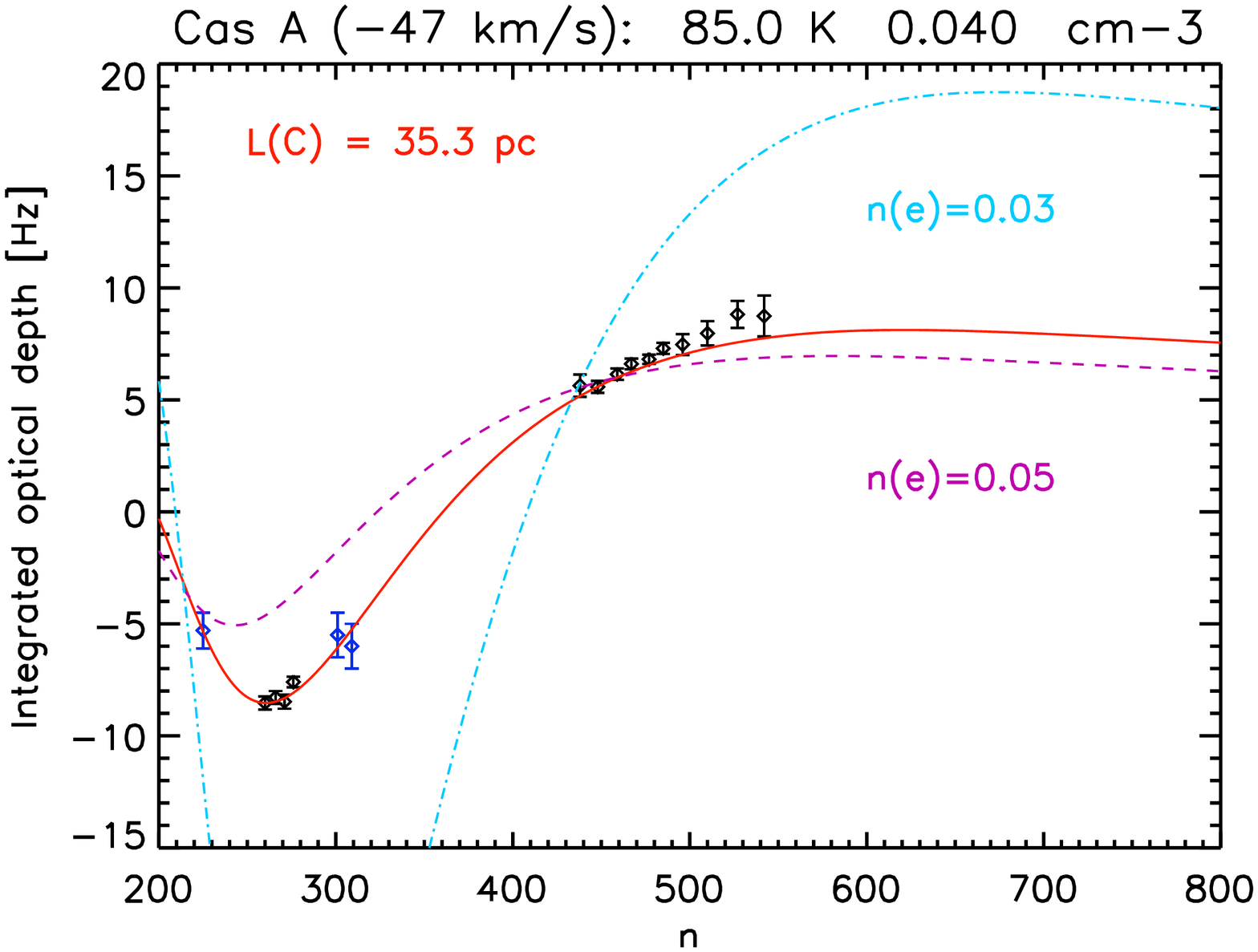}\hspace{0.5cm}
    \includegraphics[width=0.36\textwidth, angle=90]{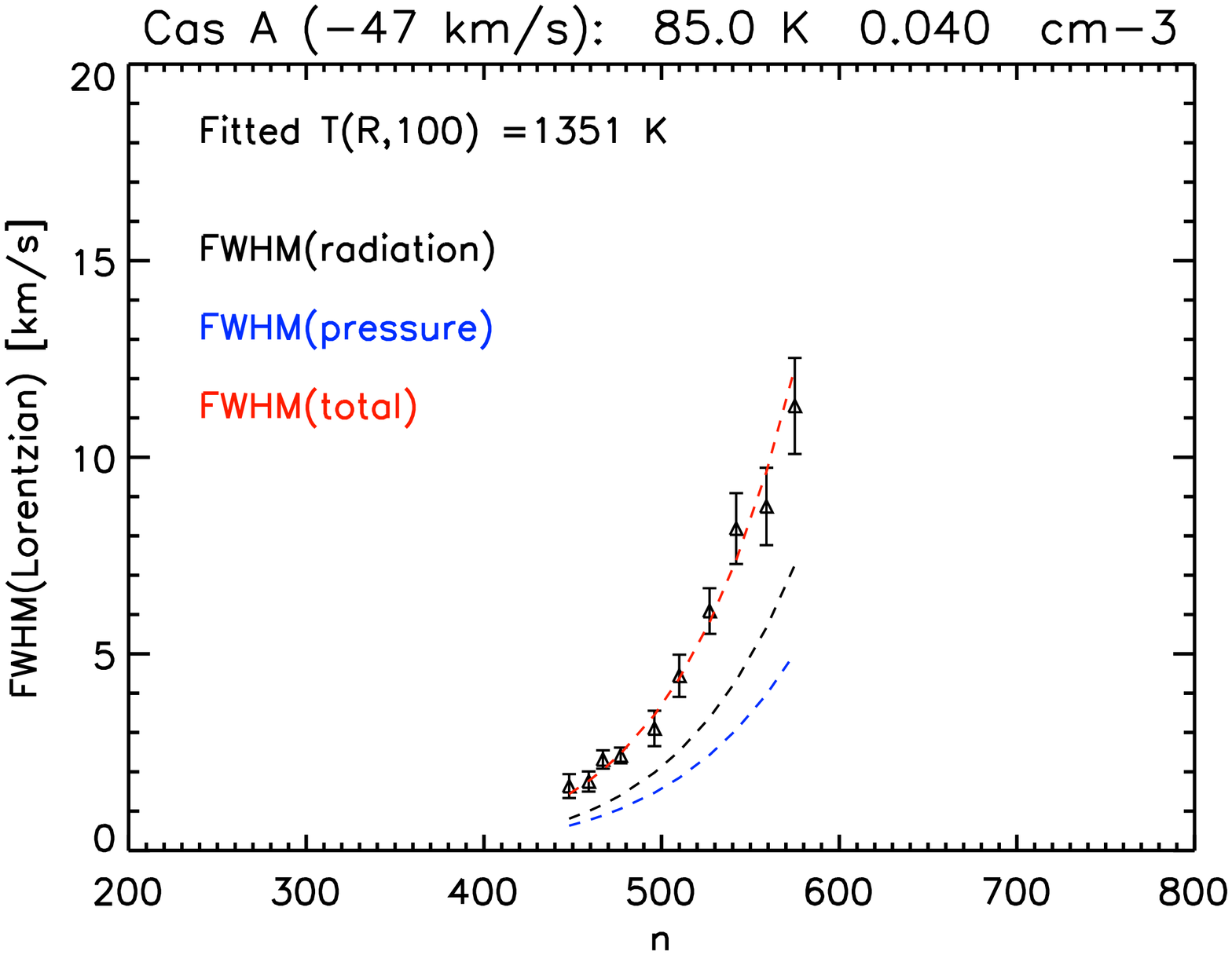}
  \vspace{0.5cm}
  \caption{Best-fit CRRL optical depth and line width models overlaid on the measurements for the -47~km~s$^{-1}$ component. Our LOFAR and WSRT data is shown in black. In addition we show the literature data that we have used in blue. The literature measurements are taken from Kantharia et al. (1998) and Payne et al. (1989) for n = 225, 301 and 309. (left) The red curve shows the best-fit optical depth model with L$_{C}$=35.3~pc for T$_{\rm{R,100}}$=1400~K. In addition we show two optical depth models for the same best-fit T$_{e}$ but with a 25 percent difference in n$_{e}$ (dot-dash: n$_{e}$=0.03~cm$^{-3}$ and dashed: n$_{e}$=0.05~cm$^{-3}$). (right) The red curve shows the best-fit line width model with T$_{\rm{R,100}}$=1351~K. The red curves in both panels have the same $T_{e}$ and $n_{e}$ values that were taken from the best-fit CRRL optical depth model, see Fig.~\ref{f_47_crrl_comb}.}\label{f_47_bf}
\end{figure*}

\begin{figure*}\vspace{0cm}
    \includegraphics[width=0.36\textwidth, angle=90]{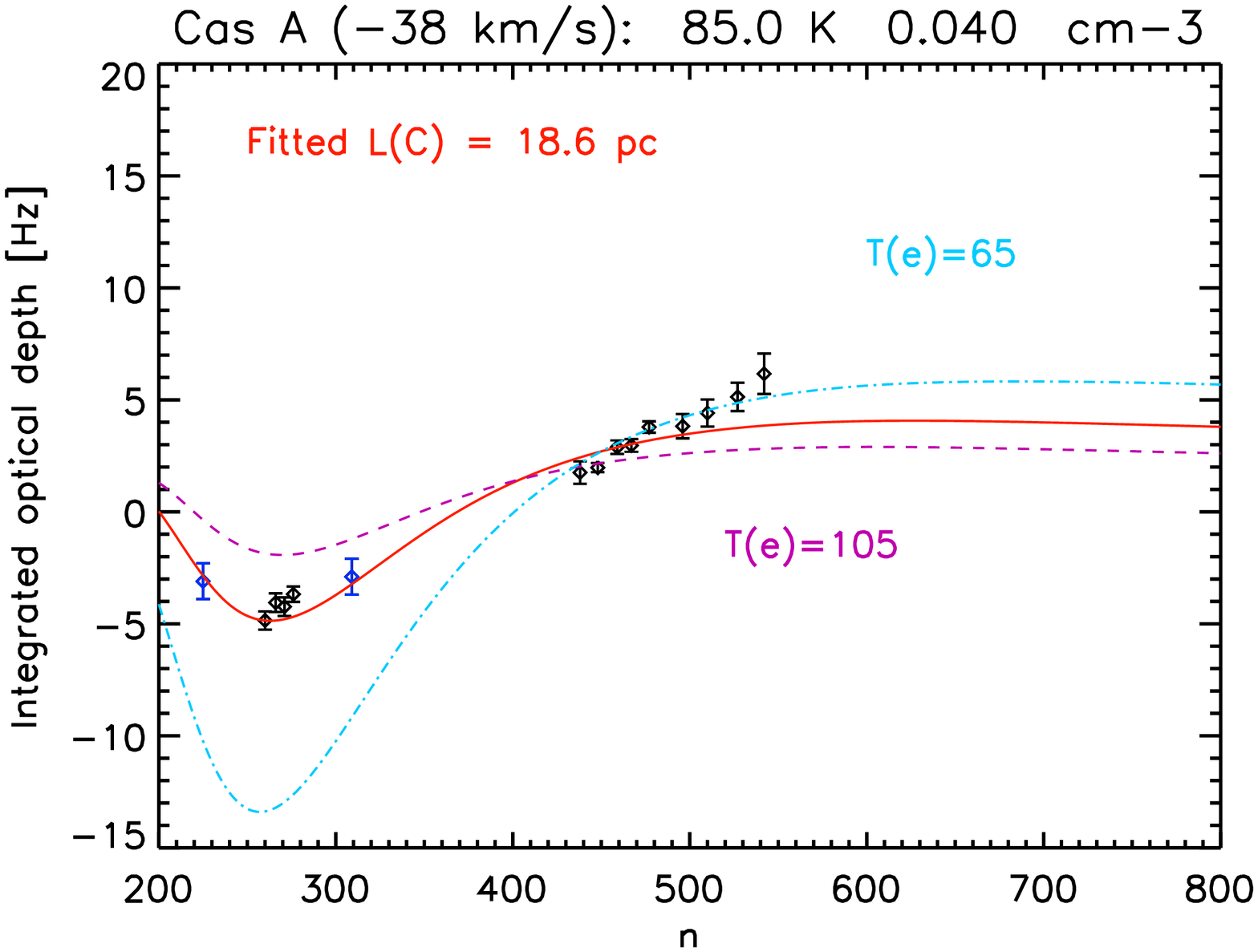}\hspace{0.5cm}
    \includegraphics[width=0.36\textwidth, angle=90]{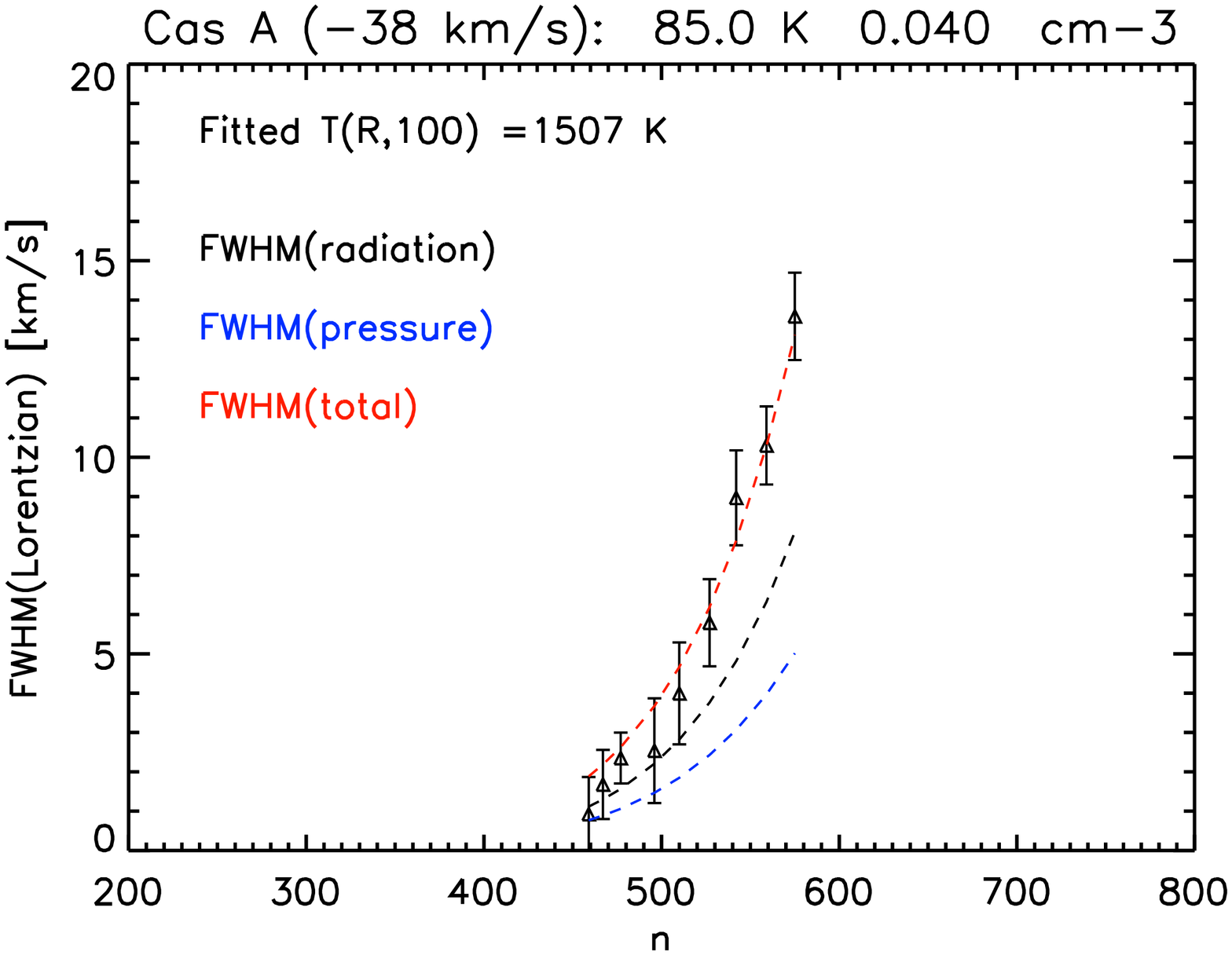}
  \vspace{0.5cm}
  \caption{Best-fit CRRL optical depth and line width models overlaid on the measurements for the -38~km~s$^{-1}$ component. Our LOFAR and WSRT data is shown in black. In addition we show the literature data that we have used in the fit in blue. The literature measurements are taken from Kantharia et al. (1998) and Payne et al. (1989) for n = 225 and 309. (left) The red curve shows the best-fit optical depth model with L$_{C}$=18.6~pc for T$_{\rm{R,100}}$=1600~K. In addition we show two optical depth models for the same best-fit n$_{e}$ but with a 25 percent difference in T$_{e}$ (dot-dash: T$_{e}$=65~K and dashed: T$_{e}$=105~K). (right) The red curve shows the best-fit line width model with T$_{\rm{R,100}}$=1507~K. The red curves in both panels have the same $T_{e}$ and $n_{e}$ values that were taken from the best-fit CRRL optical depth model, see Fig.~\ref{f_38_crrl_comb}.}\label{f_38_bf}
\end{figure*}

\clearpage

\newpage

\begin{figure*}\vspace{0.5cm}
    \hspace{-1cm}\includegraphics[width=0.55\textwidth, angle=0]{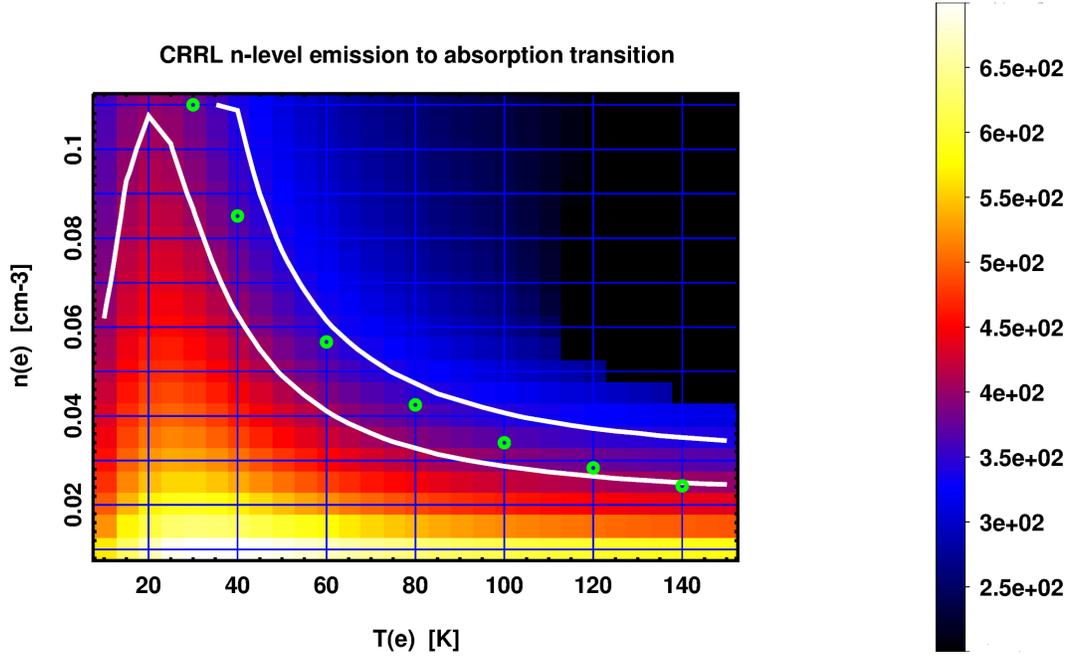}
  \vspace{0.5cm}
  \caption{CRRL n-level transition value as a function of T$_{e}$ and n$_{e}$ for our model grid. The white contours are drawn for n=340 and n=400. This n-range for the transition corresponds to the range for our measurements. The green circles highlight values of constant electron pressure p$_{e}$ = 0.04$\times$85 = 3.4~K~cm$^{-3}$. This shows that the transition n-level value is a good indicator of the electron pressure in the 20-140~K range.}\label{f_nt_emabsp}
\end{figure*}

\begin{figure*}\vspace{0.5cm}
\mbox{
    \includegraphics[width=0.5\textwidth, angle=90]{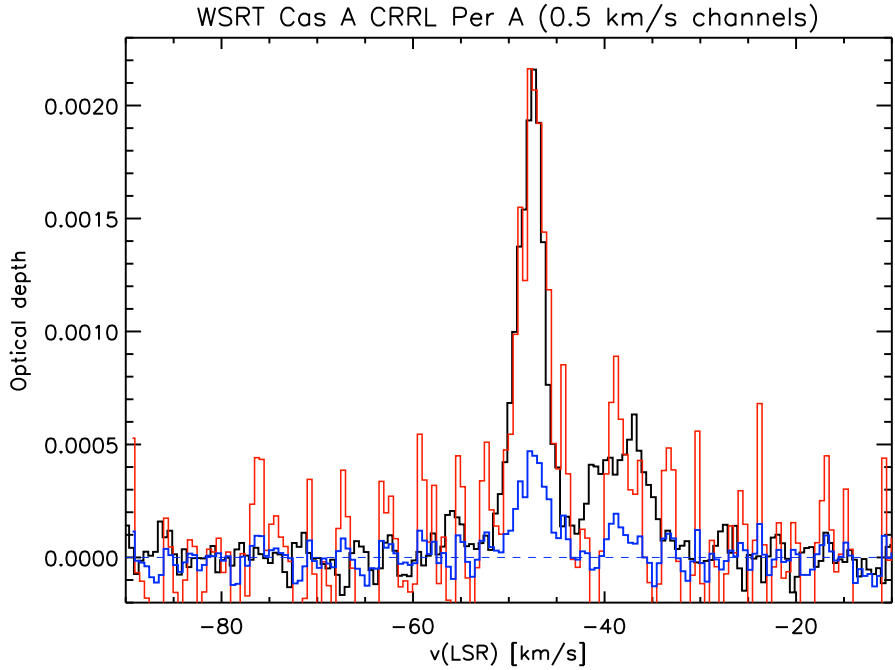}
}
  \vspace{0.5cm}
  \caption{Stacked WSRT P-band spectrum. The spectrum shows the CRRLs (black) and HRRLs (blue) overlaid. The HRRL spectrum is shifted by -149.4~km~s$^{-1}$ to match CRRL spectrum. This difference corresponds to the difference in rest frequencies for the HRRLs and CRRLs. In addition in red we show the HRRL spectrum scaled by a factor 4.6 to match the CRRL and HRRL peaks for the -47 component.}\label{f_wsrt_hcrrl_overlay}
\end{figure*}

\begin{figure*}\vspace{0.5cm}
    \includegraphics[width=0.45\textwidth, angle=90]{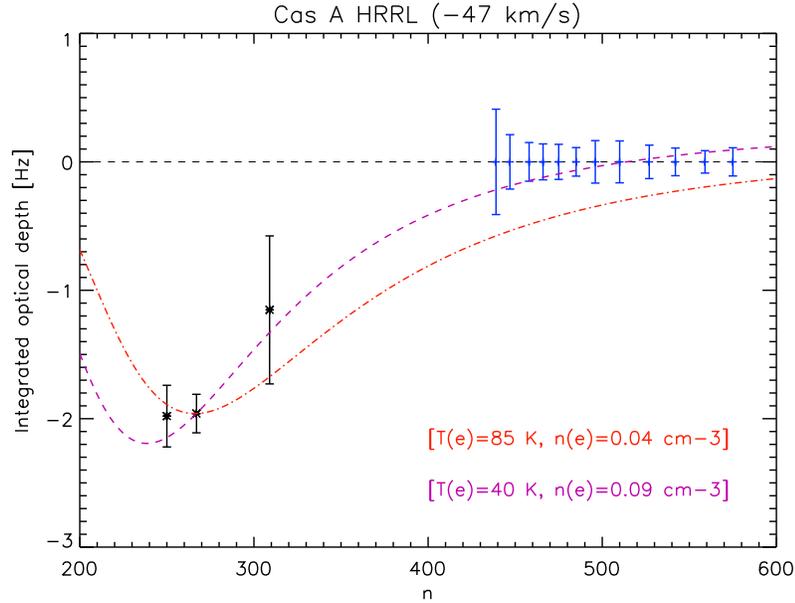}
  \vspace{0.5cm}
  \caption{Cas~A HRRL integrated optical depth vs. quantum number (n) for -47~km~s$^{-1}$ cloud. The three detections (black data points) are from this work, Oonk et al. (2015) and Sorochenko \& Smirnov (2010). The blue bars show the 3$\sigma$ HRRL upper limits obatined form our LBA measurements (Table~\ref{t_lba_hrrl} The red curve shows the scaled HRRL model upon assuming that the same physical parameters (T$_{e}$=85~K and n$_{e}$=0.04~cm$^{-3}$) as the best fit CRRL model. The scaling is done by normalizing this model at the WSRT data point and gives EM$_{H}$=0.0036~cm$^{-6}$~pc. If we set L$_{H}$=L$_{C}$ we find n$_{HII}$/n$_{CII}$=0.06. Alternatively L$_{H}$/L$_{C}$=0.06 if we set n$_{HII}$=n$_{CII}$. This model does not fit the LBA upper limits. Investigating the full HRRL grid we find that temperatures and densities in the range T$_{e}$=30-50~K and n$_{e}$=0.065-0.11~cm$^{-3}$ are able to fit both the measurements and the upper limits. The purple curve shows an example with T$_{e}$=40~K, n$_{e}$=0.09~cm$^{-3}$ and has EM$_{H}$=0.0014~cm$^{-6}$~pc.}\label{f_wsrt_hcrrl spectra}
\end{figure*}

\begin{figure*}\vspace{0.5cm}
    \includegraphics[width=0.45\textwidth, angle=90]{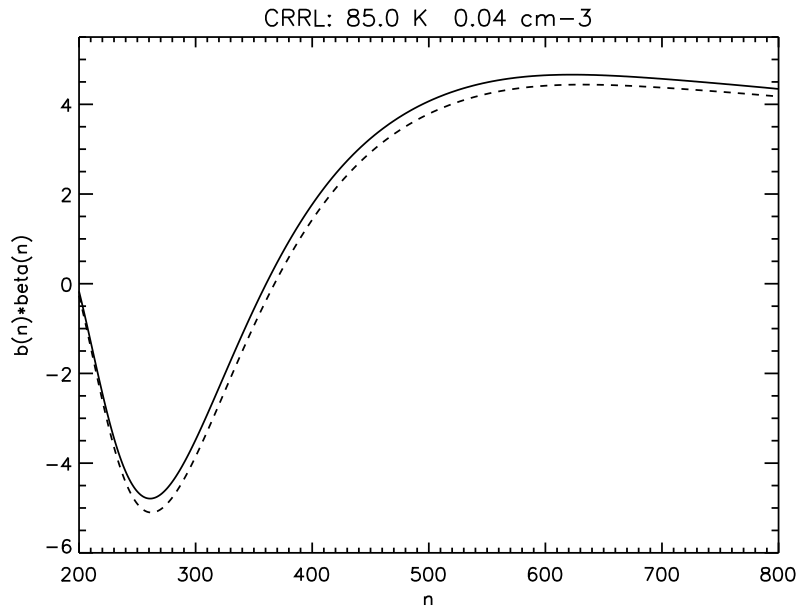}
  \vspace{0.5cm}
  \caption{Comparison of the b$_{n}\beta_{n}$ values for the ratio R of the $^{2}$P$_{3/2}$ over the $^{2}$P$_{1/2}$ population of carbon. The solid line gives the b$_{n}\beta_{n}$ value for S16a models that do not include collisions with molecular hydrogen. The dashed line gives the b$_{n}\beta_{n}$ value for S16a models that do include collisions with molecular hydrogen. Here we have used T$_{e}$=85~K, n$_{e}$=0.04~cm$^{-3}$, n$_{\rm{H_{2}}}$=n$_{\rm{HI}}$ and the collision rates from \citet{Ti85}.}\label{f_cmp_bbn_h2}
\end{figure*}


\bsp

\label{lastpage}

\end{document}